\documentclass[12pt]{article}

\usepackage{epsf,amsfonts,amsmath,amssymb,mathdots}
\usepackage{hyperref}
\usepackage{graphicx}
\usepackage{mathrsfs}
\usepackage{oldgerm}
\usepackage[usenames]{color}
\definecolor{darkblue}{rgb}{0,0,.8}
\definecolor{lightblue}{rgb}{.65,.95,1}
\definecolor{lightlightblue}{rgb}{.85,1,1}
\definecolor{rred}{rgb}{1,0,0}
\definecolor{purple}{rgb}{0.62,0.12,0.94}
\definecolor{dgreen}{rgb}{0,0.75,0}

\renewcommand{\appendix}[1]{
    \setcounter{section}{0}
    \setcounter{equation}{0}
    \renewcommand{\thesection}{\Alph{section}}
		\renewcommand{\theequation}{\Alph{section}-\arabic{equation}}
}

\newcommand\sgn {{\rm sgn}}
\renewcommand \l{\lambda}

\newcommand{\atopsum}[2]{\genfrac{}{}{0pt}{2}{#1}{#2}}
\newcommand{\Det}[1]{\,\mathrm{Det}\left(#1\right)}

\newcommand{\Tr}[1]{\,\mathrm{Tr}\:\left(#1\right)}
\newcommand{\Tro}[1]{\,\mathrm{Tr}_0\:\left(#1\right)}
\newcommand{\tro}[1]{\,\mathrm{tr}_0\:\left(#1\right)}
\newcommand{\tr}[1]{\,\mathrm{tr}\:\left(#1\right)}
\renewcommand{\cosh}[1]{\,\mathrm{cosh}\left(#1\right)}
\renewcommand{\sinh}[1]{\,\mathrm{sinh}\left(#1\right)}

\newcommand{\diag}{{\,\rm diag}}
\newcommand{\ovl}{\overline}
\def\tops#1#2{\texorpdfstring{#1}{#2}}
\newcommand{\td}[1]{{\tilde{#1}}}
\newcommand{\om}{\omega}

\newcommand{\ii}{i}
\newcommand{\e}[1]{{\rm e}^{#1}}

\newcommand{\D}{\hbox{d}}

\newcommand{\BK}[1]{\left\langle{#1}\right\rangle}
\renewcommand{\Re}{{\mathrm{Re}}}
\renewcommand{\Im}{{\mathrm{Im}}}
\renewcommand{\and}{{\qquad {\rm and} \qquad}}

\newcommand{\back}{\hspace{-4mm}}

\newcommand{\lbl}[1]{\label{#1}}
\newcommand{\beq}{\begin{equation}}
\newcommand{\eeq}{\end{equation}}
\newcommand{\bea}{\begin{eqnarray}}\newcommand{\eea}{\end{eqnarray}}
\newcommand{\beas}{\begin{eqnarray*}}\newcommand{\eeas}{\end{eqnarray*}}
\newcommand{\bacc}{\left\{ \begin{array}{l}}
\newcommand{\eacc}{\end{array}\right.}
\renewcommand{\thesection}{\arabic{section}}
\renewcommand{\theequation}{\arabic{section}-\arabic{equation}}
\makeatletter
\@addtoreset{equation}{section}
\makeatother
\newtheorem{theorem}{Theorem}[section]
\newtheorem{conjecture}{Conjecture}[section]
\newtheorem{remark}{Remark}[section]
\newtheorem{proposition}{Proposition}[section]
\newtheorem{lemma}{Lemma}[section]
\newtheorem{corollary}{Corollary}[section]
\newtheorem{definition}{Definition}[section]
\newcommand{\eq}[1]{eq.~(\ref{#1})}
\def\br{\begin{remark}\rm\small}
\def\er{\end{remark}}
\def\bt{\begin{theorem}}
\def\et{\end{theorem}}
\def\bd{\begin{definition}}
\def\ed{\end{definition}}
\def\bp{\begin{proposition}}
\def\ep{\end{proposition}}
\def\bl{\begin{lemma}}
\def\el{\end{lemma}}
\def\bc{\begin{corollary}}
\def\ec{\end{corollary}}
\def\gh{{\mathfrak h}}\def\gg{{\mathfrak g}}
\def\gb{{\mathfrak b}}
\def\gn{{\mathfrak n}}
\newcommand{\proof}[1]{{\noindent \it Proof:} {#1} $\square$}
\def\me{e}
\newcommand{\On}[1]{\mathrm{O}\left(#1\right)}
\newcommand{\Un}[1]{\mathrm{U}\left(#1\right)}
\newcommand{\Spn}[1]{\mathrm{Sp}\left(#1\right)}

\newcommand{\R}{{\mathbb R}}
\newcommand{\C}{{\mathbb C}}
\newcommand{\Z}{{\mathbb Z}}
\newcommand{\Q}{{\mathbb H}}
\renewcommand{\SS}{{\mathfrak S}}
\def\omit#1{{}}
\def\X{{X}}
\def\Y{{Y}}
\def\Zz{{Z}}
\def\x{{x}}
\def\y{{y}}
\def\z{{z}}

\def\CJ{{\cal J}}
\def\CW{{\cal W}}
\def\Q{{\mathbb H}}
\def\atopsum#1#2{\begin{subarray}{c}#1 \\ #2\end{subarray}}
\def\GO{\Omega}

\begin{document}

\def\today{\number\day\ \ifcase\month\or January \or February \or March \or
April \or May \or June \or July \or August \or September \or October \or
November \or December\fi\space\number\year}   
\def\aujour{\number\day\ \ifcase\month\or Janvier \or F\'evrier \or Mars \or
Avril \or Mai \or Juin \or Juillet \or Ao\^ut \or Septembre \or Octobre \or
Novembre \or D\'ecembre\fi\space\number\year}

\pagestyle{empty}
\hfill SPT-06/131, LPTHE-UPMC-06/116
\addtolength{\baselineskip}{0.20\baselineskip}
\addtolength{\baselineskip}{0.20\baselineskip}
\vspace{56pt}

\begin{center}
{\Large\bf 
Correlation Functions of Harish-Chandra Integrals \\ 

\vspace{12pt}
 over the Orthogonal 
and the Symplectic Groups
}

\vspace{\stretch{1}}

{\sl A.\ Prats Ferrer}\hspace*{0.05cm}${}^{\spadesuit\clubsuit}$\hspace*{0.05cm}\footnote{ E-mail: prats@lpthe.jussieu.fr }, 
{\sl B.\ Eynard}\hspace*{0.05cm}${}^{\clubsuit}$ \hspace*{0.05cm}\footnote{ E-mail: bertrand.eynard@cea.fr }, 
{\sl P. Di Francesco}\hspace*{0.05cm}${}^{\clubsuit}$\hspace*{0.05cm}\footnote{ E-mail: philippe@spht.saclay.cea.fr }, 
{\sl J.-B. Zuber}\hspace*{0.05cm}${}^{\spadesuit\clubsuit}$\hspace*{0.05cm}\footnote{ E-mail: zuber@lpthe.jussieu.fr }\\

\vspace{\stretch{2}}

{\bf Abstract}
\end{center}

The Harish-Chandra correlation functions, i.e. integrals over compact groups 
of invariant monomials $\prod \tr{X^{p_1} \GO Y^{q_1}\GO^\dagger X^{p_2}\cdots}$
with the weight $\exp \tr{X\GO Y\GO^\dagger}$ 
are computed for the orthogonal and symplectic groups.
We proceed in two steps. First, the integral over the compact 
group is recast into a Gaussian integral over 
strictly upper triangular complex matrices (with some additional symmetries), 
supplemented by a summation over the Weyl group. 
This result follows from the study of loop equations in an 
associated two-matrix integral and may be viewed as the adequate version of
Duistermaat-Heckman's theorem for our correlation function integrals.
Secondly, the Gaussian integration over 
triangular matrices is carried out and leads to compact 
determinantal expressions.

\vspace{\stretch{1}}

{\small \noindent $\clubsuit$ Service de Physique Th\'{e}orique de Saclay,
 CEA/DSM/SPhT - CNRS/SPM/URA 2306,\\
F-91191 Gif-sur-Yvette Cedex, France.\\
$\spadesuit$ LPTHE Tour 24-25 5\`eme \'etage, 
Universit\'e Pierre et Marie Curie--Paris6, CNRS UMR 7589, \\
4 Place Jussieu,
F 75252 Paris Cedex 5, France}

\newpage
\pagestyle{plain}
\setcounter{page}{1}

\def\pre#1{ ({\tt
#1})}
\def\htm#1{ {\tt #1})}

\def\He{Heckman} %
\def\DH{Duistermaat--\He}
\def\HC{Harish-Chandra}
\def\oh{{\frac{1}{2}}}
\def\smmat#1{\mbox{\small{\mbox{$\begin{matrix}#1\end{matrix}$}}}}
\def\smat#1{\mbox{\small{\mbox{$\begin{pmatrix}#1\end{pmatrix}$}}}}
\def\smaller{\small}\def\goth#1{{#1}} 
\def\gg{{\mathfrak g}} 
\def\gh{{\mathfrak h}} 
\def\I{{\cite{EPF05}}} %
\def\ve{\varepsilon}

\section{Introduction}

In the study of matrix integrals %
\cite{DFGZJ,BEC,GMGW,JPhysA03}%
, one frequently  encounters  integrals of the form 
\beq
Z^G=\int d\Omega\, \e{-\tr {\X \Omega \Y \Omega^{-1}} }
\label{a.a}
\eeq
over  some compact matrix group $G$, with $\X$ and $\Y$ two given matrices.
By the left and right invariance 
of the Haar measure $  d\Omega$, this integral is invariant under
\beq
\X \to \Omega_1 \X\Omega_1^{-1}\ ,\Y\to \Omega_2 \Y\Omega_2^{-1}\ ,
\label{a.b}
\eeq
and is thus insensitive to the choice of the representative of the orbits 
of $\X$ and of $\Y$ under the (adjoint) action of the group. 
This enables one to bring the matrices $\X$ and $\Y$ to some canonical
form, as we shall see below.

The case of reference is the so-called Harish-Chandra--{Itzykson--Zuber}
(HCIZ) integral~\cite{HC57,IZ80}, 
where the integration is performed over the unitary group
$\Omega\in$ U$(n)$ and  $\X$ and $\Y$ are two (anti)Hermitian matrices.
By the previous argument, we may with no loss of generality assume 
that $\X$ and $\Y$ are two diagonal (anti)Hermitian matrices of size $n$,
 $\X=\diag(\X_i)_{i=1,\cdots,n}$, 
and likewise for $\Y$. 
\beq
\int_{U(n)} d\Omega\, \e{ -\tr {\X \Omega \Y \Omega^\dagger} }= {\rm const.} 
\frac{\left(\det e^{-\X_i \Y_j}\right)_{1\le i,j\le n}}{\Delta(\X)
\Delta(\Y)}=  {\rm const.} \sum_{\pi\in \SS_n} \epsilon_\pi
\frac{\e{-\tr{ \X \Y^\pi} }}{\Delta(\X)\Delta(\Y) }
\ ,
\label{hciz}
\eeq
where 
\beq
\Delta(\X)=\prod_{i<j} (\X_i-\X_j) \label{VdM} 
\eeq
 is the Vandermonde determinant 
of the eigenvalues $\X_i$ of $\X$ and likewise for 
$\Delta(\Y)$;\ $\Y^\pi=\diag(\Y_{\pi(i)})_{i=1,\cdots,n}$. 
 Further examples are 
provided by more general Harish-Chandra-type integrals, where $\X$ and $\Y$ 
live in (a matrix representation of) the Lie algebra $\gg$ of $G$~\cite{HC57}.
For example, $G=O(n)$, $\X$ and $\Y$  {\it anti}symmetric real
matrices of size $n$.
In all these cases, explicit formulae are known, following from 
a diversity of methods, see below and 
for example \cite{ZJZ02}  for a review and references.

It is desirable to extend these formulae  to 
the ``correlation functions'' of the integral (\ref{a.a}), i.e. 
integrals of the form
\beq
\int d\Omega\, F(\X,  \Omega \Y \Omega^{-1})\,
\e{ -\tr {\X \Omega \Y \Omega^{-1}} }
\label{a.c}
\eeq
with $F$ invariant under (\ref{a.b}). Such correlation functions
 provide a deeper probe of these
integrals and, in a physical context,  give often access to  quantities
of interest. %
They also act as generating functions of integrals of the form 
\beq
\int d\Omega\, \GO_{i_1j_1}\GO_{i_2j_2}\cdots \GO_{i_pj_p}
\GO^{-1}_{k_1l_1}\cdots \GO^{-1}_{k_pl_p}\, 
\e{ -\tr {\X \Omega \Y \Omega^{-1}} }
\label{a.cprim}
\eeq
{\it i.e.} of {\it moments}  of $\GO$ and $\GO^\dagger$ with 
the \HC\ weight.


Kogan {\it et al} \cite{Koganetal},
Morozov \cite{Mor}, and Shatashvili \cite{Shat}
 made some attempts at computing correlation functions of the
HCIZ integral, i.e. for the unitary group.
Morozov's formula may be recast into a very compact expression~\cite{Eynardmor}
but it is only good at computing correlators quadratic  in $\Omega$, 
whilst Shatashvili's formula allows in principle to
compute all correlators, but is not of  easy use.  In paper \cite{EPF05} 
two of us have shown how to recast
the computation of correlators of the type (\ref{a.c})  for the unitary group
into a totally different setting. The method consists in two steps. In step
one, the integral is rewritten as a sum of  integrals over 
{\it upper triangular complex} matrices. 
The formulae of \I\ are in some sense a generalization of Morozov's, and
allow to compute all correlators for the $U(n)$ group in a  very simple
formula.
The initial observation was that a Gaussian integral and its polynomial
moments in an hyperplane of dimension $d$ of $M_n(\C)^2$ does not depend
on the hyperplane (up to multiplication by a constant jacobian),
and thus, one can use either the hyperplane $M_1=M_1^\dagger,
M_2=M_2^\dagger$ or the hyperplane $M_1=M_2^\dagger$, which have the same
dimension.
In the first hyperplane, after diagonalisation of $M_1$ and $M_2$, the
integral separates into a radial part and an angular part proportional to
moments of U($n$) with the HCIZ measure,
while in the second hyperplane, after Schur decomposition of $M_1$, the
integral separates into a radial part (identical to the  one in the first
hyperplane), a trivial angular part, and a Gaussian integral over
triangular matrices. 
As a result, the authors of \I\ were able to identify all moments of the HCIZ
integral with a Gaussian integral over complex strictly {upper} triangular matrices.
The formula reads:
\bea
\int_{U(n)}\!\!\! dU F(X,U Y U^\dagger)\, \e{-\tr{ X U Y U^\dagger}} =
\qquad\qquad\qquad\qquad \nonumber\\
=\frac{c_n}{\Delta(X)\Delta(Y)}\,\sum_{\sigma\in\Sigma_n}\, (-1)^\sigma\,\e{-\tr{ X
Y_\sigma}}\,\int_{T_n} dT F(X+T,Y_\sigma +T^\dagger)\,\e{-\tr{T T^\dagger}}\ ,
\eea
{for $X$ and $Y$ two real diagonal matrices.}

In a second step, the Gaussian triangular integrals in the right hand side
were computed in \I, using Wick's theorem.
The computation can be performed explicitely due to the nilpotent
properties of $T$, which ensures that most Wick's pairings actualy vanish.
The computation is most easily done by recursion on the size $n$ of the
matrix, i.e. by integrating out the last column of $T$.
An appropriate basis of all possible polynomial moments $F$ was introduced in
\I, and in that basis, it was found that:
\beq
\int_{T_n} dT F(X+T,Y_\sigma +T^\dagger)\,\e{-\tr{ T T^\dagger}}
= \prod_{i=1}^n {\cal{M}}(X_i,Y_{\sigma(i)})\ ,
\eeq
where $F$ and ${\cal M}(x,y)$ are matrices of some size $R!$.
The universal matrices ${\cal M}(x,y)$ have many remarkable 
properties, in particular they commute with one another 
\beq
[{\cal M}(x,y),{\cal M}(x',y')]=0 \ .
\eeq
The purpose of the present article is to generalize the computations of \I\
to other classical Lie groups. 
  Specifically, we address the computation
of (\ref{a.c}) for $G$ the orthogonal group O($n$) or the symplectic group
Sp($2n$), with $\X$ and $\Y$ in the Lie algebra of those groups.
First, we relate H-C correlators over those groups to Gaussian integrals over
some set of triangular matrices, then we compute the latter Gaussian
triangular integrals 
using an appropriate basis, and finally we find that the result can again be
written as products of the same matrices ${\cal M}(x,y)$ which appeared
for U($n$). Our main results are stated in Theorems 
\ref{th:MainOrt}
, \ref{th:MainSp} 
and \ref{the:matrixM} 
below and in section \ref{FinalExp}, {(\eq{finalOrte}, 
\eq{finalOrto} and  \eq{finalSp})}.
{The results for the unitary, orthogonal and symplectic groups
may be expressed in a unified way, in terms of the 
Weyl group $\cal{W}$, Borel subalgebra $\gb$ and positive roots $\alpha$, 
in the following}
\def\Xp{{X}}\def\Yp{{Y}}
\vskip -3 ex
\begin{theorem}
\bea\label{unifiintro}
&&\int_{{G}}\D \Omega\,  
{F(\X^a ,\Omega \Y^a \Omega^{-1})\,\, \e{ -\tr {\X^a \Omega \Y^a 
\Omega^{-1}} }}
\nonumber\\
&&\phantom{\int_{{G}}\D \Omega}= c \sum_{w\in \CW} 
\frac{\e{{+}\tr{\Xp w(\Yp)}}}{\prod_{\alpha>0}\alpha(\Xp) \alpha(w(\Yp))}  
\int_{\gn_+=[\gb,\gb]}\D T F(i\Xp+T,iw(\Yp)+T^\dagger) \e{-\tr{TT^\dagger}}
\eea
\end{theorem}
{for any polynomial function $F$ and for some $F$-independent constant $c$; 
here 
 $X^a$ and $Y^a$ are taken in a Cartan subalgebra, 
and should thus be thought of as {\it anti}-Hermitian matrices with extra symmetries
depending on $G$, (see sect.2 for more details), while $i\Xp$ 
and $i\Yp$ are the purely imaginary diagonal matrices with the same 
eigenvalues as
 $X^a$ and $Y^a$. In that form, the derived Borel subalgebra $\gn_+$
is made of complex strictly upper triangular matrices, also subject 
to symmetries. 
It is thus natural 
to expect these results to extend to any simple compact group $G$, 
see Conjecture 8.1.


It is our hope that these results should provide a new insight 
on the common features of all these integrals. 

Our paper is organized as follows. 
In section \ref{Overview}, we review the known results by Harish-Chandra 
and Duistermaat-\He\ and set up the notations. 
In section \ref{analytical}, we show how Gaussian integrals over two matrices
with reality properties, either antisymmetric real or antiselfdual
real quaternionic, may be equated to Gaussian integrals over one 
complex matrix constrained by some symmetry requirements. This is 
established by use of loop equations, on which we provide details
in Appendix \ref{app:I}. Section \ref{HSOrt} and \ref{HSSymp} then show how separation of
the angular variables by diagonalization or Schur decomposition
 leads us to the desired integrals, which are thus related to 
integrals over complex triangular matrices (with additional 
symmetry requirements). Section \ref{TI}, supplemented by 
Appendices \ref{app:matrixM} and \ref{app:bijection},  is devoted to the actual computation of these
integrals over triangular matrices, by means of a recursive method
using a diagrammatic method. 
The final expressions are displayed in section \ref{FinalExp}, while section 
\ref{remarks} contains our concluding remarks and suggestions of further 
directions worth exploring. Two other appendices make our notations explicit
on quaternions (App. \ref{app:Q}) or  give additional details  on the 
calculation of some Jacobians (App.~\ref{app:JACS}). 


\section{Overview of known results}\label{Overview}
\subsection{The Harish-Chandra theorem \texorpdfstring{\cite{HC57}}{}}\label{HCthm}

Following Harish-Chandra, for $G$   a compact connected Lie group, we
denote by 
${\rm Ad}$ the adjoint action of $G$  on its Lie algebra $\gg$,
by $(X,Y)$ the nondegenerate invariant inner product 
on $\gg$, which we take to be 
the trace of the product $XY$ in our matrix representation, 
 by $\gh\subset \gg$ the Cartan subalgebra, and
by $\alpha(X)$ the linear action of a root $\alpha$ on $X\in \gh$.  
If $\X$ and $\Y \in \gh$ 
\beq
\Delta(\X)\Delta(\Y)\int_G d \Omega \,\exp -(\X, {\rm Ad}(\Omega) \Y)=
\frac{ <\!\!\pi,\pi\!\!>}{|\CW|} \sum_{w\in \CW} \epsilon(w) \exp -( \X, w(\Y))\ ,
\label{harish}
\eeq
where $w$ is summed over the Weyl group $\CW$, 
$\epsilon(w)=(-1)^{\l(w)}$, $\l(w)$ is 
the number of reflections generating $w$, 
 and 
\beq\label{generVdM}
{ \Delta(X)=\prod_{\alpha >0} \alpha(X) \ , }
\eeq
a product over the positive roots of $\gg$. \
$\Delta(X)$ may be called a {\it generalized Vandermonde} determinant, since 
in the case of $U(n)$,  it reduces to  \eq{VdM} 
(and  (\ref{harish})
 reduces to (\ref{hciz})), while the expressions for the orthogonal
and symplectic groups will be given below. 
The constant $<\!\!\pi,\pi\!\!>$ in  the right hand side of (\ref{harish}) 
is computed as follows. Write all the positive roots in an orthonormal 
basis $\ve_i$ of root space, $i=1,\cdots\ell$, with 
$\ell$ the rank of $\gg$. Regard $\pi=\prod_{\alpha>0} \alpha$  
as a polynomial in the positive roots and expand it 
on symmetrized tensor products of the $\ve$, 
$\pi = \sum_{m_i\ge 0} p(m_1,m_2,\cdots, m_\ell) \ve_1^{m_1}
\cdots\ve_\ell^{m_\ell}$. Then $<\!\!\pi,\pi\!\!>= \sum_{m_i\ge 0} 
( p(m_1,m_2,\cdots, m_\ell))^2 \prod_{i=1}^\ell m_i!$. For $\gg=su(n)$, 
one finds $<\!\!\pi,\pi\!\!>= \prod_{j=1}^n j!$, while the expression for
the other classical groups will be given below.
   

\subsection{The Duistermaat-Heckman theorem \texorpdfstring{\cite{DH}}{}}\label{DHthm}

The Duistermaat-\He\ theorem states that if ${\cal M}$ is a symplectic manifold,
invariant under a U$(1)$ flow generated by a Hamiltonian $H$, then for
the integral $\int_{{\cal M}} e^{iHt}$, 
the stationary phase approximation is exact: the sum 
of the values of the integrand at its critical points, 
weighted by the  Gaussian (`one-loop') fluctuations  around them, 
gives the exact integral. 

For $\X$ and $\Y$ in the Lie algebra, consider the integral \eq{a.a} 
$$ Z=\int_G d\Omega\, \e{ -\tr{\X \Omega \Y \Omega^{-1}}}\ .$$
In such integrals, we first pick a convenient representative of the 
orbits of elements of the Lie algebra under the adjoint action of $G$. 
A theorem of Cartan asserts that any element of the Lie algebra
is the conjugate (under the adjoint action) of an element of
the Cartan algebra~\cite{Bu04}. 
Thanks to this theorem  and to the left and right invariance of 
the Haar measure $d\Omega$, one may always assume that $\X$ and $\Y$ lie in the
Cartan subalgebra $\gh$. This assumption matches that of
 Harish-Chandra's theorem. Moreover
the integration is then reduced to $G/T$, $T$ a maximal abelian subgroup
(Cartan torus) commuting with $\Y$, or alternatively, the integration 
is carried out on the orbit of $\Y$ under the action of
this quotient. This is a 
symplectic manifold, to which \DH's theorem applies \cite{others}.

We thus first look for the critical points of the `action' 
$\tr {\X \Omega \Y \Omega^{-1}}$ when $\Omega \in G/T$. In other words, 
we look for solutions in $\Omega$  of   
$$\delta\tr {\X \Omega \Y \Omega^{-1}}= \tr{\delta \Omega \Omega^{-1} [\X, \Omega \Y \Omega^{-1}]}=0\ ,$$
Since $A:= \delta \Omega \Omega^{-1}$ is 
arbitrary in $ \gg\backslash\gh$, this implies that the component
of $[\X, \Omega \Y \Omega^{-1}]$ in  $ \gg\backslash\gh$ vanishes. 
 On the other hand,  the component of 
$[\X, \Omega \Y \Omega^{-1}]$ in  $\gh$ also vanishes, since  
if $B:= [\X, \Omega \Y \Omega^{-1}]$ were in $ \gh$, then 
$\tr B^2=\tr{ B\, [\X, \Omega \Y \Omega^{-1}]}=\tr{ [B,\X] \Omega \Y \Omega^{-1}}=0$
since $\X$ and $B\in \gh$ commute. We thus conclude that $B=0$, {\it i.e.} that 
\beq
[\X, \Omega \Y \Omega^{-1}]=0\ .
\label{critical}
\eeq
 The critical points are thus 
the points $\Omega_c \in G/T$ such that  (\ref{critical})
is satisfied, which for generic $\X\in \gh$ means 
$\Omega_c \Y \Omega_c^{-1}\in \gh$, {\it i.e.} $\Omega_c$
takes the element $\Y\in \gh$ 
to an element $ \Omega_c \Y \Omega_c^{-1}\in \gh$. 
If we denote by $\CW$ the  normalizer of the Cartan torus $T$
quotiented by $T$, the previous discussion has just proved 
that  the critical points $\Omega_c$ of the action
are in one-to-one correspondence with elements  $w$ of the group $\CW$.
The group $\CW$ is known to be the Weyl group of $G$ (\cite{Bu04}, 
Prop. 15.8). 
In the sequel, we denote $\Y^w=\Omega_c \Y \Omega_c^{-1}$ for $w\in \CW$.

At this stage, \DH's theorem thus tells us that
\beq
\label{DHun}
\int_G d\Omega \e{-\tr{X\Omega Y \Omega^{-1}}}
=\sum_{w\in \CW}  \int_{\gg\backslash \gh} dA \, \e{-[\tr{X e^AY^w e^{-A}}]_2}
\eeq
where $[\cdots]_2$ means that we retain only up to the quadratic terms in 
the expansion in powers of $A\in \gg\backslash \gh$.

The final step in the application of \DH\ theorem is thus 
to compute the second order variation 
of the action at one of these critical points. 
For $\Omega=e^A$, $A\in \gg\backslash \gh$, 
\beq
- \tr {\X e^A \Y^w e^{-A}}=- \tr {\X  \Y^w} 
+\oh \tr{[A,\X] [A,\Y^w]}+{\rm o}(A^2)\ .
\eeq
We then have to carry out the Gaussian integration 
$$\int d^d A\,\,  \e{\,\textstyle{\oh \tr{[A,\X] [A,\Y^w]}}}$$
over the $d$-dimensional vector $A$. This (real) 
dimension $d=\dim \gg-\dim \gh=2r $ is even 
and equal to the number of roots of $G$.
We now expand $A$, $\X$ and $\Y^w$ in the standard basis
$A=\sum A_\alpha E_\alpha$, $\X=\sum_i \X_i H_i$ and $\Y^w=\sum_i \Y^w_i H_i$ 
and use the standard commutation relations and traces 
$\tr{H_i H_j}=\delta_{ij}$, $\tr{E_\alpha E_\beta}=\delta_{\alpha+\beta,0}$
to get 
$$\tr{[A,\X] [A,\Y^w]}= \sum_{\alpha, \beta,i,j}
A_\alpha A_\beta \X_i \Y^w_j\alpha^{(i)} \beta^{(j)} 
\tr{E_\alpha E_\beta}=-\sum_{i,j,\alpha}  
 A_\alpha A_{-\alpha} \sum_i \X_i \alpha^{(i)} \sum_j  \Y^w_j \alpha^{(j)} $$
{\it i.e.} 
$$ \tr {[A,\X] [A,\Y^w]}=-\sum_{\alpha} 
A_\alpha\, A_{-\alpha} \alpha(\X) \alpha(\Y^w)
 $$
with a sum over positive and negative roots. This quadratic form has a 
signature $(+^r, -^r)$, and upon a suitable contour rotation,  
the integration over $A$ yields 
$$\int d^d A\,\, \e{\, \textstyle{\oh \tr{[A,\X] [A,\Y^w]}}}
= \frac{\rm constant}{\prod_{\alpha>0} 
 \alpha(\X) \alpha(\Y^w)
 }\ .
$$
Putting everything together, we see that we have reconstructed the 
Harish-Chandra formula. 


\subsection{Explicit formulae}\label{Iexplicit}

It is of course a good exercise to repeat these steps and to write explicit
expressions for each of the 
classical groups U$(n)$, O$(n)$ and Sp$(n)$. The result for U$(n)$
is well known and has been recalled above. 
We shall content ourselves 
in giving the final result for the two latter cases.
In the orthogonal case  O$(n)$, we have to distinguish the $n=2m$ and 
$n=2m+1$ cases. 
In the even case, $G={\rm O}(2m)$,
 we take the $\X$ and $\Y$ matrices in the block diagonal form
\beq\label{blodiagev}
 \X=\diag\left(\smat{0 & \X_j\cr -\X_j & 0}_{j=1,\cdots,m}\right),
\qquad \Y=\diag\left(\smat{0 & \Y_j\cr -\Y_j & 0}_{j=1,\cdots,m}\right)\ .
\eeq
Then the critical points $\Omega_c$ are the product of a permutation $\tau$ 
of the $m$ blocks of $\Y$ by a diagonal matrix of signs $\diag(t_1,\dots,t_m)$, where $t_j=\pm \mathrm{Id}_{2}$, or in other   words,  the set 
 $\CW$ is  $\SS_m\times \Z_2^m$. Note that $\CW$ is larger than the ordinary
Weyl group of $D_m=so(2m)$ type, which is $W= \SS_m\times \Z_2^{m-1}$: this is
because changing the sign of {\it one} $Y_j$, say $Y_1$, is performed by 
conjugation by a matrix made of $2\times 2$ blocks,  
$\Omega=\diag(\sigma_1,\mathrm{Id}_{2},\cdots,\mathrm{Id}_{2})$, which is in 
O($2m$) but {\it not} in SO($2m$). As a result, only an even 
number of signs may be changed in the latter case, whence the factor 
$\Z_2^{m-1}$ in the Weyl group.  For the O($2m$) group that we consider here, 
we  thus have
\bea 
\!\!\!\!\!\!\!\!\!
Z^{(\On{2m})}&=& {\rm const.}
\sum_{w=(\tau,\{t_i\})} \frac{\e{2 \sum_i \X_i \Y^w_i}}
{\Delta(\X)\Delta(\Y^{w})}
= {\rm const.} \sum_{\tau\in \mathfrak{S}_m}
\varepsilon_\tau 
\frac{\prod_i \left(\e{2 \X_i \Y_{\tau_i}}+\e{-2 \X_i \Y_{\tau_i}}\right)
}{ \Delta(\X)\Delta(\Y)}\nonumber \\
&=&{\rm const.} \frac{\det (2 \cosh{2 \X_i \Y_j})_{i,j=1,\cdots,m} }{\Delta(\X)\Delta(\Y)}\ ,
\eea
where $\varepsilon_\tau$ is the signature of the permutation $\tau$,
and
\beq\label{VdMorthev}
\Delta(\X)= \prod_{i<j} (\X_i^2-\X_j^2)\ .
\eeq 

For $n=2m+1$, the calculation proceeds along the same line. 
We write 
\beq\label{blodiagod}
 \X=\diag\left(\smat{0 & \X_j\cr -\X_j & 0}_{j=1,\cdots,m}, 0\right)
\qquad \Y=\diag\left(\smat{0 & \Y_j\cr -\Y_j & 0}_{j=1,\cdots,m},0\right)\ .
\eeq
The critical points $\Omega_c$ are again
the product of a permutation $\tau$ of the $m$ blocks of $B$ 
by a matrix of signs, $t_j=\pm \mathrm{Id}_2$, $j=1,\cdots, m$, and
\bea\label{blocdiago}
Z^{(\On{2m+1})}&=&  {\rm const.}
\sum_{w=(\tau,\{t_i\})} \frac{\e{2 \sum_i \X_i \Y^w_i}}{ \Delta(\X)\Delta(\Y^w)}
= {\rm const.}\sum_{\tau\in \mathfrak{S}_m} \varepsilon_\tau \frac{\prod_i\left(
\e{2  \X_i \Y_{\tau_i}}-\e{-2 \X_i \Y_{\tau_i}}\right)} 
{ \Delta(\X)\Delta(\Y)}\nonumber
\\ &=&
{\rm const.} \frac{\det (2 \sinh{ 2 \X_i \Y_j})_{i,j=1,\cdots,m} 
}{\Delta(\X)\Delta(\Y)}\ .
\eea
with now 
\beq\label{VdMorthod}
\Delta(\X)= \prod_{i<j} (\X_i^2-\X_j^2)\,\prod_{i=1}^m \X_i  \ .
\eeq

Finally for the symplectic group Sp$(2m)$, it is convenient to use
quaternionic notations for matrices, {\it i.e.} 
to regard the matrix elements as
quaternions\footnote{We refer the reader to appendix \ref{app:Q} 
 for more details on our notations on quaternions.}, with coordinates 
in the standard quaternion basis, 
$\me_0^2=1\,;\quad \me_i^2=-1\,,\quad i=1,\dots,3\,,\quad \me_1 \me_2=\me_3$\,; 
 alternatively, the matrices may be regarded as made of $2\times 2$ blocks 
written in terms of the identity matrix ${\rm Id}_2$ and of Pauli
 matrices $\vec \sigma$ (with the identification
$\me_0 \leftrightarrow {\rm Id}_2\,,\ 
\me_j \leftrightarrow -i \sigma_j$, $j=1,2,3$). 
The Lie algebra $C_m$ of $\Spn{2m}$ is thus generated by quaternionic real 
and antihermitean  (also called antiselfdual quaternionic real, see appendix
\ref{app:Q})
$m\times m$ matrices $\X$, $\X_{ij}= \X^0_{ij} 
+\vec \X_{ij}\, \vec \me$, $\X^\alpha_{ij}\in \R$, 
$\X=-\X^\dagger$. 
Consider the Cartan algebra generated by the $m$ 
matrices $\diag(\X_j \me_2)$, 
$j=1,\cdots m$. We thus take our matrices $\X$ and $\Y$ of that form
\beq\label{CartanQ}
\X=\diag(\X_j \me_2)_{j=1,\cdots,m}\ ,\qquad  \Y=\diag( \Y_j \me_2)_{j=1,\cdots,m}\ .
\eeq
Then, the critical values $\Omega_c$ are again 
the product of a permutation $\tau$ of the $m$ blocks of $\Y$ 
by a diagonal matrix of signs, $t_j=\pm 1$. This leads to 
\bea
\!\!\!\!\!\!\!\!\!\!\!\!\!\!\!\!\!\!\!
Z^{(\Spn{2m})}&=&{\rm const.} \sum_{w=(\tau,\{t_i\})} 
\frac{\e{2 \sum_i \X_i \Y^w_i}}{\Delta(\X)\Delta(\Y^{w})}
={\rm const.} \sum_{\tau\in \SS_m} \varepsilon_\tau \frac{\prod_i\left(
\e{2  \X_i \Y_{\tau_i}}-\e{-2 \X_i \Y_{\tau_i}}\right)} 
{\Delta(\X)\Delta(\Y) }
\\ &=&
{\rm const.} 
\frac{\det (2 \sinh{2 \X_i \Y_j})_{i,j=1,\cdots,m}}
{\Delta(\X)\Delta(\Y) }
\ ,
\eea
with the same expression for $\Delta(\X)$ as in (\ref{VdMorthod}).
Thus $Z^{(\Spn{2m})}$ has  the same form as the integral over O$(2m+1)$.



\subsection{List of notations}
For the sake of the reader, we list hereafter the non standard notations
in the order they appear in the text.
\beas
\Delta(X) & {\rm Vandermonde\ determinant\ and\ generalizations\ } & 
(\ref{VdM}),(\ref{generVdM}),(\ref{VdMorthev}),(\ref{VdMorthod}) \\
 {\cal A}_n &  n\times n\ {\rm real\ antisymmetric\ matrices} &  (\ref{MEASA})\\
J & {\rm\ antidiagonal\ identity\ matrix} & (\ref{defJ})\\
J{\cal A}_n &  n\times n\ J-{\rm antisymmetric\ complex\ matrices} & {\rm sect.}\ \ref{Jantisym}\ {\rm and}\ (\ref{MEASJ}) \\ 
Q{\cal A}_m &  m\times m\ {\rm real\ quaternionic\ antiselfdual\ matrices} & {\rm sect.}\ \ref{rquaterad}\ {\rm and}\ (\ref{MEASQ})\\
\td{J}  & {\rm antidiagonal\ symplectic\ matrix }& (\ref{deftdJ})\\
\td{J}{\cal A}_{2m} &  2m\times 2m\ \td{J}-{\rm antisymmetric\ complex\ matrices} & {\rm sect.}\ \ref{complJ}\ {\rm and}\ (\ref{MEAStdJ}) \\ 
D^a_n(\R)  &   {\rm  real\ } 2\times2\ {\rm block-diagonal\ antisymmetric}\ n\times n\ {\rm matrices} & {\rm sect.}\ \ref{SEC1}\\
 M_n(\C) & n\times n\ {\rm complex\ matrices} & {\rm sect.}\ \ref{SEC2}\\
 T_n & n\times n\ {\rm strictly\ upper\ triangular\ complex\ matrices} 
 & {\rm sect.}\ \ref{SEC2}\\
 D_n(\C) &  n\times  n\ {\rm complex\ diagonal\ matrices} &{\rm sect.}\  \ref{SEC2}\\
{\rm U}^J(n) & {\rm twisted\ orthogonal\ matrices} &  (\ref{twistorth})
\\
T^J_n &  \begin{array}{c} n\times n \ {\rm  strictly\ upper\ triangular\ } \\
                          J{\rm-antisymmetric\ complex\ matrices} \end{array}  &  (\ref{TnJ})\\
 D_n^J(\C)\,,\   D_n^J(\R) 
&  n\times  n\ {\rm complex,\ resp.\ real,\ }J{\rm-antisymmetric\  
diagonal\ matrices} & %
 (\ref{Jantisdiag})\,,\ (\ref{EQ3})\\
D^{aR}_{m}(\Q)  &  \begin{array}{c} \ m\times m\ {\rm  real\  quaternionic \ 
diagonal\ matrices} \\ {\rm with\ elements\ proportional\ to\ } e_2 \end{array} & {\rm sect.}\ \ref{sec:DiagQ}\\
{\rm U}^{\td{J}}(2m) & {\rm twisted\ symplectic\ matrices} &  (\ref{twsympl})
\\ %
T^{\td{J}}_{2m} &  \begin{array}{c} 2m\times 2m \ {\rm  strictly\ upper\ triangular\ } \\
 \td{J}{\rm-antisymmetric\ complex\ matrices} \end{array}&  (\ref{TtdJ})
\eeas


\section{Analytical continuation for two-matrix integrals.}\label{analytical}

In this section, we follow the same strategy as used in \I\ for the unitary 
group: the integrals of interest (\ref{a.c}) are regarded as the 
 ``angular part'' of two-matrix integrals over the   classical  Lie algebras $so(2m+1)$, 
$so(2m)$ and $sp(2m)$, and the latter
 may be analytically continued to integrals over 
complex matrices with  special symmetries.

\subsection{Real antisymmetric two matrix integral and complex \texorpdfstring{$J$}{J}-antisymmetric matrix integral.}

\subsubsection{Real antisymmetric two matrix integral.}\label{rantisym}

Consider first the set  ${\cal A}_n$ 
of $n\times n$ real antisymmetric matrices and
consider the measure on ${\cal A}_n\times {\cal A}_n$
\bea\label{MEASA}
\D \mu(A_1,A_2)&=&\e{-\tr{\frac{\alpha_1}{2}A_1^2+\frac{\alpha_2}{2}A_2^2+\gamma A_1 A_2}} \D A_1 \D A_2 \nonumber\\
\D A_k&=&\prod_{1\le i<j\le n} \D \left(A_k\right)_{i,j}\, ;\quad \, k=1,2
\eea
Then the real antisymmetric two matrix partition function 
and the associated correlation functions are defined as
\bea\label{2RApf}
Z_{2RA}&=&\int_{{\cal A}_n\times{\cal A}_n}\D\mu(A_1,A_2)\nonumber
\eea
\bea\label{2RAcf}
\BK{F(A_1,A_2)}_{2RA}&=&\frac{1}{Z_{2RA}}\int_{{\cal A}_n\times{\cal A}_n} F(A_1,A_2)\D\mu(A_1,A_2) \nonumber
\eea 
using the measure $\D\mu(A_1,A_2)$ given in \eq{MEASA}.
The partition function is the product of $n(n-1)/2$
 uncoupled and equal integrals
over the pairs of matrix elements $((A_1)_{ij},(A_2)_{ij})$, $i<j$. 
Each integral, of the form $\int \D x\, \D y\, \exp \{ (x,y) Q (x,y)^T\}$,
$Q=\left(\begin{matrix} \alpha_1 & \gamma\\ \gamma & \alpha_2\end{matrix}\right)$, 
is absolutely convergent if the real part of the quadratic form 
$(x,y) Q (x,y)^T:=
\alpha_1 x^2 + \alpha_2 y^2 +2 \gamma x y$ is negative definite, 
which holds true if 
$\Re\, \alpha_1 \Re\, \alpha_2 -(\Re\,\gamma)^2 >0$ and $\Re\,\alpha_1$, 
$\Re\,\alpha_2 <0$ when $x$ and $y$ are integrated over the real line.
Then the partition function is  easily computed to be
\bea
Z_{2RA}&=&\left(\frac{\pi}{\sqrt{\delta}}\right)^{\frac{n(n-1)}{2}}
\eea
where $\delta=\alpha_1\alpha_2-\gamma^2$.
Likewise, for polynomial $F(A_1,A_2)$, 
the correlation function $\BK{F}_{2RA}$ is by Wick theorem
a {\it polynomial} in the matrix elements of the propagator $Q^{-1}$, 
namely $\frac{\alpha_1}{\delta}$ , $\frac{\alpha_2}{\delta}$  and 
$\frac{\gamma}{\delta}$. 


\subsubsection{Complex \texorpdfstring{$J$}{J}-antisymmetric matrix integral}\label{Jantisym}
\vskip10pt
Define now the $n\times n$ antidiagonal matrix $J=J^{-1}$
\beq\label{defJ}
J=\left(\begin{array}{ccc}
0 & \cdots  & 1 \\
 \vdots & \iddots & \vdots \\
 1 & \cdots & 0 
 \end{array}\right).
\eeq
Any matrix $M$ with the property $JM^T=-MJ$ is said to be {\it $J$-antisymmetric}.
Such a matrix is antisymmetric with respect to the second diagonal, 
{\it i.e.} $M_{i,j}=-M_{n+1-j,n+1-i}$
\begin{displaymath}
\left(\begin{array}{ccccc}
M_{1,1} & M_{1,2} & \cdots & M_{1,n-1} & 0 \\
M_{2,1} &         & \cdots & \iddots          & -M_{1,n-1} \\
 \vdots &         & \iddots&           & \vdots \\
 M_{n-1,1} & \iddots &        &           & -M_{1,2} \\ 
 0      & -M_{n-1,1} & \cdots & -M_{2,1}  & -M_{1,1}
 \end{array}\right)
\end{displaymath}
and in particular, $M_{ij}=0$ whenever $i+j=n+1$. 

\noindent On the set  $J{\cal A}_n$ of complex $J$-antisymmetric matrices, 
we consider the measure 
\bea\label{MEASJ}
\D \mu(M)&=&\e{-\tr{\frac{\alpha_1}{2}M^2+\frac{\alpha_2}{2}M^{\dagger 2}+\gamma M^\dagger M}} \,\D M \nonumber\\
\D M&=&\prod_{i+j<n+1} \, \D \Re{M}_{i,j}\, \D \Im{M}_{i,j}\ .
\eea

Then  the complex $J$-antisymmetric matrix 
partition function and the associated correlation functions are defined as
\bea
Z_{1JA}&=&\int_{J{\cal A}_n}\D\mu(M)\nonumber
\eea
\bea
\BK{F(M,M^\dagger)}_{1JA}
&=&\frac{1}{Z_{1JA}}\int_{J{\cal A}_n}\D \mu(M)\, F(M,M^\dagger)\nonumber
\eea 
using the measure $\D\mu(M)$ given in \eq{MEASJ}.
The partition function is again the product of the $n(n-1)/2$ uncoupled 
and equal integrals over the complex independent matrix elements $M_{ij}$, 
$i+j<n+1$. It is absolutely convergent 
if  $\Re\alpha_1,\Re\alpha_2>0$,  $\Re\gamma<0$ and $\Re\gamma^2>\Re \alpha_1\alpha_2$
and is then  given by 
\bea
Z_{1JA}&=&\left(\frac{\pi}{2\sqrt{-\delta}}\right)^{\frac{n(n-1)}{2}}
\eea
with $\delta=\alpha_1\alpha_2-\gamma^2$ as before. For polynomial 
$F(M,M^\dagger)$, 
the correlation functions $\BK{F}_{1JA}$ are again given by 
polynomials in $\frac{\alpha_1}{\delta}$ , $\frac{\alpha_2}{\delta}$  and 
$\frac{\gamma}{\delta}$.


\subsubsection{Analytic continuation.}\label{analc1}

The two families of integrals just studied have close connections, even
though their original domains of convergence may not overlap.  
The first trivial observation is that in both cases we have the same 
number of variables, as already manifest in 
the computations of the partition functions,  namely  $n(n-1)$ 
integration variables in both integrals. 
The second and more important observation for our purpose is that both %
integrals share the same loop equations. This will be proved in  appendix \ref{app:I}.
Third, as already stressed above, 
the correlations of polynomial invariant functions both in ${\cal A}_n\times{\cal A}_n$ and in $J{\cal A}_n$ are polynomials in the variables $\frac{\alpha_1}{\delta}$, $\frac{\alpha_2}{\delta}$ and $\frac{\gamma}{\delta}$, and thus analytic functions.

From all these observations, we formulate the following
\begin{theorem}\label{MainOrt}
The polynomial correlation functions of the two real antisymmetric matrix integral are equal to the correlation functions of the complex $J$-antisymmetric matrix integral in the sense of analytic continuation, {\it i.e.}
\beq\label{EQMainOrt}
\BK{F(A_1,A_2)}_{2RA}=\BK{F(M,M^\dagger)}_{1JA}\ .
\eeq
\end{theorem}

\proof{Note that the loop equations given in appendix \ref{app:I} are in 
fact recursion relations on the polynomial degree of the correlation functions.
 Then the fact that polynomial invariant correlation functions are 
polynomials in $\frac{\alpha_1}{\delta}$, $\frac{\alpha_2}{\delta}$ 
and $\frac{\gamma}{\delta}$ and the fact that the loop equations and their
initial condition (namely $\BK{1}=1$) are the same for both integrals imply 
that the polynomials generated from the recursion are the same.}

Although the correlation functions are not originally defined in the same
region in parameter space, the fact that they are polynomials 
allows one to analyticly continue them and to identify them.


\subsection{Real quaternionic antiselfdual two matrix integrals and 
complex \texorpdfstring{$\td{J}$}{`tilde-J'}-antisymmetric matrix integral.}

In this section we consider another pair of matrix integrals, 
related to the symplectic group, for which similar considerations hold true.

\subsubsection{Real quaternionic antiselfdual two matrix integrals }
\label{rquaterad}

Consider first the set  $Q{\cal A}_m$ 
of real quaternionic antiselfdual (antihermitian) $m\times m$ matrices, 
whose definition has been recalled in sect.~\ref{Iexplicit} and appendix
\ref{app:Q}. 

On $Q{\cal A}_m\times Q{\cal A}_m$, we consider the measure given by
\bea\label{MEASQ}
\D \mu(Q_1,Q_2)&=&\e{-\tro{\frac{\alpha_1}{2}Q_1^2+\frac{\alpha_2}{2}Q_2^2+\gamma Q_1 Q_2}} \D Q_1 \D Q_2 \nonumber\\
\D Q_k&=&\left(\prod_{i<j}\prod_{\alpha =0}^3 \D \left(Q_k^{(\alpha)}\right)_{i,j}\right)\,\left(\prod_{i=1}^m\prod_{\alpha=1}^3 \D\left(Q_k^{(\alpha)}\right)_{i,i}\right)\, ;\quad \, k=1,2
\eea
where $\tro{\dots}=2\Re\tr{\dots}$ is a scalar (while $\tr{\dots}$ 
is in general a quaternion number, see appendix
\ref{app:Q}). 
The quadratic form in this `Gaussian' measure is thus
\bea  %
&{}&\!\!\!\!\!\!\!\!\!\!\!\!\!\!\!\!
-\tro{\frac{\alpha_1}{2}Q_1^2+\frac{\alpha_2}{2}Q_2^2+\gamma Q_1 Q_2}
=2\sum_{1\le i<j\le m} \sum_{\alpha =0}^3
\left(\alpha_1 (Q^{\alpha}_1)_{ij}^2+\alpha_2 (Q^{\alpha}_2)_{ij}^2+
2\gamma (Q^{\alpha}_1)_{ij}(Q^{\alpha}_2)_{ij}\right)\nonumber\\
&{}&\qquad + \sum_{i=1}^m \sum_{\alpha =1}^3
 \left( \alpha_1 (Q^{\alpha}_1)_{ii}^2+\alpha_2 (Q^{\alpha}_2)_{ii}^2+
2\gamma (Q^{\alpha}_1)_{ii}(Q^{\alpha}_2)_{ii}\right) \ .
\eea
The real quaternionic antiselfdual two matrix
partition function and the associated correlation functions are defined as
\bea
Z_{2QA}&=&\int_{Q{\cal A}_m\times Q{\cal A}_m}\D\mu(Q_1,Q_2)
\eea
\bea
\BK{F(Q_1,Q_2)}_{2QA}&=&\frac{1}{Z_{2QA}}\int_{Q{\cal A}_n\times Q{\cal A}_n}\, F(Q_1,Q_2) \D\mu(Q_1,Q_2)\ . \nonumber
\eea
The partition function is readily computed to be 
\bea
Z_{2QA}=2^{3m}\left(\frac{\pi}{2\sqrt{\delta}}\right)^{2m^2+m}
\eea
and once again correlation functions of polynomials in
$Q_1, Q_2$ are polynomials in $\frac{\alpha_1}{\delta}$, 
$\frac{\alpha_2}{\delta}$,  and $\frac{\gamma}{\delta}$.


\subsubsection{Complex \texorpdfstring{$\td{J}$}{`tilde-J'}-antisymmetric matrix integral}\label{complJ}

We now introduce a $2m\times 2m$ matrix $\td{J}=-\td{J}^{-1}$ of the form
\beq\label{deftdJ}
\td{J}=\left(\begin{array}{cc}
0 & J \\
-J & 0 
 \end{array}\right)
\eeq
written in terms of $J$ defined above in (\ref{defJ}).
Any matrix $M$ with the property $\td{J}M^T=-M\td{J}$ is said to be 
{\it $\td{J}$-antisymmetric}. 
Such a matrix possesses a peculiar symmetry with respect to the 
second diagonal: we can write it as
\begin{displaymath}
M=\left(\begin{array}{cc}
 A & B \\
 C & D 
 \end{array}\right)
\end{displaymath}
where  $A$, $B$, $C$, $D$ are $m\times m$ matrices satisfying
\beq
A=-JD^TJ \, ; \quad \, JB^T=BJ \, ; \quad \, JC^T=CJ
\eeq
Thus, under the reflection with respect to the second
diagonal, $B$ and $C$ are invariant,  while $A$ and $-D$ are exchanged. 

On the set  $\td{J}{\cal A}_{2m}$ of complex $\td{J}$-antisymmetric matrices
 we consider the measure 
\bea\label{MEAStdJ}
\D \mu(M)&=&\e{-\tr{\frac{\alpha_1}{2}M^2+\frac{\alpha_2}{2}M^{\dagger 2}+\gamma M^\dagger M}} \D M \nonumber\\
\D M&=&\prod_{i+j\leq 2m+1} \, \D \Re{M}_{i,j}\, \D \Im{M}_{i,j}\ .
\eea
Then
the complex $\td{J}$-antisymmetric matrix 
partition function  and the associated correlation functions are defined as
\bea 
Z_{1\td{J}A}&=&\int_{\td{J}{\cal A}_{2m}}\D \mu(M)
\eea
\bea
\BK{F(M,M^\dagger)}_{1\td{J}A}&=&\frac{1}{Z_{1\td{J}A}}\int_{\td{J}{\cal A}_{2m}}\D \mu(M) F(M,M^\dagger) \nonumber 
\eea
using the measure $\D\mu(M)$ given in \eq{MEAStdJ}.
The $\td{J}$-antisymmetric partition function reads 
\bea
Z_{1\td{J}A}=2^{2m}\left(\frac{\pi}{2\sqrt{-\delta}}\right)^{2m^2+m}\ , 
\eea
and once again, correlation functions of polynomials in $M$ and  $M^\dagger$
are polynomials in the parameters   $\frac{\alpha_1}{\delta}$, 
$\frac{\alpha_2}{\delta}$,  and $\frac{\gamma}{\delta}$.


\subsubsection{Analytic continuation}\label{analc2}

Again the observations  made in sect.~\ref{analc1} extend to this case. 
The two matrix integrals of \ref{rquaterad}, \ref{complJ} have the 
same number of integration variables equal to $2m(2m+1)$,
they satisfy the same loop equations (see appendix \ref{app:I}), 
and their correlation functions of invariant polynomials
have polynomial dependence on 
$\frac{\alpha_1}{\delta}$, $\frac{\alpha_2}{\delta}$ 
and $\frac{\gamma}{\delta}$. 

These observations allow us to formulate an analogous analytic 
continuation theorem for these two matrix integrals
\begin{theorem}\label{MainSymp}
The polynomial correlation functions of the two real quaternionic antiselfdual matrix 
integral are equal to the correlation functions of the complex $\td{J}$-antisymmetric matrix integral in the sense of analytic continuation, {\it i.e.}
\beq\label{EQ0SP}
\BK{F(A_1,A_2)}_{2QA}=\BK{F(M,M^\dagger)}_{1\td{J}A}\ .
\eeq
\end{theorem}

\proof{The proof goes exactly as the one in theorem \ref{MainOrt}.}


\section{Correlation functions over the orthogonal group}\label{HSOrt}

In this section we {exploit} %
the relation found in theorem \ref{MainOrt} 
by performing a separation between ``angular'' and ``radial'' 
variables of matrices in the two sides of equation \eq{EQMainOrt}. 

\subsection{Block-diagonalization of antisymmetric matrices}\label{SEC1}

We first consider the case of antisymmetric matrices and of the 
orthogonal group O$(n)$ equiped with its Haar measure $\D O$ (normalized
to  $\int \D O =1$).

As recalled in sect.1, Cartan's theorem asserts that any antisymmetric matrix 
$A$ may be brought to the block diagonal form (\ref{blodiagev}) or (\ref{blodiagod}) 
by an orthogonal transformation of O$(n)$ (for this standard result, see also 
\cite{Mehta, Knapp, Fulton}).
Denote by $D^a_n(\R)$ the set of such real 
block-diagonal antisymmetric $n\times n$ matrices, with the Lebesgue measure:
\beq
\D X:= \prod_{i}^{m} \D \X_i \ .
\eeq
By an abuse of language, we shall refer to the $\X_i$ as the ``eigenvalues'' of $A$.

In the new variables $\{O, \X \}$, 
the Lebesgue measure in ${\cal A}_n$ reads 
\beq\label{diagoAA}
\D A = \mathrm{Jac}^O_n\,\Delta^2(\X) \, \D O\, \D \X
\eeq
where the Jacobian is (see appendix \ref{app:JACS} for details)
\beq\label{eq:JacO}
\mathrm{Jac}_n^O=
\left\{\begin{array}{c l} 
\frac{\pi^{m(m-1)} 2^{m(m-1)}}{m!\prod_{j=1}^{m-1}(2j)!} \quad
& \textrm{if $n=2m$}\\ 
\frac{\pi^{m^2} 2^{m^2}}{m!\prod_{j=1}^{m}(2j-1)!} \quad
& \textrm{if $n=2m+1$\ .} 
\end{array}\right. 
\eeq
We recall that $\Delta(X)$ takes two different forms 
(\ref{VdMorthev}) and (\ref{VdMorthod}) depending on the parity of $n$. 

This decomposition is unique up to a permutation of the $m$ ``eigenvalues'', 
a change of signs of each eigenvalue independently, and a multiplication 
of $O$ by a $2\times2$ block-diagonal matrix whose diagonal blocks belong to $\On{2}$.
In other words, $A=O\X O^T$ establishes a mapping between ${\cal A}_n$ and
$\On{n}\times D^a_n(\R)/(\On{2}^m\times\SS_m\times{\Z_2^m})$. This overcounting has already been taken into account in $\mathrm{Jac}_n^O$.


\subsection{Schur decomposition of complex \texorpdfstring{$J$}{J}-antisymmetric matrices}\label{SEC2}

A less standard result, (see \cite{Mehta, Ginibre, Knapp, Fulton} for instance),
is that any complex matrix $M\in M_n(\C)$    can be written as:
\beq\label{Schurdec}
M^\prime = U^\prime (Z^\prime+T^\prime) U^{\prime\dagger}
\eeq
where $U^\prime\in \Un{n}$ is a unitary matrix, $T^\prime\in T_n$ a strictly 
upper triangular complex matrix and $Z^\prime\in D_n(\C)$ a complex diagonal matrix.
 We can apply this Schur decomposition to a $J$-antisymmetric matrix.	
This will induce further constraints on the unitary and triangular matrices.

Define ${\mathrm U}^J(n)$ to be the subgroup of $\Un{n}$ satisfying the condition 
\beq\label{twistorth}
U^{-1}=U^{\dagger}=JU^TJ
\eeq
 with the induced normalized Haar measure. We will call these 
matrices {\it twisted orthogonal} matrices.
Define also $T^J_n$ to be the set of $n\times n$ strictly upper triangular 
$J$-antisymmetric complex matrices, with the Lebesgue measure:
\beq\label{TnJ}
\D T:= \prod_{\begin{subarray}{c}i<j\\i+j<N+1\end{subarray}} \D \Re T_{ij} \,\D\Im T_{ij}
\eeq
and $D^J_n(\C)$ to be the set of $n\times n$ complex $J$-antisymmetric diagonal matrices
\beq\label{Jantisdiag}
Z=\diag(\Zz_1,\cdots, \Zz_m,-\Zz_m,\cdots, -\Zz_1)\quad {\rm or}\quad 
\diag(\Zz_1,\cdots, \Zz_m,0, -\Zz_m,\cdots, -\Zz_1)
\ ,
\eeq
depending on the parity of $n$, with the Lebesgue measure:
\beq\label{DJC}
\D Z:= \prod_{i}^{m} \D\Re \Zz_{i} \,\D\Im \Zz_{i}\ .
\eeq
Finally we define for these matrices of  $D^J_n(\C)$ 
\beq\label{VdMZ}
\Delta(\Zz)=\prod_{i<j}(\Zz_i^2-\Zz_j^2)\,
\times \left\{
\smmat{1\cr  \prod_{i}^m \Zz_i \cr} 
\right.
\qquad 
\begin{array}{cc}{\rm if}\ & n=2m\\ {\rm if}\ & n=2m+1\end{array}
\ .
\eeq

With these notations one can prove
\begin{proposition} 
Any $J$-antisymmetric matrix $M$ may  be written as:
\beq\label{JordanJ}
M = U (Z+T) U^\dagger
\eeq
where $U\in {\rm U}^J(n)$, $T\in T^J_n$ and $Z\in D^J_n(\C)$.
The Lebesgue measure in $J{\cal A}_n$ is then:
\beq\label{shdif}
\D M = \mathrm{Jac}^{U^J}_n\, |\Delta(\Zz)|^2 \, \D U\, \D T\, \D Z
\eeq
where the Jacobian is 
\beq\label{eq:JacJ}
\mathrm{Jac}^{U^J}_n=\mathrm{Jac}^{O}_n \times \left\{\begin{array}{c l}
2^{m-m^2}\quad & \textrm{if $n=2m$}\\
2^{-m^2}\quad & \textrm{if $n=2m+1$} \end{array}\right.
\ .
\eeq
\end{proposition}

\proof{
Consider the Schur decomposition (\ref{Schurdec}) of the matrix $M$.
Noticing that
\beq
\det\left(\lambda - M\right)=\det\left(\lambda - JMJ\right)=\det\left(\lambda + M\right)
\eeq
we immediately see that the non-vanishing
 eigenvalues come in pairs $(\lambda,-\lambda)$. By a possible 
redefinition of $U$, we may always order the eigenvalues 
in a $J$-antisymmetric diagonal form $Z$ as in (\ref{Jantisdiag}). 
The  constraints  on $U$ and $T$ follow from the $J$-antisymmetry of $M$ 
and $Z$. The measure can be computed using 
the same method as in the appendices of \cite{Mehta}.}

This decomposition is  unique up to a permutation of the 
$m$ different eigenvalues,  to changes of sign of the $m$ eigenvalues 
and  to multiplication of $U$ by a diagonal matrix $V\in {\rm U}^J(n)$ whose 
elements are on the unit circle.
In other words, $M=U(Z+T)U^\dagger$ provides a $1$-to-$1$ mapping between 
$J{\cal A}_n(\C)$ and ${\rm U}^J(n)\times T^J_n\times D^J_n(\C)/(\Un{1}^m\times \SS_m\times{\Z_2^m})$. The overcounting is included in \eqref{shdif}.


\subsection{Orthogonal and triangular matrix integrals}

\subsubsection{Radial and angular integrals}
Consider the block-diagonal decomposition of the real antisymmetric matrices and the Schur decomposition of the $J$-antisymmetric complex matrices. Using these decompositions we will rewrite both sides of \eq{EQMainOrt}.

\begin{theorem}\label{TH1}
A matrix integral over ${\cal A}_n\times {\cal A}_n$ can be decomposed into a ``radial'' and an ``angular'' part using the block-diagonal decomposition $A = O\X O^T$.

\bea\lbl{eq:OTIII}
\int_{{\cal A}_n\times {\cal A}_n} \back \D \mu(A_1,A_2) \, F(A_1,A_2) &=&(\mathrm{Jac}^O_n)^2
\int_{D^a_n(\R)\times D^a_n(\R)}\back \D \X\D \Y \Delta^2(\X)\Delta^2(\Y)
\nonumber\\
 \e{-\tr{\frac{\alpha_1}{2}\X^2+\frac{\alpha_2}{2}\Y^2}}&\int_{\On{n}}&\D O \, F(\X,O\Y O^T) \e{-\gamma\tr{\X O\Y O^T}}  
\eea
with the notations of (\ref{VdMorthev}) and  (\ref{VdMorthod}). 
\end{theorem}
\proof{The theorem follows from the results of section \ref{SEC1}.} 

Notice that one of the two orthogonal matrices decouples and 
so this part of the integral gives $1$. The remaining orthogonal 
integral represents the relative angular variables.

\begin{theorem}\label{TH2}
A matrix integral over $J{\cal A}_n$ can be decomposed into a ``radial'', an ``angular'' and a ``triangular'' part using the Schur decomposition $M=U(Z+T)U^\dagger$.
\bea\label{EQ2.1}
&&\int_{J{\cal A}_n} \D \mu(M) \, F(M,M^\dagger) =
\mathrm{Jac}^{{\rm U}^J}_n\int_{D^J_n(\C)}\D Z |\Delta(\Zz)|^2 \e{-\tr{\frac{\alpha_1}{2}Z^2+\frac{\alpha_2}{2}Z^{* 2}}}
 \times\nonumber\\
&&\qquad\,\e{-\gamma\tr{Z^*Z}}\int_{T^J_n}\D T F(Z+T,Z^*+T^\dagger) 
\e{-\gamma\tr{T^\dagger T}}
\eea
with the notations of (\ref{VdMZ}). 
\end{theorem}

\proof{The theorem follows from the results of section \ref{SEC2}.} 

Notice that the twisted orthogonal matrix decouples and so this part of the integral gives $1$.
 Only the triangular and radial parts remain. Notice also that the measure for the triangular part factors out from that of the radial part and only a Gaussian measure remains for the triangular part.


\subsubsection{Relating integrals over orthogonal and  triangular matrices}

In this subsection, we relate the HC integral over the orthogonal group
$\On{n}$
\beq
{\cal I}_F^{\On{n}} 
:=\int_{\On{n}} \D O \e{-\gamma\tr { \X^a O\Y^a O^T}} F(X^a, OY^aO^T)\ ,
\eeq
with $\X^a,\ \Y^a\,\in D_n^a(\R)$,  
to an integral over complex upper triangular matrices of $T_n^J(\C)$.
 Note first 
that ${\cal I}_F^{\On{n}} $ is a completely symmetric and even 
function of the ``eigenvalues'' $\X_i$ and of the $\Y_i$,
 $i=1,\cdots, m=\lfloor n/2\rfloor$.
This is because any permutation or sign changing matrix
acting on either $\X^a$ or  $\Y^a$ may be absorbed into a redefinition
of the orthogonal matrix $O$. In contrast, the integral over complex 
triangular matrices will have to be symmetrized {\it by hand}.

To obtain the desired relation between HC-type integrals over the orthogonal 
group and integrals over the triangular matrices of $T_n^J$, we shall
follow the same steps as in \I, in particular of  Lemma A.1 there,
which asserts that for any polynomial $\omega$ in two variables, one has the 
relation
\bea\label{Foundation}
\frac{{\int_\C}\,\D \z\,\om(\z,\z^*) \e{-\tr{\frac{\alpha_1}{2}\z^2 + \frac{\alpha_2}{2}\z^{*2} + \gamma \z^* \z}}} {{\int_\C}\,\D \z\, \e{-\tr{\frac{\alpha_1}{2}\z^2 + \frac{\alpha_2}{2}\z^{*2} + \gamma \z^* \Zz}}} =
\frac{\int_{\R\times\R} \D \x\,\D \y\,\om(\x,\y)\,\e{-\tr{\frac{\alpha_1}{2}\x^2 + \frac{\alpha_2}{2}\y^2 + \gamma \x \y}}} {\int_{\R\times \R} \D \x\,\D \y\,\,\e{-\tr{\frac{\alpha_1}{2}\x^2 + \frac{\alpha_2}{2}\y^2 + \gamma \x \y}}}\nonumber\\
\eea
where we have one complex variable integration on the left hand side and two real variables on the right hand side.
 This relation may be promoted into the following equality between 
integrals over {\it diagonal} matrices 
\bea\label{EQ3}
\frac{\int_{D^J_n(\C)}\,\D \Zz\,\om(\Zz,\Zz^*) \e{-\tr{\frac{\alpha_1}{2}\Zz^2 + \frac{\alpha_2}{2}\Zz^{*2} + \gamma \Zz^* \Zz}}} {\int_{D^J_n(\C)}\,\D \Zz\, \e{-\tr{\frac{\alpha_1}{2}\Zz^2 + \frac{\alpha_2}{2}\Zz^{*2} + \gamma \Zz^* \Zz}}} =
\frac{\int_{D^J_n(\R)\times D^J_n(\R)} \D \X\,\D \Y\,\om(\X,\Y)\,\e{-\tr{\frac{\alpha_1}{2}\X^2 + \frac{\alpha_2}{2}\Y^2 + \gamma \X \Y}}} {\int_{D^J_n(\R)\times D^J_n(\R)} \D \X\,\D \Y\,\,\e{-\tr{\frac{\alpha_1}{2}\X^2 + \frac{\alpha_2}{2}\Y^2 + \gamma \X \Y}}}\ . \nonumber\\
\eea
 
We now  apply theorems \ref{TH1} and \ref{TH2} to the two sides 
of (\ref{EQMainOrt})
\bea
&{}&\langle F\rangle_{2RA}=\frac{1}{Z_{2RA}}
\int_{{\cal A}_n\times{\cal A}_n} \D A_1 \D A_2 
\e{-\tr{\frac{\alpha_1}{2} A_1^{2} + \frac{\alpha_2}{2} A_2^{2} + \gamma 
A_1 A_2}} F(A_1,A_2) \\ \label{second}
 &{}&\qquad = \frac{(\mathrm{Jac}^{O}_n)^2}{ Z_{2RA}}\int_{D^a_n(\R)\times D^a_n(\R)}
\D \X^a \D \Y^a \Delta^2(\X^a)\Delta^2(\Y^a) 
\e{-\tr{\frac{\alpha_1}{2}\X^{a2} + \frac{\alpha_2}{2}\Y^{a2}}}\\
&{}&\qquad\qquad\qquad\times  \int_{\On{n}} \D O \e{-\gamma\tr { \X^a O\Y^a O^T}} F(X^a, OY^aO^T)
\nonumber \\
&=&\langle F\rangle_{1JA}= \frac{1}{Z_{1JA}} 
\int_{J{\cal A}_n} \D M \e{-\tr{\frac{\alpha_1}{2} M^2 + 
\frac{\alpha_2}{2} M^{\dagger2} + \gamma M M^\dagger}} F(M,M^\dagger)
 \\
&=& \frac{\mathrm{Jac}_n^{U^J}Z_{D_n^J(\C)}}{ Z_{1JA}}
\frac{1}{Z_{D_n^J(\C)}}\int_{D_n^J(\C)}
\D Z |\Delta(Z)|^2 \e{+\tr{\frac{\alpha_1}{2}\Zz^2 + \frac{\alpha_2}{2}\Zz^{*2} {-} \gamma \Zz \Zz^* }} \\
&{}& \qquad\qquad\qquad \times
 \int_{T^J_n} \D T\,\, \e{-\gamma\tr{T T^\dagger}}
F(\ii Z+T ,-\ii Z^*+T^\dagger)\ .
\eea
In the last line, we have performed a change of variables $Z\to \ii Z$
for reasons that will appear soon. We then apply (\ref{EQ3}) to get
\bea
\langle F\rangle_{1JA}&=& 
 \frac{\mathrm{Jac}_n^{U^J}Z_{D_n^J(\C)}}{Z_{1JA}Z_{D^{J}_n(\R)\times D^{J}_n(\R)}}
\int_{D^{J}_n(\R)\times D^{J}_n(\R)} \D X \D Y \Delta(X) \Delta(Y) 
\e{\tr{\frac{\alpha_1}{2}\X^{2} + \frac{\alpha_2}{2}\Y^{2}
-\gamma X Y}} \nonumber \\
&{}& \qquad\qquad\qquad\qquad\qquad
\times \int_{T_n^J} dT\, \e{-\gamma\tr{T T^\dagger}}F(\ii\X +T,-\ii \Y+T^\dagger)
\\ \label{last}
&=&\frac{\mathrm{Jac}_n^{U^J}Z_{D_n^J(\C)}}{Z_{1JA}Z_{D^{J}_n(\R)\times D^{J}_n(\R)} 2^m m!}
\int_{D^{J}_n(\R)\times D^{J}_n(\R)} \D X \D Y \Delta^2(X) \Delta^2(Y)
\e{\tr{\frac{\alpha_1}{2}\X^{2} + \frac{\alpha_2}{2}\Y^{2}}}
 \nonumber\\ 
&{}& \qquad
\times\sum_{\atopsum{\tau\in \SS_m}{t\in\Z_2^m}}  \frac{\e{-\gamma\tr{X Y_{(\tau,t)}}}}{\Delta (X)\Delta(Y_{(\tau,t)})}
 \int_{T_n^J} dT\, \e{-\gamma\tr{T T^\dagger}}F(\ii\X+T,-\ii\Y_{(\tau,t)}+T^\dagger)
\eea

In the last line, we have symmetrized the integral over triangular matrices
for the reason explained at the beginning of this subsection.
In these expressions, 
$Z_{D_n^J(\R)\times D_n^J(\R)}=(\pi/\sqrt{\delta})^m$ 
 and $Z_{D_n^J(\C)}=(\pi/2\sqrt{-\delta})^m $.
We finally compare the integrands of the second and the last lines
(\ref{second}) and (\ref{last}) of the previous equation that we rewrite as
\bea\label{rewrite}
&{}&\int_{D^a_n(\R)\times D^a_n(\R)}
\D \X^a \D \Y^a \Delta^2(\X^a)\Delta^2(\Y^a) 
\e{-\tr{\frac{\alpha_1}{2}\X^{a2} + \frac{\alpha_2}{2}\Y^{a2}}}\nonumber
\\
&{}& 
\qquad\times  
\frac{(\mathrm{Jac}^{O}_n)^2}{ Z_{2RA}}
\int_{\On{n}} \D O \e{-\gamma\tr { \X^a O\Y^a O^T}} F(X^a, OY^aO^T)\nonumber\\
&=&
\int_{D^{J}_n(\R)\times D^{J}_n(\R)} \D X \D Y \Delta^2(X) \Delta^2(Y)
\e{\tr{\frac{\alpha_1}{2}\X^{2} + \frac{\alpha_2}{2}\Y^{2}}}
\\ 
&{}& \quad
\times
\frac{\mathrm{Jac}_n^{U^J}Z_{D_n^J(\C)}}{Z_{1JA}Z_{D^{J}_n(\R)\times D^{J}_n(\R)} 2^m m!}
\sum_{\atopsum{\tau\in \SS_m}{ t\in\Z_2^m}}  \frac{\e{-\gamma\tr{X Y_{(\tau,t)}}}}{\Delta (X) \Delta(Y_{(\tau,t)})}
 \int_{T_n^J} dT\, \e{-\gamma\tr{T T^\dagger}}F(\ii \X+T,-\ii\Y_{(\tau,t)}+T^\dagger)\nonumber
\eea

Note the sign difference in the two quadratic  forms: 
if $\X\in D_n^J(\R)$ has eigenvalues $\X_i$, and  $\X^a\in D_n^a(\R)$
is of the form (\ref{blodiagev}) or (\ref{blodiagod}), then 
$\tr{\X^{a2}}=-2\sum_{i=1}^m \X_i=-\tr \X^2$, and likewise for $Y$, 
so that the Gaussian measures match. This
justifies {\it a posteriori} our change of $\Zz\to \ii\Zz$. 

In order to identify the two integrands (the second and fourth lines of
(\ref{rewrite})), we notice that by definition these integrands belong to 
$L^2(\R^{2m})$  with respect to the measure given by the 
first and third lines of \eq{rewrite}. Now
 we proceed as in \I: by multiplying $F(X,Y)$ by arbitrary 
polynomials of $X$, resp. $Y$,  we may multiply the integrands on 
both sides by arbitrary symmetric even polynomials of the $X_i$ or 
of the $Y_j$.
By projecting onto the orthogonal polynomials basis of $L^2(\R^{2m})$ with respect to the measure, we deduce that the integrands must be equal. This gives the 

\begin{theorem}\lbl{th:MainOrt}
For any invariant polynomial function $F(.,.)$ and any 
$X^a,Y^a \in D_n^a(\R)$, and $X,Y$ the corresponding matrices
 in $D^J_n(\R)$, one has:
\beq\label{thm33}
\int_{\atopsum{\On{n=2m}}{\On{n=2m+1}}} 
\D O \e{-\gamma\tr{\X^aO\Y^aO^T}}\, F(\X^a,O\Y^aO^T)
=\eeq \vspace{-10pt} \beq
=c_n \sum_{\tau \in \SS_m}\sum_{t \in \Z_2^m}\varepsilon_\tau 
\frac{ \e{\gamma  \tr{ X Y_{(\tau,t)}}}}{\Delta(\X)\Delta(\Y)}
\int_{T^J_n}\D T \, \e{-\gamma\tr{T^\dagger T}}\,
F(\ii\X+T,\ii\Y_{(\tau,t)}+T^\dagger)
\times\left\{ \begin{array}{c} 1\\ \prod_i t_i \end{array} \right.
 \nonumber
\eeq
where 
\beq\label{eqb}
c_n=\frac{\mathrm{Jac}_n^{U^J}Z_{2RA}Z_{D_n^J(\C) }}{(\mathrm{Jac}^{O}_n)^2Z_{1JA}Z_{D^{J}_n(\R)\times D^{J}_n(\R)} 2^m m!}=\frac{2^{\frac{n(n-1)}{2}}}{4^m m! \mathrm{Jac}_n^O} 
.
\eeq
\end{theorem}
In (\ref{thm33})  the dependence on $\tau$ and the signs $t_i$ has been 
made more explicit, and all $t_i$ changed into their opposite.


\subsubsection{Examples}

Take as an example the case $F(A,B)=1$. Then
\beq
\int_{\atopsum{\On{2m}}{\On{2m+1}}}\D O \e{-\gamma\tr{\X^aO\Y^aO^T}}=c_n \sum_{\tau \in \SS_m}\sum_{t \in \Z_2^m}
\varepsilon_\tau\frac{\prod_{i=1}^m\, \e{2\gamma \X_{i}\Y_{(\tau,t)(i)}}} {\Delta(\X)\Delta(\Y)}\int_{T^J_n}\D T \e{-\gamma\tr{T^\dagger T}}\ 
\times  \left\{ \begin{array}{c} 1\\ \prod_i t_i\end{array} \right.
. \nonumber
\eeq

Merging the constant $c_n$ and the triangular integral which decouples 
into a constant $K_n$ we get just a summation	 over permutations and signs. 

\bea
\int_{\atopsum{\On{2m}}{\On{2m+1}}}
\D O \e{-\gamma\tr{\X^aO\Y^aO^T}}&=& K_n \sum_{\tau \in \SS_m}
\sum_{t \in \Z_2^m}\varepsilon_\tau
\frac{\prod_{i=1}^m\, \e{2\gamma \X_{i}\Y_{(\tau,t)(i)}}} {\Delta(\X)\Delta(\Y)}
\times  \left\{ \begin{array}{c} 1\\ \prod_i t_i\end{array} \right.
\nonumber\\
&=&\frac{K_n}{\Delta(\X)\Delta(\Y)}\sum_{\tau \in \SS_m}\varepsilon_\tau
\prod_{i=1}^m \left[\e{2\gamma \X_{i}\Y_{\tau(i)}}\pm \e{-2\gamma \X_{i}\Y_{\tau(i)}}\right]\nonumber\\
&=&\frac{K_n}{\Delta(\X)\Delta(\Y)}\times
\left\{ \begin{array}{c} \det{2 \cosh{2\gamma \X_{i}\Y_j}}\\
\det{2 \sinh{2\gamma \X_{i}\Y_j}} \end{array}  \right.
\eea
with
\beq
K_n=c_n \int_{T^J_n}\,\e{-\gamma\tr{T^\dagger T}}\ .
\eeq
Then with
\beq
\int_{T^J_n}\,\e{-\gamma\tr{T^\dagger T}}=
\left\{\begin{array}{c l}
\left(\frac{\pi}{2\gamma}\right)^{m(m-1)}\quad & \textrm{if $n=2m$}\\
\left(\frac{\pi}{2\gamma}\right)^{m^2}\quad & \textrm{if $n=2m+1$}
\end{array}\right.
\eeq
we obtain
\beq
K_n=\left\{\begin{array}{cl}
\frac{\prod_{j=1}^{m-1}(2j)!}{2^{m}\gamma^{m(m-1)}}&\quad \textrm{if $n=2m$}\\
\frac{\prod_{j=1}^{m}  (2j-1)!}{ 2^{m}\gamma^{m^2}}&\quad \textrm{if $n=2m+1$\ .}
\end{array}\right.
\eeq
This is exactly what was obtained by the \DH\ theorem in sect.~1.3 
and serves as a check of our formulae.


\section{Correlation functions over the symplectic group}\label{HSSymp}

In this section we repeat the analysis made in section \ref{HSOrt}, 
in the case related to the symplectic group  $\Spn{2m}$ of $2m\times 2m$ symplectic matrices and 
to theorem \ref{MainSymp}. Following the same steps we perform the 
separation between ``angular'' and ``radial'' variables 
of matrices in both sides of \eq{EQ0SP}.

\subsection{Diagonalization of real quaternion antiselfdual matrices}
\label{sec:DiagQ}

We consider the set  $D_m^{aR}(\Q)$ 
of real quaternion diagonal $m\times m$ matrices whose diagonal elements 
are real quaternions  proportional to $e_2$ (see (\ref{CartanQ})), with the Lebesgue measure:
\beq
\D \X:= \prod_{i}^{m} \D \X_i 
\eeq
Appealing again to Cartan's theorem, as we did in sect.~1.4, 
any real antiselfdual quaternion matrix $Q\in Q{\cal A}_{m}$ 
may be written under the form
\beq
Q = S \X S^\dagger
\eeq
where $S\in \Spn{2m}$ and $\X \in D^{aR}_{2m}(\Q)$.

The Lebesgue measure in $Q{\cal A}_m$ is then:
\beq
\D Q = \mathrm{Jac}^{Sp}_{2m}\,\Delta^2(\X) \, \D S\, \D \X
\eeq
with (see appendix \ref{app:JACS})
\beq\label{eq:JacQ}
\mathrm{Jac}_{2m}^{Sp}=\frac{\pi^{m^2} 2^m}{m!\prod_{j=1}^m(2j-1)!}
\eeq
and 
$\Delta(\X)$ as in (\ref{VdMorthod}).

This decomposition is unique up to a permutation of 
the $m$ eigenvalues, up to a change of sign of each eigenvalue independently, 
and up to multiplication of $S$ by a diagonal quaternion matrix 
$V\in \Spn{2m}$ whose diagonal elements $v_i$ satisfy
\beq
v_i=\cos\theta_i+\sin\theta_i e_2\,;\quad \theta_i\in[0,2\pi)\ .
\eeq
The latter matrices generate a group isomorphic to $\On{2}^m$.
In other words, $Q=S\X S^\dagger$ provides a $1$-to-$1$ mapping between $Q{\cal A}_{m}$ and
$\Spn{2m}\times D_m^{aR}(\Q)/(\On{2}^m\times\SS_m\times\Z_2^m)$.


\subsection{Schur decomposition of complex \texorpdfstring{$\td{J}$}{`tilde-J'}-antisymmetric matrices}
\label{sec:SchurtdJ}

Let ${\rm U}^{\td{J}}(2m)$  be the subgroup of $\Un{2m}$ unitary group 
satisfying the condition 
\beq\label{twsympl}
U^{-1}=U^{\dagger}=\td{J}U^T\td{J}^{-1}\ .
\eeq 
with the induced normalized Haar measure.
We will call these matrices {\it twisted symplectic matrices}.

Define also $T^{\td{J}}_{2m}$ to be the set of $2m\times 2m$ strictly upper triangular $\td{J}$-antisymmetric complex matrices, with the Lebesgue measure:
\beq\label{TtdJ}
\D T\,:= \prod_{\begin{subarray}{c}i<j\\i+j\leq 2m+1\end{subarray}} d\Re T_{ij} \,d\Im T_{ij}
\eeq
 The $2m\times 2m$  $J$-antisymmetric complex diagonal 
matrices of $D^J_{2m}(\C)$,  (see sect.~\ref{SEC2}), 
are also $\td{J}$-antisymmetric and come with the Lebesgue measure 
\eq{DJC}. 
Then we prove
\begin{proposition}
Any $\td{J}$-antisymmetric complex matrix $M$ may always be written as
\beq\label{JordantdJ}
M = U (Z+T) U^\dagger
\eeq  
where $U\in {\rm U}^{\td{J}}(2m)$, $T\in T^{\td{J}}_{2m}$ and $Z\in D^{J}_{2m}(\C)$.

The Lebesgue measure in $\td{J}{\cal A}_n$ is then:
\beq\label{eq:JactdJ}
\D M = \mathrm{Jac}^{{\rm U}^{\td{J}}}_{2m}\,  |\Delta(\Zz)|^2 \, \D U\, \D T\,\, \D Z
\eeq
where again $\Delta(\Zz)=\prod_{i<j}(\Zz_i^2-\Zz_j^2) \prod_{i}^m \Zz_i$ 
and 
\beq\label{eq:JactdJ2}
\mathrm{Jac}^{{\rm U}^{\td{J}}}_{2m}= 2^{-2m}\mathrm{Jac}^{Sp}_{2m}
\eeq
\end{proposition}

\proof{Similarly to the $J$-antisymmetric case one can see that
\beq
\Det{\lambda - M}=\Det{-\lambda - \td{J}M\td{J}}=\Det{\lambda + M}
\eeq
so that the eigenvalues come in pairs ($\lambda$,$-\lambda$). 
One may reorder them to make $Z$ as well as $M$ 
$\td{J}$-antisymmetric 
and then the constraints on $U$ and $T$ follow. Again the computation of 
the measure follows the lines of \cite{Mehta}.}

This decomposition is unique up to a permutation of the $m$ different eigenvalues, up to changes of sign of the $m$ eigenvalues and up to multiplication of $U$ by a diagonal matrix $V\in {\rm U}^{\td{J}}(2m)$ whose elements are on the unit circle. 

In other words, $M=U(Z+T)U^\dagger$ provides a $1$-to-$1$ mapping between $\td{J}{\cal A}_{2m}$ and
${\rm U}^{\td{J}}(2m)\times T^{\td{J}}_{2m}\times D^{\td{J}}_{2m}(\C)/({\rm U}(1)^m\times \SS_m\times\Z_2^m)$.


\subsection{Symplectic and triangular matrix integrals}

\subsubsection{Radial and angular integrals}
Consider the diagonal decomposition of the real antiselfdual quaternion matrices and the Schur decomposition of the $\td{J}$-antisymmetric complex matrices. Using them we will rewrite both sides of \eq{EQ0SP}. 
With the measure $\D \mu(Q_1,Q_2)$ defined in \eq{MEASQ}, we have the

\begin{theorem}\label{TH1SP}
A matrix integral over $Q{\cal A}_{m}\times Q{\cal A}_{m}$ can be decomposed into a ``radial'' and an ``angular'' part using the diagonal decomposition $Q = S\X S^\dagger$.
We have:
\bea
\int_{Q{\cal A}_{m}\times Q{\cal A}_{m}} \back \back \D \mu(Q_1,Q_2) \, F(Q_1,Q_2) &=& (\mathrm{Jac}^{Sp}_{2m})^2
\int_{D_m^{aR}(\Q)\times D_m^{aR}(\Q)}\back \back \D \X\D \Y 
\Delta^2(\X)\Delta^2(\Y) \times \nonumber\\
&& \e{-\tro
{\frac{\alpha_1}{2}\X^2+\frac{\alpha_2}{2}\Y^2}}
\int_{\Spn{2m}}\D S \, F(\X,S\Y S^\dagger) \e{-\gamma\tro{\X S\Y S^\dagger}}\ .
\nonumber
\eea

\end{theorem}
\proof{The theorem follows from what is explained in section \ref{sec:DiagQ}.}

Notice that one of the two symplectic matrices decouples and so this part of the integral gives $1$. The remaining symplectic integral runs over the relative angular variables. 

\begin{theorem}\label{TH2SP}
A matrix integral over $\td{J}{\cal A}_{2m}$ can be decomposed into a ``radial'', an ``angular'' and a ``triangular'' part using the Schur decomposition $M=U(Z+T)U^\dagger$.
We have:
\bea
\int_{M^{\td{J}}_{2m}(\C)} \D \mu(M) \, F(M,M^\dagger)
 &=& \mathrm{Jac}^{{\rm U}^{\td{J}}_{2m}}
\int_{D^{\td{J}}_{2m}(\C)}\D Z |\Delta(\Zz)|^2 
\times\nonumber\\
&&\e{-\tr{\frac{\alpha_1}{2}Z^2+\frac{\alpha_2}{2}Z^{* 2}}}\,\e{-\gamma\tr{Z^*Z}}\int_{T^{\td{J}}_{2m}}\D T F(Z+T,Z^*+T^\dagger) 
\e{-\gamma\tr{T^\dagger T}}\ .\nonumber
\eea
\end{theorem}

\proof{The theorem follows from what is explained in section 
\ref{sec:SchurtdJ}.}

Notice that the twisted symplectic matrix decouples and so this part of the integral gives $1$. Only the triangular and radial parts remain. Notice also that the measure for the triangular and the radial part factors out and so only a Gaussian measure remains for the triangular part.


\subsubsection{Relating integrals over symplectic and triangular matrices}

Just as in the orthogonal case, we first observe that the 
integral over the symplectic group %
\beq
{\cal I}^{\Spn{2m}}:=\int_{\Spn{2m}} \D S\, \e{-\gamma \tr X S Y S^\dagger}
F( X ,S Y S^\dagger) 
\eeq
with $X,Y\in D_m^{aR}(\Q)$, is a completely symmetric and even function
of the variables $X_i$ and of the $Y_i$, $i=1,\cdots, m$, since permutation 
and sign changing matrices may be absorbed into the symplectic matrix $S$.  
Then the same considerations as in sect.~4.3.2  apply 
when we want to use \eq{EQ3}. The Gaussian measure still gets the wrong sign, and the same change of variables $\Zz\to \ii\Zz$ must be used. Then 

\begin{theorem}\lbl{th:MainSp}
For any polynomial invariant function {$F(.,.)$, for any 
$X^{a},Y^{a}\in D_{m}^{aR}(\Q)$ and $X,Y$ the associated matrices in 
$D_{2m}^{{J}}(\R)$}, one has: 
\bea
&&\back\back\int_{\Spn{2m}}\D S F(\X^{a},S\Y^{a}S^\dagger)\e{-\gamma\tro{\X^{a}S\Y^{a}S^\dagger }}= 
\td{c}_{2m} \sum_{\tau \in \SS_m}\sum_{t \in \Z_2^m}\varepsilon_\tau 
\prod_{j=1}^m t_j
\frac{\e{\gamma \tr{ \X \Y_{(\tau,t)}}}}
{\Delta(\X)\Delta(\Y)}\nonumber\\
&&\qquad\qquad\times\int_{T^{\td{J}}_{2m}}\D T F(\ii\X+T,\ii \Y_{(\tau,t)}+T^\dagger)\,\e{-\gamma\tr{T^\dagger T}} \nonumber
\eea
with 
\beq
\td{c}_{2m}=
\frac{(-1)^m{\rm Jac}^{{\rm U}^{\td{J}}_{2m}} Z_{2QA}Z_{D_{2m}^J(\C)} }
{({\rm Jac}^{Sp}_{2m})^2 Z_{1\td{J}A} Z_{D_m^{aR}(\Q)\times D_m^{aR}(\Q)}
2^m m!}=
\frac{1}{2^m m!Jac^O_{2m}}\frac{1}{4^m}\ .
\eeq
\end{theorem}
\proof{The proof starts from \eq{EQ0SP}, makes use of 
theorems \ref{TH1SP} and \ref{TH2SP}, and then follows the same  
steps as the proof of theorem \ref{th:MainOrt}, including a 
change of variables $\Zz\to\ii\Zz$, a symmetrization in
the variables $X_i$ and $Y_j$ and the use of orthogonal polynomials.}

\subsubsection{Examples}

Let's take as an example the case $F(A,B)=1$. Then 
\bea
\int_{\Spn{2m}}\D S \e{-\gamma\tr{S\X^{a}S^\dagger \Y^{a}}}&=&
\td{c}_{2m} \sum_{\tau \in \SS_m}
\sum_{t \in \Z_2^m}\varepsilon_\tau \prod_{j=1}^m t_j
\frac{\prod_{i=1}^m\, \e{2\gamma t(i)\X_{i}\Y_{\tau(i)}}}{\Delta(\X)\Delta(\Y)
}\times\nonumber
\int_{T^{\td{J}}_{2m}}\D T \e{-\gamma\tr{T^\dagger T}} \nonumber
\\ &=& \td{K}_{2m} \sum_{\tau \in \SS_m}
\sum_{t \in \Z_2^m}\varepsilon_\tau \prod_{j=1}^m t_j
\frac{\prod_{i=1}^m\, \e{2\gamma t(i)\X_{i}\Y_{\tau(i)}}}{\Delta(\X)\Delta(\Y)
}=\td{K}_{2m} \frac{\det{2\sinh{2\gamma \X_i \Y_j}}}{\Delta(\X)\Delta(\Y)}
\nonumber
\eea
with
\beq
\td{K}_{2m}=\td{c}_{2m} \int_{T^{\td{J}}_{2m}}\,\e{-\gamma\tr{T^\dagger T}}\ .
\eeq
Then with
\beq
\int_{T^{\td{J}}_{2m}}\,\e{-\gamma\tr{T^\dagger T}}=2^m\left(\frac{\pi}{2\gamma}\right)^{m^2}
\eeq
we get
\beq
\td{K}_{2m}=2^{-(m^2+2m)}\frac{\prod_{j=1}^m(2j-1)!}{2^{m}}
\eeq
which reproduces again the \DH\ result. 

In the examples considered in this section and in sect.~3.3.3, 
the triangular integrals are just constants, 
which is not the case in general. 
We are going to present an explicit formula to compute them.



\section{\texorpdfstring{$J/\td{J}$}{J/'tilde-J'}-antisymmetric triangular integrals}\label{TI}

In order to compute the correlation functions in the orthogonal and the symplectic group we need to compute explicitely various kinds of triangular integrals. For the orthogonal, resp. symplectic, case we have to compute integrals 
over $J$-, resp. $\td{J}$-, antisymmetric strictly upper triangular complex matrices. 
We shall unify both kinds of integrals into one formalism and
explicitely perform the integration.

\subsection{Preliminaries to the integration}

The type of integrals we are interested in are of the form
\beq\lbl{eq:int}
\int_{\cal J}^{(n)}F(\X+T,\Y+T^\dagger)\, \e{-\gamma\tr{T^\dagger T}}\, \D T
\eeq
where $n$ is the matrix size,  ${\cal J}$ stands for $J$ or for $\td{J}$, 
and $\int_{\cal J}$ refers to whether we integrate over $J$ or $\td{J}$-antisymmetric triangular matrices, and $X$ and $Y$ are $J$- (or $\td{J}$-) antisymmetric {\it diagonal} real matrices. 
Since the measure is Gaussian it is more convenient to normalize the integrals
\bea
\BK{F(\X+T,\Y+T^\dagger)}_{\cal J}&=&\frac{\int_{\cal J}^{(n)}F(\X+T,\Y+T^\dagger)\, \e{-\gamma\tr{T^\dagger T}}\, \D T}
{\int_{\cal J}^{(n)}\e{-\gamma\tr{T^\dagger T}}\, \D T}.
\eea
From now on we set $2\gamma=1$, in order to  make the propagators simpler.

The typical functions $F(A,B)$ we want to use are constructed from resolvents $\frac{1}{\x-A}$ and twisted resolvents ${\cal J}\left(\frac{1}{\x-A}\right)^T{\cal J}$. These functions are not allowed in general by the analytical continuation theorems, but this is not a problem if we consider that their $x$ series
expansions are generating functions of invariant polynomials.
An example of such a function is 
\bea
&&\tr{\frac{1}{\x_1-(\X+T)}{\cal J}\left(\frac{1}{\y_1-(\Y+T^\dagger)}\right)^T{\cal J}}
\times\nonumber\\
&&\tr{\frac{1}{\x_2-(\X+T)}{\cal J}\left(\frac{1}{\y_2-(\Y+T^\dagger)}\right)^T
\left(\frac{1}{\x_3-(\X+T)}\right)^T{\cal J}\frac{1}{\y_3-(\Y+T^\dagger)}}\ .\nonumber
\eea

The procedure we use to compute this integral consists in integrating over
the last column (and by symmetry, over the first row) of the triangular matrices, so as to find a recursion on the size $n$ of the matrices, which 
takes $n$ to $n-2$. 

Define the submatrices $\hat{\X}$ and $\hat{\Y}$ by
\bea
\X=\diag{(\alpha,\hat{\X},-\alpha)}\, &;& \quad \Y=\diag{(\beta,\hat{\Y},-\beta)},
\eea
and the  $\CJ$-antisymmetric upper-triangular  matrices $\hat{T}$
of size $n-2$
\beq 
T=\left(\begin{array}{ccc} 
0 & T_{12}\qquad \dots  & T_{1n}\\
\vdots & \left(\begin{array}{ccc}
0 & & \raisebox{-1.5ex}{$\back\back\hat T$} \\
\vdots & \ddots & \\
0 & \dots & 0 
\end{array}\right) & \genfrac{}{}{0pt}{0}{\vdots}{T_{n-1\,n}} \\
0 & \dots & 0 
\end{array}\right)\ .
\eeq
With these definitions and the relations
\beq
\frac{1}{\x-(\X+T)}=\frac{1}{\x-\X}\sum_{n=0}^\infty\left(T\frac{1}{\x-\X}\right)^n=\sum_{n=0}^\infty\left(\frac{1}{\x-\X}T\right)^n\frac{1}{\x-\X}\ ,
\eeq
we can expand the resolvent of size $n$ in terms of the resolvent 
of size $n-2$ and of the variables to be integrated out 
\bea\label{expanres}
\left(\frac{1}{\x-(\X+T)}\right)_{i,j}&=&
\delta_{i,1}\delta_{j,1}\frac{1}{\x-\alpha}+\delta_{i,n}\delta_{j,n}\frac{1}{\x+\alpha}
+\delta_{i,1}\delta_{j,n}\frac{1}{\x-\alpha}T_{1,n}\frac{1}{\x+\alpha}\nonumber\\
&&+(1-\delta_{i,1}-\delta_{i,n})(1-\delta_{j,1}-\delta_{j,n})\left(\frac{1}{\x-(\hat{\X}+\hat{T})}\right)_{i,j}\nonumber\\
&&+\delta_{i,1}(1-\delta_{j,1}-\delta_{j,n})\frac{1}{\x-\alpha} \left[\sum_{k=2}^jT_{1,k} \left(\frac{1}{\x-(\hat{\X}+\hat{T})}\right)_{k,j}\right]  \nonumber\\
&&+(1-\delta_{i,1}-\delta_{i,n})\delta_{j,n}\left[ \sum_{l=i}^{n-1}\left(\frac{1}{\x-(\hat{\X}+\hat{T})}\right)_{i,l}T_{l,n} \right] \frac{1}{\x+\alpha} \nonumber \\
&&+\delta_{i,1}\delta_{j,n}\frac{1}{\x-\alpha} 
\left[\sum_{2\le k<l\le n-1}
T_{1,k}\left(\frac{1}{\x-(\hat{\X}+\hat{T})}\right)_{k,l}T_{l,n}\right] \frac{1}{\x+\alpha}
\ . 
\eea
Notice that $T_{1,n}$ in the $J$-antisymmetric case is identically zero, 
a fact that will be accounted for in the following.
In both cases, the only independent integration 
variables are the matrix elements of the first row.
Their propagators are read off the Gaussian weight, which is, in the 
$J$-antisymmetric case 
\beq
\e{-\frac{1}{2}\sum_{i=2}^{n-1}\left(|T_{1,i}|^2+|T_{i,n}|^2\right)}=
\e{-\sum_{i=2}^{n-1}|T_{1,i}|^2}
\eeq
while for the $\td{J}$-antisymmetric case it is
\beq
\e{-\frac{1}{2}\sum_{i=2}^{n-1}\left(|T_{1,i}|^2+|T_{i,n}|^2\right)-\frac{1}{2}|T_{1,n}|^2}=
\e{-\sum_{i=2}^{n-1}|T_{1,i}|^2-\frac{1}{2}|T_{1,n}|^2}.
\eeq
The independent nonzero  propagators are thus 
\bea\label{propagators}
\BK{T_{1,i}T^\dagger_{j,1}}&=&\delta_{i,j}\quad,\qquad
2\le i,j \le n-1\nonumber\\
\BK{T_{1,n}T^\dagger_{n,1}}&=& 1+b
\eea
where $b=-1$ (resp. $b=+1$) for the $J$-antisymmetric (resp. $\td{J}$-antisymmetric) case, so that 
in the $b=-1$ case, the propagator for $T_{1,n}$ is zero, as it should. 
The other propagators  encountered in the integration result
from the symmetry properties
\beq\label{prop}
\BK{T_{i,n}T^\dagger_{n,j}}=\delta_{i,j} \qquad
\BK{T_{1,i}T^\dagger_{n,j}}=-{\cal J}_{i,j}\qquad
\BK{T_{i,n}T^\dagger_{j,1}}=-\left({\cal J}^{-1}\right)_{i,j}\ .
\eeq

This is what is needed to perform the first step in the recursive computation
 of the triangular integrals. Let's take the simplest mixed case.


\subsection{Example: Morozov-like formula}\label{morozo}

The simplest case involves two resolvents. Define the two functions
\bea
F_{+}^{(n)}(\x,\y,A,B)&=&\tr{\frac{1}{\x-(\X+T)}\,\,\frac{1}{\y-(\Y+T^\dagger)}}+1\nonumber\\
F_{-}^{(n)}(\x,\y,A,B)&=&\tr{\frac{1}{\x-(\X+T)}{\cal J}
\left(\frac{1}{\y-(\Y+T^\dagger)}\right)^T{\cal J}}+b\ .\nonumber
\eea 
{The second one, $F_-^{(n)}$, is {\it twisted} by the action of $\CJ$.}
Using that,  for $A$ and $B$ two $\CJ$-antisymmetric matrices (such as
$X$, $Y$, $T$ or $\hat{T}$),
\bea\lbl{eq:twist}
J\left(\frac{1}{\x-A}\right)^TJ&=&\frac{1}{\x+A}\nonumber\\
\td{J}\left(\frac{1}{\x-A}\right)^T(-\td{J})&=&\frac{1}{\x+A}\ ,
\eea
i.e. $\CJ\left(\frac{1}{\x-A}\right)^T \CJ=\frac{b}{-\x-A}$, 
one sees that $F_{-}^{(n)}(\x,\y,A,B)=bF_{+}^{(n)}(\x,-\y,A,B)$, 
thus it suffices to carry out the integration over the last column 
and first row of $F_+$ only.

The computation goes as follows
\bea\label{intstep}
&&\back\back\BK{\tr{\frac{1}{\x-(\X+T)}\,\frac{1}{\y-(\Y+T^\dagger)}}+1}_{(n)}=
\left[1+\frac{1}{\x-\alpha}\frac{1}{\y-\beta}+\frac{1}{\x+\alpha}\frac{1}{\y+\beta}\right]\BK{1}_{(n-2)}\nonumber\\
&&\qquad\qquad\qquad+\frac{1}{\x^2-\alpha^2}\frac{1}{\y^2-\beta^2}\BK{1}_{(n-2)}\BK{T_{1,n}T^\dagger_{n,1}} +\BK{\tr{\frac{1}{\x-(\hat{\X}+\hat{T})}\frac{1}{\y-(\hat{\Y}+\hat{T}^\dagger)}}}_{(n-2)}\nonumber\\
&&\qquad\qquad+\frac{1}{\x-\alpha}\frac{1}{\y-\beta}\BK{\left(\frac{1}{\x-(\hat{\X}+\hat{T})} 
\frac{1}{\y-(\hat{\Y}+\hat{T}^\dagger)}\right)_{k,l}}_{(n-2)}\BK{T_{1,k}T^\dagger_{l,1}}\nonumber\\
&&\qquad\qquad+\frac{1}{\x+\alpha}\frac{1}{\y+\beta}\BK{\left(\frac{1}{\y-(\hat{\Y}+\hat{T}^\dagger)}\frac{1}{\x-(\hat{\X}+\hat{T})}\right)_{k,l}}_{(n-2)}\BK{T_{l,n}T^\dagger_{n,k}}\\
&&\qquad+\frac{1}{\x^2-\alpha^2}\frac{1}{\y^2-\beta^2}\BK{\left(\frac{1}{\x-(\hat{\X}+\hat{T})}\right)_{k,l}\left(\frac{1}{\y-(\hat{\Y}+\hat{T}^\dagger)}\right)_{k^\prime,l^\prime}}_{(n-2)}\BK{T_{1,k}T_{l,n}T^\dagger_{n,k^\prime}T_{l^\prime,1}^\dagger}\ .\nonumber
\eea
Inserting the propagators  given above we find
\bea\label{eq:comp1}
&&\BK{\tr{\frac{1}{\x-(\X+T)}\frac{1}{\y-(\Y+T^\dagger)}}+1}_{(n)}=\nonumber\\
&&\qquad\qquad\left[1+\frac{1}{\x-\alpha}\frac{1}{\y-\beta}\right]\left[1+\frac{1}{\x+\alpha}\frac{1}{\y+\beta}\right]\BK{\tr{\frac{1}{\x-(\hat{\X}+\hat{T})}\frac{1}{\y-(\hat{\Y}+\hat{T}^\dagger)}}+1}_{\!\!(n-2)}\nonumber\\
&&\qquad\qquad+\left[\frac{1}{\x^2-\alpha^2}\frac{1}{\y^2-\beta^2}\right]
\BK{\tr{\frac{1}{\x-(\hat{\X}+\hat{T})}{\cal J}\left(\frac{1}{\y-(\hat{\Y}+\hat{T}^\dagger)}\right)^T
{\cal J}}+b}_{\!\!(n-2)}\ .
\eea
We have split the two terms in the $\BK{T_{1,n}T^\dagger_{n,1}}=1+b$ propagator in the following way: the weight $1$ goes together with the untwisted minimal cycle and the weight $b$ with a twisted one. This is a general rule as we shall see later.

Notice that we need both functions $F_{\pm}$ in order to close the recursion relation. Defining the column vector ${\cal V}_n=(F_{+}^{(n)},F_{-}^{(n)})^{T}$, 
we obtain a recursion formula for the two functions in the form
\beq\lbl{eq:rec.1}
{\cal V}_n=\ovl{{\cal M}}(\x,\y,\alpha,\beta){\cal V}_{n-2}
\eeq
where
\beq
\ovl{{\cal M}}(\x,\y,\alpha,\beta)=\left(\begin{array}{cc}
\left(1+\frac{1}{\x-\alpha}\,\frac{1}{\y-\beta}\right)\left(1+\frac{1}{\x+\alpha}\,\frac{1}{\y+\beta}\right) & \frac{1}{\x^2-\alpha^2}\frac{1}{\y^2-\beta^2} \\
\frac{1}{\x^2-\alpha^2}\frac{1}{\y^2-\beta^2} & \left(1+\frac{1}{\x-\alpha}\,\frac{1}{(-\y)-\beta}\right)\left(1+\frac{1}{\x+\alpha}\,\frac{1}{(-\y)+\beta}\right) 
\end{array}\right)
\eeq

This structure will appear in the general case. 

\subsection{Last-row/first-column integration: general case}

{In the general case, the recursion involve combinations of correlation functions conveniently labelled
by graphs. } 


\subsubsection{Basis of correlation functions}\label{sssec:Basis}

In the example above, we had to mix correlation functions with two 
resolvents with correlation functions with a lesser 
number of resolvents, in order to write recursion relations. 
This  will still be necessary in the general case, and 
we shall see that the basis of correlation functions that we have to consider
is conveniently 
labelled by {\it tetrads}  $\omega=\{\sigma,\tau,s,t\} \in \SS_R\times\SS_R\times\Z_2^R\times\Z_2^R$, made of two permutations of  $\SS_R$ of $R$ objects 
and two sets of $R$ signs. This integer $R$ will turn out to be the 
maximal number of resolvents of $X$-type appearing in the correlation function.
\\
To each such tetrad, we first associate an oriented bicolored graph $G$ in the 
following way: 
$G$ has $2R$ vertices, $R$ of each color; the $i$th black (resp. white) vertex carry a sign $s(i)$ (resp. $t(i)$), $i=1,\cdots,R$.
An oriented edge connects each $i$th black (resp. white) vertex to the $\sigma(i)$th white (resp. $\tau^{-1}(j)$th black) vertex. The graph is thus made of  bicolored cycles. 

According to these rules, the graph representation of the tetrad 
\beq
\omega=\{(13)(24),(1)(243),(+,+,+,-),(+,-,-,+)\}
\eeq
 is
\begin{center}
\includegraphics{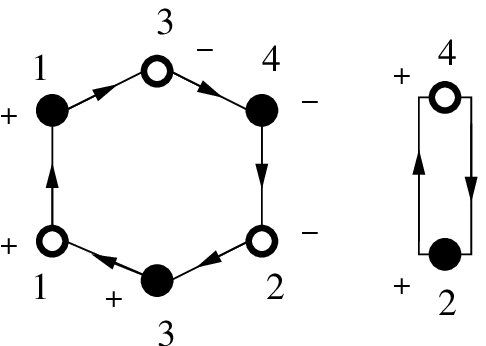}  
\end{center}
For our purposes we are only interested in the relative signs between vertices in each cycle. We say that two tetrads $\omega$ and $\omega^\prime$ are equivalent if they are equal up to independent global signs {\it in each cycle}. We take as representatives of the equivalence classes the tetrads with $s(i)=+$ for the black vertex with the smallest $i$ in each cycle. 

There exists also a graphical representation for the equivalence classes. 
We call  diagrams these new objects. Since we only care about the relative signs between vertices in the same cycle, we replace the $\Z_2$ variables by symbols representing changes of sign
\footnote{This operation is accompanied by a change of orientation of edges, 
and the cycles are no longer oriented.  But 
the original class of oriented graphs may be reconstructed 
from these diagrams.}.
We represent this with a short bar across the edges indicating the change of 
orientation of the edge. In the diagrams, the vertices will be called dots and the edges will be called links. 
The equivalence class 
\beq
[\omega]=[\{(13)(24),(1)(243),(+,+,+,-),(+,-,-,+)\}]
\eeq
is thus represented by the diagram
\begin{center}
\includegraphics{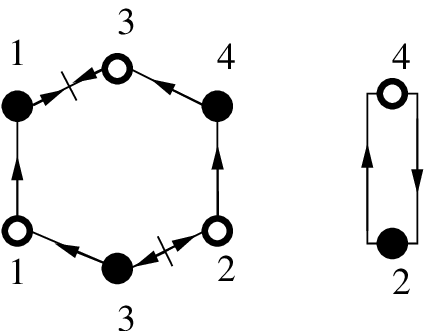} 
\end{center}
\vskip-12mm
In appendix \ref{app:bijection} we prove that the space of equivalence classes of tetrads is isomorphic to the space of permutations of size $2R$, so we can label the equivalence classes $[\omega]$ by an associated permutation $\pi\in\SS_{2R}$. This  very important  bijection will allow us later to relate the $\On{n}$ and the $\Spn{2m}$ integrals with the $U_n$ group integrals studied in \I.

The correlation functions we need turn out to be labelled by these equivalence classes, or equivalently by permutations of size $2R$. We define in general the $[\omega]$ component of the basis of correlation functions as follows.
Consider $[\omega]=[\{\sigma,\tau,s,t\}]$ and consider two sets of 
variables, $\{\x_1,\dots,\x_R\}$ for $\X$-type resolvents and 
$\{\y_1,\dots,\y_R\}$ for $\Y$-type resolvents. 
Then we define the following function,
\bea\label{BASIS1}
&&\qquad\qquad F^{J}_{\{\sigma,\tau,s,t\}}(\{\x\},\{\y\},A,B):=\nonumber\\
&&\prod_{k=1}^p\Bigg(t(j_{k,1})\delta_{R_k,1}+\tr{\prod_{l=1}^{R_k} \left(\frac{1}{\x_{i_{k,l}}-A}\right)^{\zeta^{\x}(i_{k,l})}
J^{\eta^{\x\to \y}_{k,l}}
\left(\frac{1}{\y_{j_{k,l}}-B}\right)^{\zeta^{y}(j_{k,l})}
J^{\eta^{\y\to \x}_{k,l}}}\Bigg)
\eea
for the $J$-antisymmetric integrals and
\bea\label{BASIS2}
&&\qquad\qquad F^{\td{J}}_{\{\sigma,\tau,s,t\}}(\{\x\},\{\y\},A,B):=\nonumber\\
&&\prod_{k=1}^p\Bigg(\delta_{R_k,1}+\tr{\prod_{l=1}^{R_k} 
\left(\frac{1}{\x_{i_{k,l}}-A}\right)^{\zeta^{\x}(i_{k,l})}
\td{J}^{\eta^{\x\to \y}_{k,l}}
\left(\frac{1}{\y_{j_{k,l}}-B}\right)^{\zeta^{\y}(j_{k,l})}
\td{J}^{\eta^{\y\to \x}_{k,l}}}\Bigg)
\eea
for the $\td{J}$-antisymmetric ones,
where $p$ is the number of cycles of the permutation 
$\sigma\tau^{-1}$, and $R_k$ is the length of the $k$th cycle.

The permutations  $\sigma$ and $\tau \in \SS_m$ yield the ordering  of the labels
\beq
\sigma(i_{k,l})=j_{k,l}\,,\,\, \mathrm{and} \quad \tau^{-1}(j_{k,l})=i_{k,l+1}\,,\,\, 
\mathrm{with}\quad i_{k,R_k+1}=i_{k,r}<i_{k,r}\,\quad r=2,R_k,
\eeq
and the signs $s$, satisfying the constraints
\bea\lbl{eq:const}
s(i_{k,1})=+1,
\eea
together with the signs $t$ define the functions
\bea
\eta^{\x\to \y}_{k,l}=\left\{
\begin{array}{ll} 
0 & \mathrm{if}\, s(i_{k,l})=t(j_{k,l})\\
1& \mathrm{if}\, s(i_{k,l})=-t(j_{k,l}) 
\end{array}\right.\nonumber\\
\eta^{\y\to \x}_{k,l}=\left\{
\begin{array}{ll} 
0 & \mathrm{if}\, t(j_{k,l})=s(i_{k,l+1})\\
1& \mathrm{if}\, t(j_{k,l})=-s(j_{k,l+1}) 
\end{array}\right.\nonumber
\eea
and the operations
\bea
\zeta^{\x}(i_{k,l})=\left\{
\begin{array}{ll} 
T\mathrm{(ranspose)} & \mathrm{if}\, s(i_{k,l})=-1\\
I\mathrm{(dentity)}& \mathrm{if}\, s(i_{k,l})=1
\end{array}\right.\nonumber\\
\zeta^{\y}(j_{k,l})=\left\{
\begin{array}{ll} 
T\mathrm{(ranspose)} & \mathrm{if}\, t(j_{k,l})=-1\\
I\mathrm{(dentity)}& \mathrm{if}\, t(j_{k,l})=1 
\end{array}\right.\nonumber
\eea
that perform the twisting of resolvents.

The structure of these functions is easily understood 
from the diagrams associated to $\pi$ (equivalently, $[\omega]$). Each cycle in the diagram represents by a trace; 
to each dot is attached  a resolvent if the dot is traversed 
clockwise by an
arrow, and a transposed resolvent if it is counterclockwise; finally
each change of orientation in the links corresponds to
 a $J$ or a $\td{J}$ matrix. 

The functions are invariant under
an independent  global twist inside each trace,   
so {the claim that our prescription depends only on} the equivalence classes defined {above} is justified. 

Finally the terms $\delta_{R_k,1}$ and $t(j)\delta_{R_k,1}$ 
in the definition of the functions are the analogues of the $1$ and $b$ appearing in $F_{\pm}$ in the example of sect. \ref{morozo}. Here too they come 
only with the traces containing two resolvents, {\it i.e.} $R_k=1$.

As an example the $[\{(1)(2)(3),(1)(23),(+,+,-),(-,-,+)\}]$ component of the basis for the orthogonal case would be 
\bea
&&\left(-1+\tr{\frac{1}{\x_1-(\X+T)}J\left(\frac{1}{\y_1-(\Y+T^\dagger)}\right)^TJ}\right)\times\nonumber\\
&&\tr{\frac{1}{\x_2-(\X+T)}J\left(\frac{1}{\y_2-(\Y+T^\dagger)}\right)^T
\left(\frac{1}{\x_3-(\X+T)}\right)^TJ\frac{1}{\y_3-(\Y+T^\dagger)}}\nonumber
\eea
which is represented by the diagram \\
\begin{center}
\includegraphics[scale=0.8]{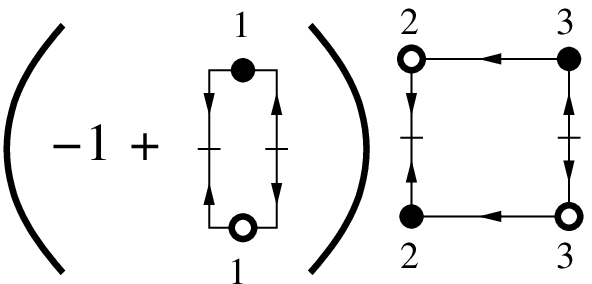}
\end{center}

There is a unified representation for the two bases of correlation functions
corresponding to the orthogonal and symplectic cases. 
Define $N_{\cal J}=\sum_{k,l} (\eta^{\x\to \y}_{k,l}+\eta^{\y\to \x}_{k,l})$ (which is the total number of ${\cal J}$ matrices appearing in the traces), 
and $\Pi(s)=\prod_i^R s(i)$ for $s\in\Z_2^R$. Then we have\footnote{Here we have used \eq{eq:twist} in order to remove the ${\cal J}$ and introduce the signs.}
\bea\label{BASIS3}
&&\qquad F^{\cal J}_{\{\sigma,\tau,s,t\}}(\{\x\},\{\y\},A,B)=\nonumber\\
&&=\frac{1}{n}\tr{{\cal J}^{N_{\cal J}}}\Pi(s)\Pi(t)\prod_{k=1}^p\Bigg(\delta_{R_k,1}+\tr{\prod_{l=1}^{R_k} 
\frac{1}{s(i_{k,l}) \x_{i_{k,l}}-A} \frac{1}{t(j_{k,l}) \y_{j_{k,l}}-B}}\Bigg)\nonumber\\
&&=\frac{1}{n}\tr{{\cal J}^{N_{\cal J}}}\Pi(s)\Pi(t)
\left(F^{U}_{\pi,e\pi e}(\{\x\}_{2R},\{\y\}_{2R},A,B)\right)^{\frac{1}{2}}\ .
\eea
where $\pi\in\SS_{2R}$ is the associated permutation following appendix \ref{app:bijection}, $e$, $\{\x\}_{2R}$ and $\{\y\}_{2R}$ are also defined in that appendix, 
and $F^{U}_{\pi,\pi^\prime}$ is the basis of 
correlation functions found in \I\ 
for the unitary case. The sign can be computed through the limit 
$A,B\to\infty$. We prove the last equality in appendix \ref{app:bijection}.


\subsubsection{Recursion relation}

Using this basis we have the following theorem 
\begin{theorem}\lbl{the:matrixM}
The functions defined in \eq{BASIS1} and \eq{BASIS2} or equivalently in \eq{BASIS3} satisfy the recursion relation
\bea
&&\BK{F^{\cal J}_{\{\sigma,\tau,s,t\}}(\{\x\},\{\y\},\X+T,\Y+T^\dagger)}_{(n)}=\nonumber\\
&&\sum_{\{\sigma^\prime,\tau^\prime,s^\prime,t^\prime\}}
\ovl{\cal M}_{(R)\{\sigma,\tau,s,t\}}^{\phantom{(R)}\{\sigma^\prime,\tau^\prime,s^\prime,t^\prime\}}
(\{\x\},\{\y\},\alpha,\beta)
\BK{F^{\cal J}_{\{\sigma^\prime,\tau^\prime,s^\prime,t^\prime\}}
(\{\x\},\{\y\},\hat{\X}+\hat{T},\hat{\Y}+\hat{T}^\dagger)}_{(n-2)}
\eea
where $\alpha$ and $\beta$ are the first eigenvalues of $\X$ and $\Y$ respectively, and $\hat{A}$ is the submatrix of size $n-2$ resulting 
from erasing the first and the last rows and columns of $A$, and where 
\bea
\ovl{\cal M}_{\phantom{(R)}\{\sigma,\tau,s,t\}}^{(R)\{\sigma^\prime,\tau^\prime,s^\prime,t^\prime\}}
(\{\x\},\{\y\},\alpha,\beta)&=&
\left(\prod_{i=1}^R(\delta_{\sigma(i),\sigma^\prime(i)} \delta_{s(i),s^\prime(i)}\delta_{t(\pi(i)),t^\prime(\pi(i))}
+\frac{1}{s(i) \x_i+\alpha}\frac{1}{t(\sigma(i)) \y_{\sigma(i)}+\beta})\right)\nonumber\\
&&\left(\prod_{i=1}^R(\delta_{\tau(i),\tau^\prime(i)}\delta_{s(i),s^\prime(i)}\delta_{t(\tau(i)),t^\prime(\tau(i))}
+\frac{1}{s(i) \x_i-\alpha}\frac{1}{t(\tau(i)) \y_{\tau(i)}-\beta})\right)\nonumber
\eea
\end{theorem}

\proof{The proof of this theorem is given in appendix \ref{app:matrixM}.}

The first thing to notice is that (see appendix \ref{app:bijection} for a proof) the matrix $\ovl{\cal M}$ is again closely related to the corresponding recursion matrix ${\cal M}$ found in \I\ for the unitary case. Recall that
\beq
{\cal M}^{(2R)}_{\pi,\pi^\prime}(\{\x\}\{\y\},\xi,\eta)=
\prod_{i=1}^R\left(\delta_{\pi(i),\pi^\prime(i)}+\frac{1}{\x_i-\xi}\frac{1}{\y_{\pi(i)}-\eta}\right),
\eeq
then taking into account the bijection defined in appendix \ref{app:bijection} we find that 
\beq\label{identM}
\ovl{\cal M}_{\phantom{(R)\,}\{\sigma,\sigma^\prime,s,s^\prime\}}^{(R)\,\{\tau,\tau^\prime,t,t^\prime\}} (\{\x\}_R,\{\y\}_R,\alpha,\beta)=
{\cal M}_{\pi,\pi^\prime}^{(2R)}(\{\x\}_{2R},\{\y\}_{2R},-\alpha,-\beta)
\eeq
which again connects  the orthogonal and symplectic cases 
with the unitary case. 
The precise relation between the arguments of the two sides of this
equation is defined in appendix \ref{app:bijection}. 
Trivial consequences of this fact are the commutativity property of $\ovl{\cal M}$
\beq\label{commutM}
\left[\ovl{\cal M}^{(R)}(\{\x\},\{\y\},\alpha,\beta),\ovl{\cal M}^{(R)}(\{\x\},\{\y\},\xi,\eta)\right]=0,
\eeq
and the symmetry ${\cal M}={\cal M}^T$.

The recursion relation we just found is %
valid for any value of $n$ such that $n\geq 3$. The special cases $n=2$ and $n=1$ correspond to the initial condition for the recursion relation in the even $n$ and odd $n$ case respectively.


\subsubsection{Initial conditions}
\label{IC}
Let us consider first the case $n$ even. For any $n>2$ even, the recursion relation is valid. The last step for $n=2$ requires a slightly more careful analysis. In a $2\times 2$ strictly upper triangular matrix, the only term is $T_{1,2}$. Since the ``last column/first row'' integration reduces to that of $T_{1,2}$, the procedure explained in appendix \ref{app:matrixM} is still valid, and the recursion relation can be naively applied just by considering that when $n=0$, all the traces are equal to zero in the correlation functions (or equivalently taking the strict limit where $\x_i$, $\y_i\to \infty$ for $i=1,\dots,R$). This gives us the initial condition vector
\beq
(I^{{\cal J}}_{0})_{\{\sigma,\tau,s,t\}}=
\tr{{\cal J}^{N_J}}\Pi(s)\Pi(t)\delta_{\sigma,\tau} \ .
\eeq

Explicitely, the two cases ${\cal J}=J$ and ${\cal J}=\td{J}$ are
\bea\label{neqtwo}
(I^{J}_{0})_{\{\sigma,\tau,s,t\}}=\Pi(s)\Pi(t)\delta_{\sigma,\tau}
\\
(I^{\td{J}}_{0})_{\{\sigma,\tau,s,t\}}=\delta_{\sigma,\tau}\ . \nonumber
\eea

The case $n=1$ corresponds to the case where there is no triangular matrix at all, thus no integration. The initial condition corresponds in this case to the vector
\beq\label{neqone}
(I^{J}_{1})_{\{\sigma,\tau,s,t\}}(\{\x\},\{\y\}) =
\tr{{\cal J}^{N_J}}\prod_{k=1}^p\left(t(j_{k,l}) \delta_{R_k,1}+\prod_{l=1}^{R_k} 
\frac{1}{\x_{i_{k,l}}\y_{j_{k,l}}}\right).
\eeq

\section{Correlation functions over \texorpdfstring{$\On{n}$}{O(n)} and \texorpdfstring{$\Spn{2m}$}{Sp(n)}: the final expression}\label{FinalExp}

In this section we find a determinantal formula for the correlation functions
 in sections \ref{HSOrt} and \ref{HSSymp}. We use need the matrix determinant Mdet defined as,
\bea
\mathrm{Mdet}(M)=\sum_{\sigma\in\SS(n)}(-1)^{\sigma}\prod_{i=1}^n M_{i,\sigma(i)}
\eea
where each $M_{i,j}$ is a matrix. This means that $\mathrm{Mdet}(M)$ is itself a matrix. The ordering of matrices in the product does not matter if the
matrices commute with one another.

\penalty -8000
{$\bullet$  {\bf The \texorpdfstring{$\On{2m}$}{O(2m)} case}}\\
\penalty 10000
Take theorem \ref{th:MainOrt} for $n=2m$ and the recursion relation found in section \ref{TI}. After some algebra we find
\bea\label{finalOrte}
&&\frac{\int_{\On{2m}}\D O F^{(R)}(\X,O\Y O^T)\e{-\gamma\tr{\X O\Y O^T}}}
{\int_{\On{2m}}\D O \e{-\gamma\tr{\X O\Y O^T}}}\nonumber\\
&&=\frac{\mathrm{Mdet}\left(
\e{2\gamma \X_{k}\Y_{j}}\ovl{\cal M}^{(R)}(\{\x\},\{\y\},\ii\X_{k},\ii\Y_{j})+
\e{-2\gamma \X_{k}\Y_{j}}\ovl{\cal M}^{(R)}(\{\x\},\{\y\},\ii\X_{k},-\ii\Y_{j})\right)_{k,j=1,\dots,m}
 } {\det\left({2\cosh{2\gamma \X_k \Y_j}}\right)_{k,j=1,\dots,m}}\,I_{0}^{J}  \nonumber\\
&&=\sum_{t\in \Z_2^m}\frac{\mathrm{Mdet}\left(
\e{2\gamma t_j\X_{k}\Y_{j}}\ovl{\cal M}^{(R)}(\{\x\},\{\y\},\ii\X_{k},\ii\ t_j\Y_{j})\right)_{k,j=1,\dots,m}
 } {\det\left({2\cosh{2\gamma \X_k \Y_j}}\right)_{k,j=1,\dots,m}}\,I_{0}^{J}  %
\eea
where $F^{(R)}$ is a vector with $(2R)!$ components, Mdet is a $(2R)!\times (2R)!$ matrix and $I_0^J$ is the vector defined in \eq{neqtwo}. 


%
{$\bullet$ {\bf The \texorpdfstring{$\On{2m+1}$}{O(2m+1)} case}}\\
Take now the $n=2m+1$ part of theorem \ref{th:MainOrt}. With the results of section \ref{TI} we find
\bea\label{finalOrto}
&&\frac{\int_{\On{2m+1}}\D O F^{(R)}(\X,O\Y O^T)\e{-\gamma\tr{\X O\Y O^T}}}
{\int_{\On{2m+1}}\D O \e{-\gamma\tr{\X O\Y O^T}}}\nonumber\\
&&=\frac{ \mathrm{Mdet}\left(
\e{2\gamma \X_{k}\Y_{j}}\ovl{\cal M}^{(R)}(\{\x\},\{\y\},\ii\X_{k},\ii\Y_{j})-
\e{-2\gamma \X_{k}\Y_{j}}\ovl{\cal M}^{(R)}(\{\x\},\{\y\},\ii\X_{k},-\ii\Y_{j})\right)_{k,j=1,\dots,m}
 } {\det\left({2\sinh{2\gamma \X_k \Y_j}}\right)_{k,j=1,\dots,m}} \, I_{1}^{J}(\{\x\},\{\y\}) \nonumber\\
&&=\sum_{t\in \Z_2^m}\left(\prod_{l=1}^m t_l\right) \frac{\mathrm{Mdet}\left(
\e{2\gamma t_j\X_{k}\Y_{j}}\ovl{\cal M}^{(R)}(\{\x\},\{\y\},\ii\X_{k},\ii\ t_j\Y_{j})\right)_{k,j=1,\dots,m}
 } {\det\left({2\sinh{2\gamma \X_k \Y_j}}\right)_{k,j=1,\dots,m}}\,I_{1}^{J}(\{\x\},\{\y\}) %
\eea
where $I_1^J$ is given in \eq{neqone}. 


%
{$\bullet$ {\bf The \texorpdfstring{$\Spn{2m}$}{Sp(2m)} case}}\\
Finally take theorem \ref{th:MainSp} and the corresponding part of section \ref{TI}. After some algebra 
\bea\label{finalSp}
&&\frac{\int_{\Spn{2m}}\D S F^{(R)}(\X,S\Y S^\dagger)\e{-\gamma\tr{\X S\Y S^\dagger}}}
{\int_{\Spn{2m}}\D S \e{-\gamma\tr{\X S\Y S^\dagger}}} \nonumber\\
&&=\frac{
\mathrm{Mdet}\left(
\e{2\gamma \X_{k}\Y_{j}}\ovl{\cal M}^{(R)}(\{\x\},\{\y\},\ii\X_{k},\ii\Y_{j})-
\e{-2\gamma \X_{k}\Y_{j}}\ovl{\cal M}^{(R)}(\{\x\},\{\y\},\ii\X_{k},-\ii\Y_{j})\right)_{k,j=1,\dots,m}
 } {\det\left({2\sinh{2\gamma \X_k \Y_j}}\right)_{k,j=1,\dots,m}}\,I_{0}^{\td{J}} \nonumber\\
&&=\sum_{t\in \Z_2^m}\left(\prod_{l=1}^m t_l\right) 
\frac{\mathrm{Mdet}\left(
\e{2\gamma t_j\X_{k}\Y_{j}}\ovl{\cal M}^{(R)}(\{\x\},\{\y\},\ii\X_{k},\ii t_j\Y_{j})\right)_{k,j=1,\dots,m}
 } {\det\left({2\sinh{2\gamma \X_k \Y_j}}\right)_{k,j=1,\dots,m}}\,I_{0}^{\td{J}} %
\eea
where $I_0^{\td{J}}$ is given in \eq{neqtwo}.

\section{Concluding remarks}\label{remarks}

{\bf A remark on contour deformation in matrix integrals}
\\
We first return  to the transformation of our original integrals over
two real antisymmetric, resp. two antiselfdual real quaternionic, matrices
into integrals over  $J$-, resp. $\td{J}$-, antisymmetric complex
matrices.
Define the following measure on 
$M_n(\C)\times M_n(\C)$, the set of pairs of two complex $n\times n$
matrices, 
\beq
\e{-\tr{\frac{\alpha_1}{2}M_1^2+\frac{\alpha_2}{2}M_2^2+\gamma M_1 M_2}}\D M_1\D M_2\ .
\eeq 
We will consider two hyperplanes of $M_n(\C)\times M_n(\C)$
and the measure on these hyperplanes induced by the measure above.

The first hyperplane is defined by the equations
\beq
\left. \begin{array}{rcl}
M_i&=&M_i^* \, \equiv \, A_i\\
A_i&=&-A_i^T
\end{array}\right\}\qquad i=1,2
\eeq
endowed with the induced measure \eq{MEASA} reproduces
the two real antisymmetric matrix integral of \ref{rantisym}. 
The second hyperplane, which describes  
 one complex $J$-antisymmetric matrix, i.e. $J{\cal A}_n(\C)$, is
defined by the equations
\beq
\left. \begin{array}{rcl}
M_1&=&M_2^\dagger \, \equiv \, M\\
J M&=&-M^T J
\end{array}\right.
\eeq
with the induced measure \eq{MEASJ}.

In this construction, the real antisymmetric two matrix integrals and 
the complex $J$-antisymmetric matrix integrals are 
nothing but the same integral on different hyperplanes of 
$M_n(\C)\times M_n(\C)$. 
By the counting  done above, these two hyperplanes have the same dimension, and
a plausible interpretation of theorem \ref{MainOrt} is that it results 
from a contour deformation 
taking the first set of matrix integrals into the second one.

Similarly we can consider two different hyperplanes in the space of pairs 
of quaternionic matrices, namely the one defining 
$Q{\cal A}_{m}\times Q{\cal A}_{m}$
\beq
\left. \begin{array}{rcl}
M_i&=&M_i^* \, \equiv \, Q_i\\
Q_i&=&-{Q_i}^\dagger
\end{array}\right\}\qquad i=1,2
\eeq
with the induced measure \eq{MEASQ}, and the one defining $\td{J}{\cal A}_{2m}$
\beq
\left. \begin{array}{rcl}
M_1&=&M_2^\dagger \, \equiv \, M\\
\td{J} M&=&-M^T \td{J}
\end{array}\right.
\eeq
with the induced measure \eq{MEAStdJ}.
Both hyperplanes have the same dimension 
and we may again interpret Theorem \ref{MainSymp} as resulting from a contour 
deformation. 

This apparently much simpler and intuitive approach has the drawback of neglecting convergence issues. This is the reason why we chose to prove our results by use of loop equations.

Other choices of deformations to other hyperplanes are also conceivable.

\medskip
{\bf Comparison with the Duistermaat-Heckman form}
\\
\omit{First observe that we may rewrite the main theorems \ref{th:MainOrt} and \ref{th:MainSp} together with theorem 4-1 in \I\ 
into the unified form presented in (\ref{unifiintro})
\bea\label{unifi}
&&\int_{{G}}\D \Omega\,  
{F(\X^a ,\Omega \Y^a \Omega^{-1})\,\, \e{ -\tr {\X^a \Omega \Y^a 
\Omega^{-1}} }}
\nonumber\\
&&\phantom{\int_{{G}}\D \Omega}= c \sum_{w\in \CW} 
\frac{\e{{+} \tr{\Xp w(\Yp)}}}{\prod_{\alpha>0}\alpha(\Xp) \alpha(w(\Yp))}  
 %
\int_{T_G}\D T F(i\Xp+T,iw(\Yp)+T^\dagger) \e{-\tr{TT^\dagger}}
\eea
where $w$ is summed over the Weyl group $\CW$ as in sect.
\ref{DHthm}.}
First observe that we may rewrite the main theorems \ref{th:MainOrt} and \ref{th:MainSp} together with theorem 4-1 in \I\ 
into a unified form involving an integration over a set $T_G$ of complex
triangular matrices, as already mentionned in the Introduction. 
 When ${G}$ is $O(n)$, $Sp(2m)$ or $U(n)$,
 $T_{{G}}$ corresponds respectively to the set of
$J$, $\td{J}$-antisymmetric, or unconstrained, strictly upper triangular
matrices.
Moreover, we notice that in each case, the set $T_G$ is precisely
the derived ideal $%
[\gb,\gb]=:\gn_+$ of the Borel
subalgebra $\gb$ associated with the choice of Cartan algebra
made above. This is in fact the space generated by the positive roots
$\gn_+=\oplus_{\alpha>0}\,\gg^\alpha$. 

It is thus natural to conjecture 
that an analogous formula 
holds for any compact group, with the %
identification of $T_G$ %
with the subalgebra $\gn_+$:
\begin{conjecture}
For any compact group $G$
\bea\label{unifiiintro}
\omit{
&&{\prod_{\alpha>0}\alpha(X) \alpha(Y)}  \int_{{G}}\D \Omega F(\X,Ad(\Omega)\Y) \e{-(\X,Ad(\Omega)\Y)}=\nonumber\\
&&c_n \sum_{w\in \CW}\varepsilon_w  %
{\e{{+}(\X,w(\Y))}}
\int_{\gn_+=[\gb,\gb]}\D T F(i\X+T,iw(\Y)+T^\dagger) \e{-\tr{TT^\dagger}}}
&&\int_{{G}}\D \Omega\,  
{F(\X^a ,\Omega \Y^a \Omega^{-1})\,\, \e{ -\tr {\X^a \Omega \Y^a 
\Omega^{-1}} }}
\nonumber\\
&&\phantom{\int_{{G}}\D \Omega}= c \sum_{w\in \CW} 
\frac{\e{{+}\tr{\Xp w(\Yp)}}}{\prod_{\alpha>0}\alpha(\Xp) \alpha(w(\Yp))}  
\int_{\gn_+=[\gb,\gb]}\D T F(i\Xp+T,iw(\Yp)+T^\dagger) \e{-\tr{TT^\dagger}}\ .
\eea
\end{conjecture}

This form of our result has to be confronted
with the form given by \DH's localization theorem, eq. (\ref{DHun}).
{Note that the integration over $\gn^+$ %
plays here the role played in
sect. \ref{DHthm} by the integration over the ``fluctuations''
$A\in \gg\backslash \gh$ in (\ref{DHun}). This points
to a possible much more compact and geometric derivation of our results.}

\medskip
{\bf Other comments}

First, it is remarkable that the recursion on $n$ for triangular integrals 
involves the {\it same} matrix ${\cal M}$ for all cases U($n$), O($n$), Sp($2n$).
Only the initial conditions differ. This fact needs to be understood, and it shows that the matrix ${\cal M}$ is universal. Moreover, its commutation properties suggest the existence of some underlying integrable structure.
The symmetries of the group under consideration are reflected in the symmetries of the spectral parameters at which ${\cal M}$ is evaluated.
The initial conditions  also seem to have such symmetries, and it is remarkable that those symmetries are reminiscent of root lattices of size $R$ or $2R$ (we started with a root lattice of size $n$ or $2n$).
This suggests a duality between $R$ and $n$, similar to that of the ``supersymmetric'' method of evaluation of determinantal correlation functions 
\cite{Efetov}.
Remarkably, the triangular matrix ensembles as those considered here seem to play an important role in generalizations of the so-called Razumov-Stroganov conjecture. Indeed multidegrees of the corresponding matrix varieties are solutions of the quantum Knizhnik-Zamolodchikov equation based on the root systems of type $A$, $B$, $C$ and $D$ \cite{DFZJ}, thus pointing again towards some possible integrable structure.
{It might be also interesting to attack the ``angular'' integrals 
considered in this paper with 
character expansion techniques, see \cite{Fyo} for recent references.}

Our last comment is that it would be highly desirable to know how to 
compute integrals
like (\ref{a.a}) on other orbits. For example, little is known about the
integral over the O($n$) group when $\X$  and $\Y$ are {\it symmetric} real
matrices (see however \cite{BrHi02}). %

\subsection*{Acknowledgements}

We would like to thank M. Bauer and M. Talon for very helpful discussions.
This work was supported by the Enigma European RTN network under contract
 MRT-CT-2004-5652 and by the ESF program Misgam, and partly supported by 
the ANR project G\'eom\'etrie et int\'egrabilit\'e en physique math\'ematique ANR-05-BLAN-0029-01,
and by the Enrage European network MRTN-CT-2004-005616.
B.E. thanks the CRM (Montreal QC) for its support.
J.-B. Z. thanks KITP, Santa Barbara,
for hospitality and support, where part of this 
work was carried out, with partial support
by the National Science  Foundation under Grant No. PHY99-07949.


{\subsection*{Note Added:}}

At the date of this resubmission, conjecture \ref{unifiiintro} has been recently proved in \cite{TimeTravel} giving a full meaning to each of the elements appearing in the conjecture. The role of the borel subalgebra and the complexification procedure is thus quite well understood.


\appendix

\vspace{1cm}

\noindent{\Large {\bf Appendices}}
\penalty 10000
\section{Quaternions}
\label{app:Q}

\def\({\left(}\def\){\right)}
\def\bun{{\bf 1}}
We review here some well known facts about quaternions to 
fix our notations, which follow basically those of 
\cite{Mehta}. We shall only consider the set of {\it real quaternions}, 
which is the algebra over $\R$ generated by 4 elements:
the neutral element  $e_0$, which by an abuse of notation
we often write $e_0=1$,  and  $e_i$, $i=1,2,3$, 
\beq\label{Qdef}
 q= q^{(0)} e_0+ q^{(1)} e_1 + q^{(2)} e_2 +q^{(3)} e_3\qquad q^{({\alpha})}
\in \R
\eeq
with  multiplication $e_i^2=e_1 e_2 e_3=-1$, from which it follows 
that  $e_1 e_2 =-e_2 e_1 = e_3$ and its cyclic permutations.
One may represent $e_0$ by $\mathrm{Id}_2$ the $2\times 2$ identity matrix, and
the $e_i$ in terms of $2\times 2$ Pauli matrices 
by  
\beq\label{Pauli}e_i=-i \sigma_i\ .\eeq 
The conjugate quaternion of $q$ is defined as  
$$\bar q=q^{(0)} 1- q^{(1)} e_1 - q^{(2)} e_2 -q^{(3)} e_3 \ .$$
(This is also called  {\it hermitian conjugate}, which is 
justified by  the fact that Pauli matrices are hermitian).
Note that $q \bar q:= 
|q|^2=|q^{(0)}|^2+ |q^{(1)}|^2 + |q^{(2)}|^2 +|q^{(3)}|^2$, 
the square norm of the quaternion, and hence $q\ne 0$ has an inverse
$q^{-1}=\bar q/|q|^2$. 
Real quaternions form a non-commutative field. 
Note also that conjugation %
reverses the order of factors  of a product
$ \overline{(q_1 q_2)}=\bar q_2 \bar q_1
\ .$

\medskip
{\bf Quaternionic matrices}
\\ \noindent
We now consider matrices $Q$ with real quaternionic elements
$Q_{ij}$, $i,j=1,\cdots,m$. Alternatively, using (\ref{Pauli}),
one may regard also $Q$ as a $2m\times 2m$ matrix with 
$2\times 2$ blocks made of real combinations of $\mathrm{Id}_2$ and the Pauli 
matrices. One may apply to $Q$ the
same conjugation as defined above. One may also transpose $Q$. 
The {\it dual} $Q^R$ of a quaternionic matrix  $Q$ is the matrix
\beq (Q^R)_{ij}= \bar Q_{ji}\ .\eeq
(This is also the  hermitian conjugate $Q^\dagger$ of $Q$ in the usual sense.)
A real quaternionic matrix is thus {\it self-dual} if 
\beq Q^R=Q=Q^\dagger= \(Q_{ij}\)
=\(\bar Q_{ji}\)\ . \eeq
A real quaternionic matrix is  {\it anti self-dual} if 
\beq Q^R=Q^\dagger= -Q\ ;\eeq
it is thus anti-hermitian.
In particular, its diagonal matrix elements are such that $Q^{(0)}_{ii}=0$. 

On quaternionic matrices, we may define the ordinary trace
$$\tr{Q}=\sum_{i=1}^m Q_{ii}\ , $$
which is in general a quaternion, or
\beq\label{tro}
\tro{Q}=2\sum_{i=1}^m Q^{(0)}_{ii}= \tr{Q}+ \overline{\tr{Q}}
\eeq
which is a scalar. Note that $\tro{Q}$ is nothing else 
than the trace of the corresponding $2m\times 2m$ matrix.

\bigskip
{\bf Symplectic group} $\Spn{2m}$
\\
\medskip
\noindent Let $\mathbb{H}$ be the  space of real
quaternions. 
Consider the hermitian form on
$\mathbb{H}^m$
\beq (x,y)= \sum_{i=1}^m \bar{x_i} y_i\ . \eeq 
The {\it compact} unitary symplectic group  Sp($2m$) is defined as the invariance group of that form, and 
is thus the group of $m\times m$ 
real quaternionic matrices $Q$ such that
\beq {\bar Q}^T Q=I \qquad {\rm or} \qquad Q^RQ=I \ . \eeq 
   These matrices may be called unitary real quaternionic 
matrices. The Lie algebra of  Sp($2m$) is generated by 
real quaternionic matrices $A$ satisfying the infinitesimal
version of (A-8), 
\beq {\bar A}^T + A=0 \qquad {\rm or} \qquad A^R=A^\dagger
=-A \ , \eeq 
hence by {\it antiselfdual real quaternionic} matrices.


\section{Loop equations I}\label{app:I}

In this appendix we show how to compute loop equations in the real antisymmetric two matrix integral and in the $J$-antisymmetric complex matrix integral.

\subsection{Loop equations for the 2 real antisymmetric matrix integral}

\subsubsection{Loop equations}

Schwinger--Dyson equations, also called loop equations in the case of matrix integrals, merely amount to saying that the integral of a total derivative vanishes:
\beq
0=\sum_{i<j} \int dA_1 dA_2 \,\,\frac{\partial}{ \partial A_{1ij}}\,\left( f(A_1,A_2)_{ij}\,\e{-\tr{ \frac{\alpha_1}{2}
A_1^2 +\frac{\alpha_2}{2}A_2^2+\gamma A_1 A_2}} \right)
\eeq
where $f(A_1,A_2)=-f^t(A_1,A_2)$ is any sufficiently regular matrix valued function; in particular $f$ can be any non-commutative polynomial in $A_1$ and $A_2$, and may contain also product of traces of polynomials.

The loop equation thus turns into an equality between expectation values:
\beq
\left< K_1(f) \right> = \left< \tr{ (\alpha_1 A_1 + \gamma A_2) f(A_1,A_2)} \right>
\eeq
where
\beq
K_1(f) = \sum_{i<j} \frac{\partial f(A_1,A_2)_{ij}}{ \partial A_{1ij}}\ .
\eeq
Notice that $K_1(f)$ is linear and satisfies Leibniz rule:
\beq
K_1(fg) = K_1(f) g + f K_1(g)\ .
\eeq

\medskip

The most general $f$ we shall consider is of the type:
\beq\label{apBdeff}
f(A_1,A_2) = \frac{1}{ 2}(F_0(A_1,A_2) -F_0(A_1,A_2)^T) \, \prod_{r=1}^R \tr{F_r(A_1,A_2)}
\eeq
where $F_0$ is an odd degree non commutative monomial of $A_1$ and $A_2$:
\beq\label{apBdefF0}
F_0(A_1,A_2) = A_2^{l_{0,0}} A_1^{k_{0,1}}A_2^{l_{0,1}} A_1^{k_{0,2}}\dots A_2^{l_{0,p_0-1}} A_1^{k_{0,p_0}} A_2^{l_{0,p_0}}
\eeq
and each $F_r$ with $r\geq 1$ is an even degree non commutative monomial:
\beq\label{apBdefFr}
F_r(A_1,A_2) = A_1^{k_{r,1}} A_2^{l_{r,1}} A_1^{k_{r,2}}\dots A_2^{l_{r,p_r-1}} A_1^{k_{r,p_r}} A_2^{l_{r,p_r}}
\eeq
and we call $\deg(f)$ the total number of matrices $A_1$ + the total number of matrices $A_2$.

Then compute:
\bea\label{splitruleAsym}
K_1(F_0-F_0^T)
&=& \sum_{q=1}^{p_0} \sum_{m=0}^{k_{0,q}-1} \sum_{i<j}\,
\left[
(A_2^{l_{0,0}} A_1^{k_{0,1}}A_2^{l_{0,1}} \dots A_2^{l_{q-1}} A_1^{m})_{ii} (A_1^{k_{0,q}-m-1} A_2^{l_{0,q}} \dots A_2^{l_{0,p_0}} )_{jj} \right.\\
&&\qquad\qquad\qquad
 - %
(A_2^{l_{0,0}} A_1^{k_{0,1}}A_2^{l_{0,1}} \dots A_2^{l_{q-1}} A_1^{m})_{ij} (A_1^{k_{0,q}-m-1} A_2^{l_{0,q}} \dots A_2^{l_{0,p_0}} )_{ij} \cr
&&\qquad\qquad\qquad
 - %
(A_2^{l_{0,0}} A_1^{k_{0,1}}A_2^{l_{0,1}} \dots A_2^{l_{q-1}} A_1^{m})_{ji} (A_1^{k_{0,q}-m-1} A_2^{l_{0,q}} \dots A_2^{l_{0,p_0}} )_{ji} \cr
&& \left.\qquad\qquad\qquad
+ %
(A_2^{l_{0,0}} A_1^{k_{0,1}}A_2^{l_{0,1}} \dots A_2^{l_{q-1}} A_1^{m})_{jj} (A_1^{k_{0,q}-m-1} A_2^{l_{0,q}} \dots A_2^{l_{0,p_0}} )_{ii} \right]\cr
&=& \sum_{q=1}^{p_0} \sum_{m=0}^{k_{0,q}-1}
\left[
\tr{A_2^{l_{0,0}} A_1^{k_{0,1}}A_2^{l_{0,1}} A_1^{k_{0,2}}\dots A_2^{l_{0,q-1}} A_1^{m}}  \tr{A_1^{k_{0,q}-m-1} A_2^{l_{0,q}} \dots A_2^{l_{0,p_0}}} \right.\cr
&& \!\!\!\!\!\!\!\!\!\!\!\!\!\!\!\!\!\!\!\!\!\!\!\!\!\!\!\!\!\!\!\!\!\!\!\!\!\!\!\!\!\!\!\!\!
 \left.- %
(-1)^{k_{0,1}+\dots+k_{0,q-1}+m+l_{0,0}+\dots+l_{0,q-1}}
\tr{A_1^m A_2^{l_{0,q-1}} \dots  A_1^{k_{0,2}} A_2^{l_{0,1}}  A_1^{k_{0,1}}    A_2^{l_{0,0}} A_1^{k_{0,q}-m-1} A_2^{l_{0,q}} \dots A_2^{l_{0,p_0}} }\right]
\nonumber
\eea
This equality is known as the split rule.

Then we have for any antisymmetric matrix $C$:
\bea\label{mergeruleAsym}
K_1(C\tr{F_r})
&=& \sum_{q=1}^{p_r} \sum_{m=0}^{k_{r,q}-1} \sum_{i<j}\,\sum_s
\left[
(A_1^{k_{r,1}}A_2^{l_{r,1}} A_1^{k_{r,2}}\dots A_2^{l_{r,q-1}} A_1^{m})_{si}\, C_{ij}\,(A_1^{k_{r,q}-m-1} A_2^{l_{r,q}} \dots A_2^{l_{r,p_r}} )_{js}\right. \cr
&&\qquad\qquad\qquad\quad\left.-(A_1^{k_{r,1}}A_2^{l_{r,1}} A_1^{k_{r,2}}\dots A_2^{l_{r,q-1}} A_1^{m})_{sj}\, C_{ij}\,(A_1^{k_{r,q}-m-1} A_2^{l_{r,q}} \dots A_2^{l_{r,p_r}} )_{is}
\right]
 \cr
&=& \sum_{q=1}^{p_r} \sum_{m=0}^{k_{r,q}-1}
\tr{ A_1^{k_{r,1}}A_2^{l_{r,1}} A_1^{k_{r,2}}\dots A_2^{l_{r,q-1}} A_1^{m}\, C\, A_1^{k_{r,q}-m-1} A_2^{l_{r,q}} \dots A_2^{l_{r,p_r}} }\ . %
\eea
This equality is known as the merge rule.

\medskip
Due to Leibniz rule and using repeatedly the split and merge rules,
we find that if $f$ has the form of eq.\eqref{apBdeff} then $K_1(f)$ is a linear combination of monomial invariant functions of degree $\leq \deg(f)-1$.

The loop equations read:
\bea
\left< \tr{\alpha_1 A_1 f + \gamma A_2 f }\right>=K_1(f) \cr
\left<\tr{\alpha_2 A_2 f + \gamma A_1 f }\right>=K_2(f)
\eea
or equivalently:
\bea
\left< \tr{A_1 f} \right>= \frac{\alpha_2}{\delta}K_1(f) - \frac{\gamma}{\delta}K_2(f)\cr
 \cr
\left< \tr{A_2 f} \right>= \frac{\alpha_1}{ \delta}K_2(f) - \frac{\gamma}{ \delta}K_1(f)\cr
\eea

\subsubsection{Polynomial invariant functions}

$F(A_1,A_2)$ is a monomial invariant function of  two antisymmetric matrices $A_1,A_2$, if it is either:
\beq
\left\{
\begin{array}{ll}
& F=1 \cr
\hbox{or} & F(A_1,A_2)=\tr{A_1 f(A_1,A_2)} \cr
\hbox{or} & F(A_1,A_2)=\tr{A_2 f(A_1,A_2)}
\end{array}
\right.
\eeq
where $f$ is of the following form:
\beq
f(A_1,A_2) = F_0(A_1,A_2)  \, \prod_{r=1}^R \tr{F_r(A_1,A_2)}
\eeq
where $F_0$ is an odd degree non commutative monomial of $A_1$ and $A_2$:
\beq\label{apBdefF0bis}
F_0(A_1,A_2) = A_2^{l_{0,0}} A_1^{k_{0,1}}A_2^{l_{0,1}} A_1^{k_{0,2}}\dots A_2^{l_{0,p_0-1}} A_1^{k_{0,p_0}} A_2^{l_{0,p_0}}
\eeq
and each $F_r$ with $r\geq 1$ is an even degree non commutative monomial:
\beq\label{apBdefFrbis}
F_r(A_1,A_2) = A_1^{k_{r,1}} A_2^{l_{r,1}} A_1^{k_{r,2}}\dots A_2^{l_{r,p_r-1}} A_1^{k_{r,p_r}} A_2^{l_{r,p_r}}
\eeq
and we call $\deg(F)$ the total number of matrices $A_1$ + the total number of matrices $A_2$.
Notice also that $f$ can be antisymmetrized without changing $F$, and thus $f$ can be taken of the form of eq.\eqref{apBdeff}.

Notice that if $\deg(F)$ is odd, we have:
\beq
\left< F \right>=0
\eeq

If $F=1$, i.e. if $\deg(F)=0$ we have:
\beq
\left< 1\right>=1
\eeq
and if $\deg(F)>0$, and $F=\tr{A_1 f}$, the loop equations imply:
\beq
\left< F \right> = \frac{\alpha_2}{\delta}K_1(f) - \frac{\gamma}{\delta}K_2(f)
\eeq
where the right hand side is the expectation value of a polynomial invariant function of degree $\leq \deg(F)-2$.
And if $\deg(F)>0$, and $F=\tr{A_2 f}$, the loop equations imply:
\beq
\left< F \right> = \frac{\alpha_1}{ \delta}K_2(f) - \frac{\gamma}{\delta}K_1(f)
\eeq
where again the right hand side is the expectation value of a polynomial invariant function of degree $\leq \deg(F)-2$.

\smallskip

In other words, the loop equations allow to compute every expectation of polynomial invariant functions by recursion on the degree.

\smallskip

Notice also that the expectation value of any monic monomial invariant function is a polynomial in $\frac{\alpha_1}{\delta}$, $\frac{\alpha_2}{\delta}$ and $\frac{\gamma}{\delta}$.


\subsection{Loop equations for the complex \tops{$J$}{J}-antisymmetric matrix integral}\label{JAloop}

\subsubsection{Loop equations}

Similarly to the previous section, loop equations, in the case of a complex $J$-antisymmetric matrix integral, can be written:
\bea
0&=& \sum_{i<j} \int dM \,\,\left(\frac{\partial}{\partial \Re M_{i,n+1-j}}-i\frac{\partial}{\partial \Im M_{i,n+1-j}}\right)\,\left(  f(M,M^\dagger)_{i,n+1-j}\,\e{-\tr{ \frac{\alpha_1}{2}M^2 +\frac{\alpha_2}{2}M^{\dagger 2}+\gamma M M^\dagger}} \right)\cr
\eea
where $f(M,M^\dagger)=-J f(M,M^\dagger)^T J$ is any sufficiently regular matrix valued function, in particular $f$ can be any non-commutative polynomial in $M$ and $M^\dagger$, and may contain also product of traces of polynomials.

The loop equation thus turns into an equality between expectation values:
\beq
\left< K_1(f) \right> = \left< \tr{ (\alpha_1 M + \gamma M^\dagger) f(M,M^\dagger)} \right>
\eeq
where
\beq
K_1(f) = \frac{1}{2}\sum_{i<j} \left(\frac{\partial}{\partial \Re M_{i,n+1-j}} - i\frac{\partial}{\partial \Im M_{i,n+1-j}}\right)\, f_{i,n+1-j}\ .
\eeq
Notice that $K_1(f)$ is linear and satisfies Leibniz rule:
\beq
K_1(fg) = K_1(f) g + f K_1(g)
\eeq

\medskip

The most general $f$ we shall consider is of the type:
\beq\label{apBdeffCJ}
f(M,M^\dagger) = \frac{1}{2}(F_0(M,M^\dagger) -J F_0(M,M^\dagger)^T J) \, \prod_{r=1}^R \tr{F_r(M,M^\dagger)}
\eeq
where $F_0(M,M^\dagger)$ is an odd degree non commutative monomial of $M$ and $M^\dagger$:
\beq\label{apBdefF0CJ}
F_0(A_1,A_2) = M^{\dagger l_{0,0}} M^{k_{0,1}}M^{\dagger l_{0,1}} M^{k_{0,2}}\dots M^{\dagger l_{0,p_0-1}} M^{k_{0,p_0}} M^{\dagger l_{0,p_0}}
\eeq
and each $F_r(M,M^\dagger)$ with $r\geq 1$ is an even degree non commutative monomial:
\beq\label{apBdefFrCJ}
F_r(M,M^\dagger) = M^{k_{r,1}} M^{\dagger l_{r,1}} M^{k_{r,2}}\dots M^{\dagger l_{r,p_r-1}} M^{k_{r,p_r}} M^{\dagger l_{r,p_r}}
\eeq
and we call $\deg(f)$ the total number of matrices $M$ + the total number of matrices $M^\dagger$.

We have
\bea\label{splitruleCJ}
&& K_1(F_0-JF_0^T J) \cr
&=& \sum_{q=1}^{p_0} \sum_{m=0}^{k_{0,q}-1} \sum_{i<j}
\left[
\left(M^{\dagger l_{0,0}} M^{k_{0,1}} \dots M^{\dagger l_{0,q-1}} M^{m}\right)_{i,i} \left(M^{k_{0,q}-m-1} M^{\dagger l_{0,q}}  \dots  M^{k_{0,p_0}} M^{\dagger l_{0,p_0}}\right)_{n+1-j,n+1-j} \right.\cr
&& \qquad\qquad\quad - 
\left(M^{\dagger l_{0,0}} M^{k_{0,1}} \dots M^{\dagger l_{0,q-1}} M^{m}\right)_{i,j} \left(M^{k_{0,q}-m-1} M^{\dagger l_{0,q}}  \dots  M^{k_{0,p_0}} M^{\dagger l_{0,p_0}}\right)_{n+1-i,n+1-j} \cr
&& \qquad\qquad\quad -
\left(M^{\dagger l_{0,0}} M^{k_{0,1}} \dots M^{\dagger l_{0,q-1}} M^{m}\right)_{j,i} \left(M^{k_{0,q}-m-1} M^{\dagger l_{0,q}}  \dots  M^{k_{0,p_0}} M^{\dagger l_{0,p_0}}\right)_{n+1-j,n+1-i} \cr
&& \left.  \qquad\qquad\quad +
\left(M^{\dagger l_{0,0}} M^{k_{0,1}} \dots M^{\dagger l_{0,q-1}} M^{m}\right)_{j,j} \left(M^{k_{0,q}-m-1} M^{\dagger l_{0,q}}  \dots  M^{k_{0,p_0}} M^{\dagger l_{0,p_0}}\right)_{n+1-i,n+1-i}\right] \cr
&=& \sum_{q=1}^{p_0} \sum_{m=0}^{k_{0,q}-1}
\left[\tr{ M^{\dagger l_{0,0}} M^{k_{0,1}} \dots M^{\dagger l_{0,q-1}} M^{m}} \tr{ M^{k_{0,q}-m-1} M^{\dagger l_{0,q}}  \dots  M^{k_{0,p_0}} M^{\dagger l_{0,p_0}} } \right.\\
&&\!\!\!\!\!\!\!\!\!\!
\left.
  - %
(-1)^{l_{0,0}+\dots+l_{0,q-1}+k_{0,1}+\dots+k_{0,q-1}+m} %
\tr{ M^{m} M^{\dagger l_{0,q-1}} \dots  M^{k_{0,1}}  M^{\dagger l_{0,0}} M^{k_{0,q}-m-1} M^{\dagger l_{0,q}}  \dots  M^{k_{0,p_0}} M^{\dagger l_{0,p_0}}  }
\right]\nonumber
\eea
Notice that the  split rule for $J$-antisymetric complex matrices is identical to the split rule for real antisymmetric matrices eq.\eqref{splitruleAsym}.

Similarly, we compute the merge rule (where $C=-J C^T J$ is any J-antisymmetric complex matrix):
\bea\label{mergeruleCJ}
&& K_1(C \tr{F_r}) \cr
&=& \sum_{q=1}^{p_r} \sum_{m=0}^{k_{0,q}-1} \sum_{i<j} \sum_s
\left[\left(M^{k_{r,1}} \dots M^{\dagger l_{r,q-1}} M^m\right)_{s,i}\, C_{i,n+1-j}\, \left( M^{k_{r,q}-m-1} M^{\dagger l_{r,q}}  \dots  M^{k_{r,p_r}} M^{\dagger l_{r,p_r}}\right)_{n+1-j,s}\right. \cr
&&\left.\qquad\qquad\qquad
 -%
\left(M^{k_{r,1}} \dots M^{\dagger l_{r,q-1}} M^m\right)_{s,j}\, C_{i,n+1-j}\, \left( M^{k_{r,q}-m-1} M^{\dagger l_{r,q}}  \dots  M^{k_{r,p_r}} M^{\dagger l_{r,p_r}}\right)_{n+1-i,s}\right] \cr
&=& \sum_{q=1}^{p_r} \sum_{m=0}^{k_{0,q}-1}
\tr{ M^{k_{r,1}} \dots M^{\dagger l_{r,q-1}} M^m \, C\, M^{k_{r,q}-m-1} M^{\dagger l_{r,q}}  \dots  M^{k_{r,p_r}} M^{\dagger l_{r,p_r}} } %
\eea
And again the  merge rule for $J$-antisymetric complex matrices is identical to the merge rule for real antisymmetric matrices eq.\eqref{mergeruleAsym}.

\medskip

We conclude that the expectation values of invariant polynomials of $M$ and $M^\dagger$ are entirely determined by the same recursion relations (on the degree) as the expectation values of invariant polynomials of two real antisymmetric matrices. This completes the proof of Theorem \ref{MainOrt}.

\subsection{Symplectic case}

The procedure to obtain the loop equations for the real quaternionic antiselfdual two-matrix integral and for the $\td{J}$-antisymmetric complex matrix integral and to prove Theorem \ref{MainSymp} is completely analogous to the one above.

\section{Calculation of Jacobians.}\label{app:JACS}

In this appendix we are going to detail the main steps for the computation 
of the Jacobians (\ref{eq:JacO}), (\ref{eq:JacJ}), (\ref{eq:JacQ}) and 
(\ref{eq:JactdJ}). For this purpose we will need one of the special 
limiting cases of the Selberg integral
called the Laguerre limit   (see for example \cite{Mehta}) 
\bea\label{eq:selberg}
I(\alpha,\gamma,n)&=&\int_{-\infty}^\infty\dots\int_{-\infty}^\infty \left(\prod_{1\le i<j\le n}(x_j^2-x_i^2)\right)^{2\gamma}\left(\prod_{k=1}^n x^{2\alpha-1}_k \e{-x_k^2}\right) \D x_1\dots \D x_n=\nonumber\\
&&\qquad\qquad\qquad\qquad=\prod_{j=0}^{n-1}\frac{\Gamma(1+\gamma+j\gamma)\Gamma(\alpha+j\gamma)}{\Gamma(1+\gamma)}\ .
\eea
The two values of this integral we need are
\bea
I\left(\frac{1}{2},1,m\right)=m!\frac{\left(\sqrt{\pi}\right)^m}{2^{m(m-1)}}\prod_{j=1}^{m-1}(2j)!\\
I\left(\frac{3}{2},1,m\right)=m!\frac{\left(\sqrt{\pi}\right)^m}{2^{m^2}}\prod_{j=1}^m(2j-1)!
\eea
The procedure is essentially the same for the four cases. Let us show 
in detail the first one, $\mathrm{Jac}^{O}_n$.

The following relation holds true by the block-diagonal decomposition shown in section \ref{SEC1}
\bea
\int_{{\cal A}_n}\D A \, \e{\Tr{\frac{A^2}{2}}}&=&\mathrm{Jac}^O_n\int_{-\infty}^\infty
\prod_{1\le i<j\le m}(x_j^2-x_i^2)^2
\left\{\begin{array}{cl}
\prod_{k=1}^m\e{-x_k^2} \D x_k & \textrm{if $n=2m$} \\
\prod_{k=1}^m x_k^2 \e{-x_k^2} \D x_k & \textrm{if $n=2m+1$}
\end{array}\right.\nonumber \\
&=&\mathrm{Jac}_n^O\left\{\begin{array}{cl}
I\left(\frac{1}{2},1,m\right) & \textrm{if $n=2m$} \\
I\left(\frac{3}{2},1,m\right) & \textrm{if $n=2m+1$\ .}
\end{array}\right.
\eea
Computing the Gaussian integral on the left hand side we find
\bea
\mathrm{Jac}^O_n &=& \left(\sqrt{\pi}\right)^{\frac{n(n-1)}{2}}
\left\{\begin{array}{cl}
(I\left(\frac{1}{2},1,m\right))^{-1} & \textrm{if $n=2m$}\\
(I\left(\frac{3}{2},1,m\right))^{-1} & \textrm{if $n=2m+1$}
\end{array}\right.\nonumber
\eea
which gives exactly the expression in equation \ref{eq:JacO}.

The Jacobian $\mathrm{Jac}_{2m}^{Sp}$ is computed with the same technique 
from a real quaternionic antiselfdual Gaussian integral,
\bea
\left(\frac{1}{2}\right)^{m(m-1)}\left(\sqrt{\pi}\right)^{m(2m+1)}=
\int_{Q{\cal A}_m} \D Q \e{\Tro{\frac{Q^2}{2}}}= \mathrm{Jac}^{Sp}_{2m}\, I\left(\frac{3}{2},1,m\right).
\eea

In order to compute the two remaining Jacobians $\mathrm{Jac}^{U^J}_n$ and $\mathrm{Jac}^{U^{\td{J}}}_{2m}$ we need to introduce two new matrix ensembles. 
Consider the {\it Hermitean} $J/\td{J}$-antisymmetric one-matrix model. By the $J/\td{J}$-antisymmetry of these matrices we know that they can be put into a triangular form by a twisted orthogonal or symplectic matrix respectively. By hermiticity we conclude that the triangular part of this Schur form will be zero, and the diagonal part (eigenvalues) is real. It is easy to argue that the Jacobians for these transformations have to be the same as the ones we seek in sections \ref{SEC2} and \ref{sec:SchurtdJ}. This allows us to write the following 
\bea
\int\D H \, \e{-\Tr{\frac{H^2}{2}}}&=&\left\{\begin{array}{cl}
\left(\sqrt{\pi}\right)^{\frac{n(n-1)}{2}}2^{m-m^2} & \textrm{if ${\cal J}=J$} \\
\left(\sqrt{\pi}\right)^{2m^2+m}2^{m-m^2} \textrm{if ${\cal J}=\td{J}$}
\end{array}\right.
\nonumber\\
&=&\mathrm{Jac}^{U^{\cal{J}}}_n 
\left\{\begin{array}{cl}
I\left(\frac{1}{2},1,m\right) & \textrm{if ${\cal J}=J$ and $n=2m$}\\
I\left(\frac{3}{2},1,m\right) & \textrm{if ${\cal J}=J$ and $n=2m+1$}\\
2^{2m}I\left(\frac{3}{2},1,m\right) & \textrm{if ${\cal J}=\td{J}$ and $n=2m$ }
\end{array} \right.
\eea
which gives \eq{eq:JacJ} and \eq{eq:JactdJ2}.

\section{Proof of theorem \ref{the:matrixM}}\label{app:matrixM}

In this appendix we use the graphical representation of the basis
 of correlation functions introduced  in section \ref{sssec:Basis} 
to prove theorem \ref{the:matrixM}.
The idea is to identify all possible occurrences of elements 
of the first row and last column of the $T$ matrix (and vice versa
for $T^\dagger$) by means of the decomposition of \eq{expanres};
then to use the constraints coming from (i) the triangular structure 
of these matrices, (ii) %
the contractions of indices within traces, 
(iii) the propagators \eq{propagators} and \eq{prop}, to represent the 
result of the integration in a graphical way, leading to the recursion 
formulae. 


\subsection{Last column/first row integration}

Take the functions defined in \ref{sssec:Basis} and their
graphical representation.
We first rewrite \eq{expanres} in a slightly reshuffled form 
\bea\label{EXP}
\left(\frac{1}{\x-(\X+T)}\right)_{i,j}&=&
\delta_{i,1}\delta_{j,n}\frac{1}{\x-\alpha}T_{1,n}\frac{1}{\x+\alpha}
\nonumber\\  %
&&+\delta_{i,1}\delta_{j,1}\frac{1}{\x-\alpha}+\delta_{i,n}\delta_{j,n}\frac{1}{\x+\alpha}\\   %
%
&&+\Bigg\{{\delta_{i,1}(1-\delta_{j,1}-\delta_{j,n})\frac{1}{\x-\alpha}} 
{\left[\sum_{k=2}^jT_{1,k} \left(\frac{1}{\x-(\td{\X}+\td{T})}\right)_{k,j}\right]}  
\nonumber\\  
&&+(1-\delta_{i,1}-\delta_{i,n})\delta_{j,n}\left[ \sum_{l=i}^{n-1}\left(\frac{1}{\x-(\td{\X}+\td{T})}\right)_{i,l}T_{l,n} \right] \frac{1}{\x+\alpha} \nonumber \\ 
&&+\delta_{i,1}\delta_{j,n}\frac{1}{\x-\alpha} \left[\sum_{\begin{subarray}{c}k<l \\ =2\end{subarray}}^{n-1} 
T_{1,k}\left(\frac{1}{\x-(\td{\X}-\td{T})}\right)_{k,l}T_{l,n}\right] \frac{1}{\x+\alpha}\Bigg\}\nonumber\\ %
&&+(1-\delta_{i,1}-\delta_{i,n})(1-\delta_{j,1}-\delta_{j,n})\left(\frac{1}{\x-(\td{\X}+\td{T})}\right)_{i,j} \nonumber  %
\eea
where $\alpha$ and $\beta$ are the first eigenvalues of $\X$ and $\Y$ 
respectively.
We substitute this expression for each resolvent in the integrand, 
and perform all possible ``contractions'' of the $T_{1,k}$, $T_{1,n}$ 
and $T_{k,n}$ variables by means of the propagators (\ref{propagators}). 
This can be represented as operations on the diagram associated to the 
given function.
Note that the terms in the  last four lines  of (\ref{EXP})
still contain a resolvent (of size $n-2$), while those on the first 
two lines do not. 
 Let us now enumerate the operations corresponding to each 
term in the expansion \eq{EXP}:
\begin{itemize}
\item{{\it Operation 1:} The  term on the first line, which
singles out one $T_{1n}$ variable,   
removes one resolvent from the integrand, which is represented 
by erasing a dot in the diagram. Since $T_{1,n}$ can only be contracted 
with $T^\dagger_{n,1}$,  
the appearance of this term forces the erasing of a dot of the opposite
color, by another application of Operation 1 on a $y$-type resolvent, 
somewhere in the diagram. 
Since this $T_{1n}$ appears in a trace, its left and right
neighbouring resolvents %
must have a $T^\dagger$  with one matching  index 1 or $n$.
The operation of erasing dots leaves pairs of {\it free links}
with only one dot at their end, carrying such a  $T^\dagger$ variable; 
their role and their weight will be reconsidered in {\it Operation 3}. 
 The same applies to the other erased dot.  
Let us now perform the contraction of the selected 
$T_{1,n}T^\dagger_{n,1}$ pair,
giving a factor $(1+b)$. The graphical representation is,
\bea
\left(\begin{array}{c}\includegraphics[scale=0.6]{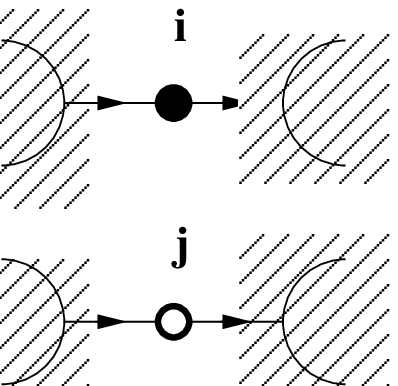}\end{array}\right)
&\Rightarrow&(1+b)\frac{1}{\x_i^2-\alpha^2}\frac{1}{\y_j^2-\beta^2}
\left(\begin{array}{c}\includegraphics[scale=0.6]{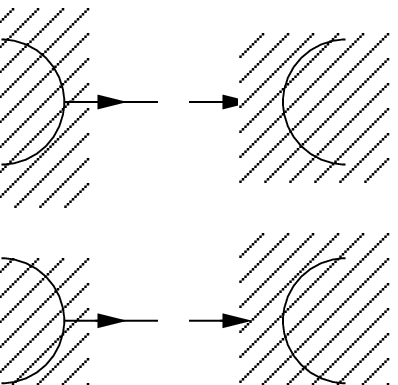}\end{array}\right)\nonumber
\eea
These pairings have to be performed in all inequivalent ways.}
\item{{\it Operation 2:} The two terms in the second line play a similar role.
 They also remove a resolvent, which is again represented by erasing a dot. 
This forces one of the neighbors to be replaced by a similar term. This will be represented by the operation of erasing a link and its two adjacent dots. The possible configurations are
\bea
\begin{array}{c}\includegraphics[scale=0.6]{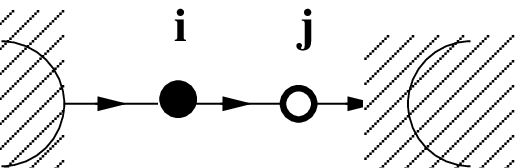}\end{array}
&\Rightarrow&\frac{1}{\x_i-\alpha}\frac{1}{\y_j-\beta}
\begin{array}{c}\includegraphics[scale=0.6]{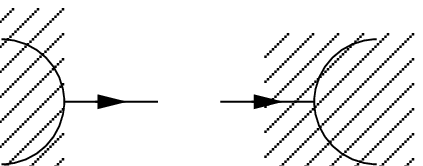}\end{array}\nonumber\\
\begin{array}{c}\includegraphics[scale=0.6]{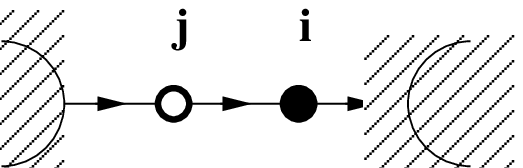}\end{array}
&\Rightarrow&\frac{1}{\x_i+\alpha}\frac{1}{\y_j+\beta}
\begin{array}{c}\includegraphics[scale=0.6]{graph2b.eps}\end{array}\nonumber\\
\begin{array}{c}\includegraphics[scale=0.6]{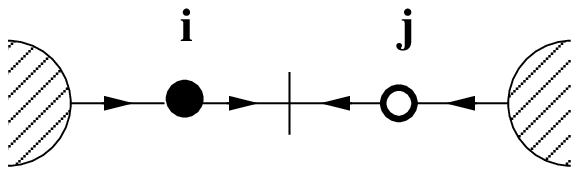}\end{array}
&\Rightarrow&-\frac{1}{\x_i-\alpha}\frac{1}{\y_j+\beta}
\begin{array}{c}\includegraphics[scale=0.6]{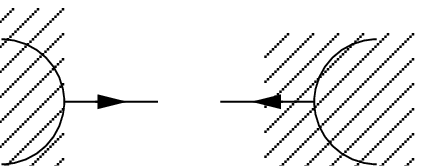}\end{array}\nonumber\\
\begin{array}{c}\includegraphics[scale=0.6]{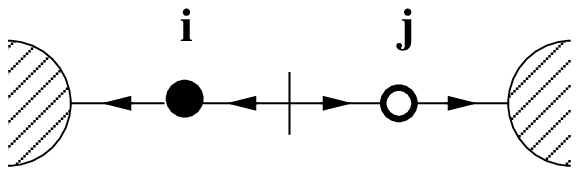}\end{array}
&\Rightarrow&-\frac{1}{\x_i+\alpha}\frac{1}{\y_j-\beta}
\begin{array}{c}\includegraphics[scale=0.6]{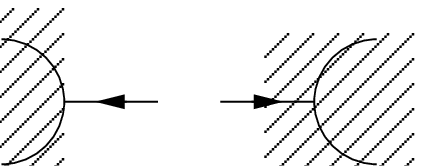}\end{array}.\nonumber
\eea
The signs in the last two equations come %
from the twist of the erased link. 
This operation also leaves some free links.
}

\item{{\it Operation 3:} All the remaining terms do not remove resolvents 
so they do not erase any dot in the diagram. Instead, they represent cuts 
in the links, since each $T_{1,k}$ and $T_{k,n}$ forces a $T^\dagger_{n,l}$ or $T^\dagger_{1,l}$ in a neighbor. 
We must also consider here  all free links created by erasing 
dots in {\it Operations 1} and {\it 2}. As discussed above, 
 these terms contain also a $T$ 
or a $T^\dagger$ variable at the end of the free link,  
and will contribute to the weight. Graphically we have
\bea
\begin{array}{c}\includegraphics[scale=0.6]{graph2.1.eps}\end{array}
&\Rightarrow&\frac{1}{\x_i+\alpha}\frac{1}{\y_j+\beta}
\begin{array}{c}\includegraphics[scale=0.6]{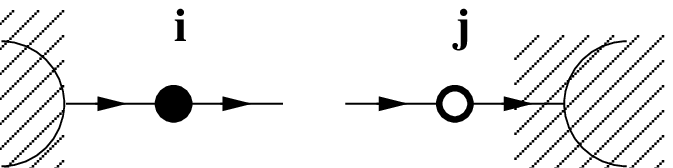}\end{array}\nonumber\\
\begin{array}{c}\includegraphics[scale=0.6]{graph2.2.eps}\end{array}
&\Rightarrow&\frac{1}{\x_i-\alpha}\frac{1}{\y_j-\beta}
\begin{array}{c}\includegraphics[scale=0.6]{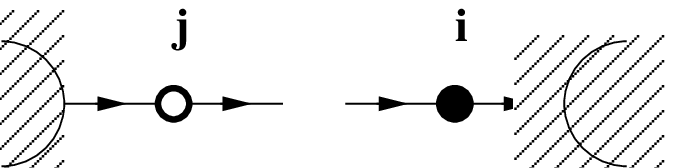}\end{array}\nonumber\\
\begin{array}{c}\includegraphics[scale=0.6]{graph2.3.eps}\end{array}
&\Rightarrow&-\frac{1}{\x_i+\alpha}\frac{1}{\y_j-\beta}
\begin{array}{c}\includegraphics[scale=0.6]{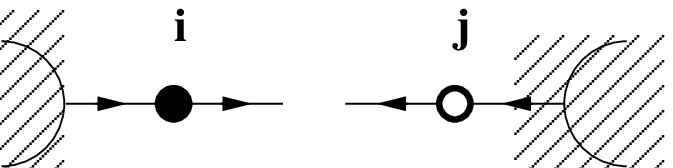}\end{array}\nonumber\\
\begin{array}{c}\includegraphics[scale=0.6]{graph2.4.eps}\end{array}
&\Rightarrow&-\frac{1}{\x_i-\alpha}\frac{1}{\y_j+\beta}
\begin{array}{c}\includegraphics[scale=0.6]{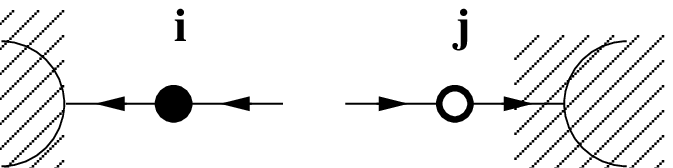}\end{array}\nonumber
\eea
for the cutting, and
\bea
\frac{1}{\x_i+\alpha}\begin{array}{c}\includegraphics[scale=0.6]{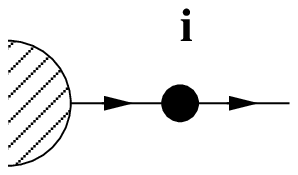}\end{array}&\;&\quad
-\frac{1}{\x_i+\alpha}\begin{array}{c}\includegraphics[scale=0.6]{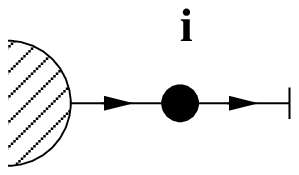}\end{array}\nonumber\\
\frac{1}{\x_i-\alpha}\begin{array}{c}\includegraphics[scale=0.6]{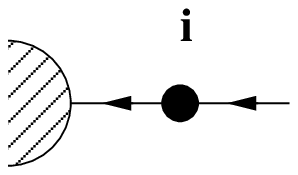}\end{array}&\;&\quad
-\frac{1}{\x_i-\alpha}\begin{array}{c}\includegraphics[scale=0.6]{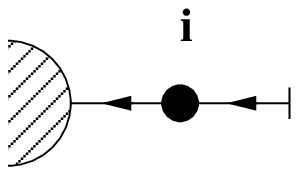}\end{array}\nonumber\\
\frac{1}{\y_j-\beta}\begin{array}{c}\includegraphics[scale=0.6]{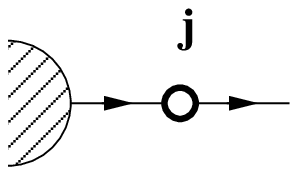}\end{array}&\;&\quad
-\frac{1}{\y_j-\beta}\begin{array}{c}\includegraphics[scale=0.6]{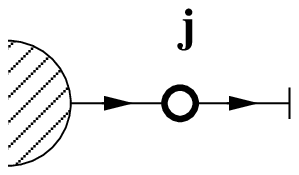}\end{array}\nonumber\\
\frac{1}{\y_j-\beta}\begin{array}{c}\includegraphics[scale=0.6]{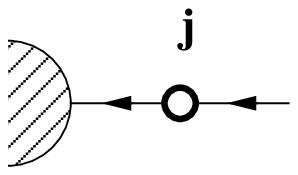}\end{array}&\;&\quad
-\frac{1}{\y_j-\beta}\begin{array}{c}\includegraphics[scale=0.6]{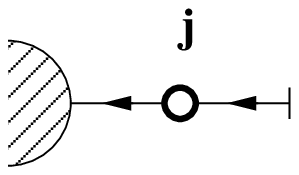}\end{array}\nonumber
\eea
for the free links coming from {\it Operations 1} and {\it 2}. In the
right column, the bar across the free link indicates the presence of
a ${\cal J}$ matrix.}

\item{{\it Operation 4:} Finally, the only term we did not consider (the one with one resolvent and no $T$ variable on the last line of (\ref{EXP}))
accounts for doing nothing to a dot.}
\end{itemize}

Substituting \eq{EXP} for each resolvent  is equivalent to performing 
Operations $1$ to $4$ on all dots/links and in all possible ways. 
After this we have diagrams with free links and missing dots. The final step 
is to join the remaining free links. This is equivalent to contracting 
the $T$ and $T^\dagger$ variables 
in all possible ways. The gluing of free links gives a trivial weight, 
so this final step is just graphical.

Let us illustrate this procedure on the example treated in subsection
\ref{morozo}.
The following equation represents the application of Operations 1 to 4 in all possible ways, with their corresponding weights.
\bea
\left(\begin{array}{c}\includegraphics[scale=0.6]{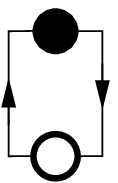}\end{array}+1\right)&=&
1+\frac{1}{x-\alpha}\frac{1}{y-\beta}
\begin{array}{c}\includegraphics[scale=0.6]{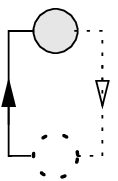}\end{array}
+\frac{1}{x+\alpha}\frac{1}{y+\beta}
\begin{array}{c}\includegraphics[scale=0.6]{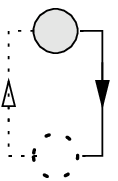}\end{array}\nonumber\\
&&+(1+b)\frac{1}{x^2-\alpha^2}\frac{1}{y^2-\beta^2}
\begin{array}{c}\includegraphics[scale=0.6]{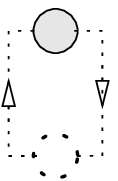}\end{array}
+\begin{array}{c}\includegraphics[scale=0.6]{examdiagr1.eps}\end{array}\nonumber\\
&&+\frac{1}{x-\alpha}\frac{1}{y-\beta}
\begin{array}{c}\includegraphics[scale=0.6]{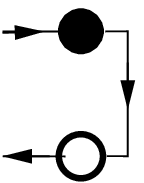}\end{array}
+\frac{1}{x+\alpha}\frac{1}{y+\beta}
\begin{array}{c}\includegraphics[scale=0.6]{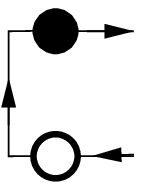}\end{array}\nonumber\\
&&+\frac{1}{x^2-\alpha^2}\frac{1}{y^2-\beta^2}
\begin{array}{c}\includegraphics[scale=0.6]{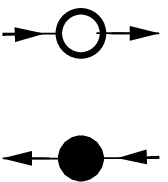}\end{array}
\eea
In this equation shaded dots and links represent erased dots and links. Notice that this intermediate formula can be identified term by term with formula \eqref{intstep}.
Gluing of the free links gives
\bea
\left(\begin{array}{c}\includegraphics[scale=0.6]{examdiagr1.eps}\end{array}+1\right)&=&
1+\frac{1}{x-\alpha}\frac{1}{y-\beta}+\frac{1}{x+\alpha}\frac{1}{y+\beta}
+(1+b)\frac{1}{x^2-\alpha^2}\frac{1}{y^2-\beta^2}\nonumber\\
&&\left(1+\frac{1}{x-\alpha}\frac{1}{y-\beta}+\frac{1}{x+\alpha}\frac{1}{y+\beta}\right)
\begin{array}{c}\includegraphics[scale=0.6]{examdiagr1.eps}\end{array}\nonumber\\
&&+\frac{1}{x^2-\alpha^2}\frac{1}{y^2-\beta^2}
\left(\begin{array}{c}\includegraphics[scale=0.6]{examdiagr1.eps}\end{array}+
\begin{array}{c}\includegraphics[scale=0.6]{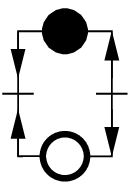}\end{array}\right)\nonumber\\
&=&\left(1+\frac{1}{x-\alpha}\frac{1}{y-\beta}\right)\left(1+\frac{1}{x+\alpha}\frac{1}{y+\beta}\right)
\left(\begin{array}{c}\includegraphics[scale=0.6]{examdiagr1.eps}\end{array}+1\right)\nonumber\\
&&+\frac{1}{x^2-\alpha^2}\frac{1}{y^2-\beta^2}
\left(\begin{array}{c}\includegraphics[scale=0.6]{examdiagr1bis.eps}\end{array}+b\right)\nonumber
\eea
which matches exactly the result \eqref{eq:comp1} in section \ref{morozo} but calculated here using the graphical rules we have defined.

\subsection{Computation of the weight for the final diagrams.}

Consider a reduced problem where no erasing of dots is allowed, %
{\it i.e.} only {\it Operation 3} and {\it 4} are taken into account. In this case only cutting and gluing is allowed and %
no difference at the graphical level appears between the 
$J$-antisymmetric and the $\td{J}$-antisymmetric cases.

To each link in the final diagram  $G^\prime$
is attached a weight coming from the different ways 
we  obtain it from the original diagram $G$.
That is, when a link in $G^\prime$ %
is part of $G$, %
we can either cut that original link and glue it again, 
or just do nothing. Instead, if the link in $G^\prime$ does not belong to 
$G$, the only way to obtain it is by gluing cut links. 
Both contributions will add up to 
\bea
\begin{array}{c}\includegraphics[scale=0.6]{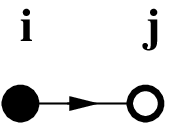}\end{array}\in G&\to&
\left\{\begin{array}{ll} \left(1+\frac{1}{\x_i+\alpha}\frac{1}{\y_j+\beta}\right) & \textrm{if the link } \in G^\prime \\
\frac{1}{\x_i+\alpha}\frac{1}{\y_j+\beta} & \textrm{if the link } \not\in G^\prime \end{array}\right.
\nonumber\\
\begin{array}{c}\includegraphics[scale=0.6]{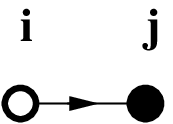}\end{array}\in G&\to&
\left\{\begin{array}{ll} \left(1+\frac{1}{\x_i-\alpha}\frac{1}{\y_j-\beta}\right) & \textrm{if the link } \in G^\prime \\
\frac{1}{\x_i-\alpha}\frac{1}{\y_j-\beta} & \textrm{if the link } \not\in G^\prime \end{array}\right.
\nonumber\\
\begin{array}{c}\includegraphics[scale=0.6]{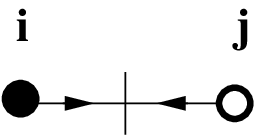}\end{array}\in G&\to&
\left\{\begin{array}{ll} \left(1-\frac{1}{\x_i+\alpha}\frac{1}{\y_j-\beta}\right) & \textrm{if the link } \in G^\prime \\
-\frac{1}{\x_i+\alpha}\frac{1}{\y_j-\beta} & \textrm{if the link } \not\in G^\prime \end{array}\right.
\nonumber\\
\begin{array}{c}\includegraphics[scale=0.6]{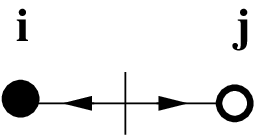}\end{array}\in G&\to&
\left\{\begin{array}{ll} \left(1-\frac{1}{\x_i-\alpha}\frac{1}{\y_j+\beta}\right) & \textrm{if the link } \in G^\prime \\
-\frac{1}{\x_i-\alpha}\frac{1}{\y_j+\beta} & \textrm{if the link } \not\in G^\prime \end{array}\right.
\nonumber
\eea

Returning now to the original complete problem, 
where the erasing of dots is allowed, %
one notices that the weight found in Operations 1 and/or 2 %
by erasing a pair of dots is the same as the one %
obtained by forming a cycle with this pair of dots with Operations 3 and 4. 
Indeed, consider  a minimal cycle 
({\it i.e.} a cycle of length 2) %
in $G^\prime$ and erase from $G$ the dots in this minimal cycle using 
{\it Operation 1} and if possible {\it Operation 2}. 
 {\it Operation 1} contributes a factor $(1+b)$; if the minimal cycle in 
$G^\prime$  is non twisted, we assign it the weight 1, while if it is, we assign 
it the weight $b$\footnote{This is the origin of the additive $\pm 1$'s coming with every minimal cycle in the basis of functions \eq{BASIS1}.}.  
Iterating this operation for all minimal cycles in $G^\prime$, 
one finds that the two procedures, erasing pairs of dots
or forming minimal cycles with the same dots, produce the same weight.
A slightly different manipulation is needed when the minimal cycle is present already in $G$. The outcome will be the same.

Here is an example of the kind of diagrams %
which have the same weight within the complete problem:
\bea
\ovl{\cal M}\left(
\begin{array}{c}
\includegraphics[scale=0.3]{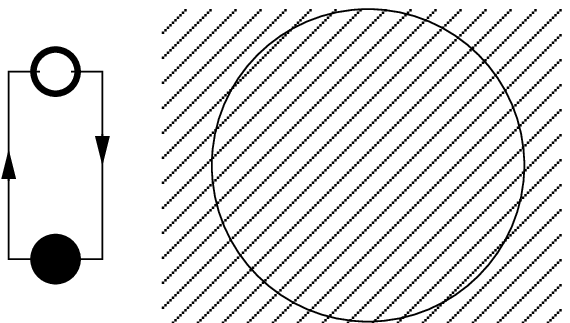}
\end{array}\right)\times
\left(\begin{array}{c}
\includegraphics[scale=0.5]{graph7.1.eps}
\end{array}+
\begin{array}{c}
\includegraphics[scale=0.5]{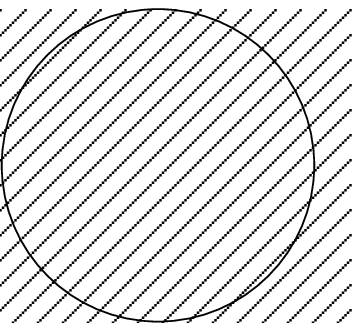}
\end{array}\right)\nonumber\\
+%
\ovl{\cal M}\left(
\begin{array}{c}
\includegraphics[scale=0.3]{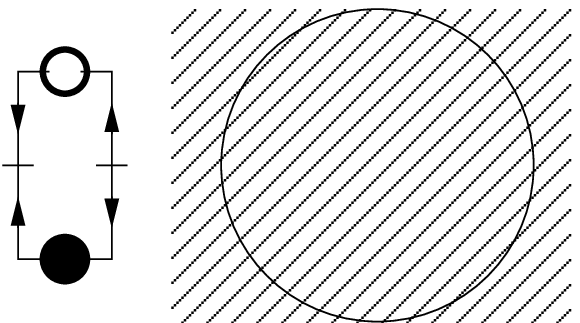}
\end{array}\right)\times
\left(\begin{array}{c}
\includegraphics[scale=0.5]{graph7.2.eps}
\end{array}+b
\begin{array}{c}
\includegraphics[scale=0.5]{graph7.3.eps}
\end{array}\right)\nonumber
\eea 
where $\ovl{\cal M}(G^\prime)\equiv\ovl{\cal M}_G^{G^\prime}$ 
is the weight associated with $G^\prime$, for  $G$ the original diagram. 

We proved that for the complete problem, the recursion matrix is the same as that of the reduced problem.
Reassembling everything together and using back the sets of signs $s$ and $s^\prime$, we get the recursion matrix
\bea
\ovl{\cal M}_{G}^{G^\prime}&=&\Bigg(\prod_{((\x_i,s(i)),(\x_{\pi(i)},s^\prime(\pi(i))))\in G'}
\left(1+\frac{1}{s(i) \x_i+\alpha}\frac{1}{s^\prime(\pi(i)) \y_{\pi(i)}+\beta}\right)\nonumber\\
&&\prod_{((\x_i,s(i)),(\y_{\pi(i)},s^\prime(\pi(i))))\not\in G'}
\left(\frac{1}{s(i) \x_i+\alpha}\frac{1}{s^\prime(\pi(i)) \y_{\pi(i)}+\beta}\right)\Bigg)\nonumber\\
&&\Bigg(\prod_{((\x_i,s(i)),(\y_{\pi^\prime(i)},s^\prime(\pi^\prime(i))))\in G'}
\left(1+\frac{1}{s(i) \x_i-\alpha}\frac{1}{s^\prime(\pi^\prime(i)) \y_{\pi^\prime(i)}-\beta}\right)\nonumber\\
&&\prod_{((\x_i,s(i)),(\y_{\pi^\prime(i)},s^\prime(\pi^\prime(i))))\not\in G'}
\left(\frac{1}{s(i) \x_i-\alpha}\frac{1}{s^\prime(\pi^\prime(i)) \y_{\pi^\prime(i)}-\beta}\right)\Bigg).\nonumber
\eea
Labelling each diagram using the labels in \eq{BASIS1}, 
\eq{BASIS2} and \eq{BASIS3} we get
\bea
\ovl{\cal M}_{\{(\pi,s),(\pi^\prime,s^\prime)\}}^{\{(\tau,t),(\tau^\prime,t^\prime)\}}&=&
\left(\prod_{i=1}^R(\delta_{\pi(i),\tau(i)}\delta_{s(i),t(i)}\delta_{s^\prime(\pi(i)),t^\prime(\pi(i))}
+\frac{1}{s(i) \x_i+\alpha}\frac{1}{s^\prime(\pi(i)) \y_{\pi(i)}+\beta})\right)\nonumber\\
&&\left(\prod_{i=1}^R(\delta_{\pi^\prime(i),\tau^\prime(i)}\delta_{s(i),t(i)}\delta_{s^\prime(\pi^\prime(i)),t^\prime(\pi^\prime(i))}
+\frac{1}{s(i) \x_i-\alpha}\frac{1}{s^\prime(\pi^\prime(i)) \y_{\pi^\prime(i)}-\beta})\right)
\eea

which completes the proof.

\section{Relations between orthogonal/symplectic and unitary recursion equations}
\label{app:bijection}
In this appendix we relate 
tetrads $\omega=\{\sigma,\tau,s,t\}$ introduced in sect. 
\ref{sssec:Basis} and permutations $\pi\in \SS_{2R}$, 
and more precisely to show the bijection between
the set of {\it equivalence classes} $[\omega]$  and $\SS_{2R}$; 
this leads to an important relation between the recursion matrix $\ovl{\cal M}$ and the basis of correlation functions $F^{\cal J}$ for the 
orthogonal/symplectic case and the recursion matrix ${\cal M}$ 
\bea
{\cal M}^{(2R)}_{\pi,\pi^\prime}(\{\x\},\{\y\},\alpha,\beta)&=&\prod_{i=1}^{2R}\left(\delta_{\pi(i),\pi^\prime(i)}
+\frac{1}{\x_i-\alpha}\frac{1}{\y_{\pi(i)}-\beta}\right)
\eea
and the basis of correlation functions $F^{U}$ 
\bea
F^{U}_{\pi,\pi^\prime}(\{\x\},\{\y\},A,B)&=&
\prod_{k=1}^p\left(\delta_{R_k,1}+\tr{\prod_{l=1}^{R_k}\frac{1}{\x_{i_{k,l}}-A}\frac{1}{\y_{j_{k,l}}-B}}\right)
\eea
found in \I\ in the unitary case. 

\subsection{Bijection between \texorpdfstring{$\SS_{2R}$}{PERM(2R)} and equivalence clases in \texorpdfstring{$\SS_{R}\times\SS_{R}\times\Z_2^R\times\Z_2^R$}{PERM(2R)xPERM(2R)xSIGN(2R)xSIGN(2R)}}

Consider %
pairs of permutations $\sigma$ and $\tau$ belonging to $\SS_R$, 
pairs of sets of $R$ signs $s$ and $t$ belonging to $\Z_2^R$. As explained
in sect. \ref{sssec:Basis}, $\sigma \circ\tau^{-1}$ represents 
a permutation of $R$ (black) points with signs $s(i)$ attached to them,  
and $\tau\circ\sigma^{-1}$ a permutation of $R$ (white) points 
with signs $t(i)$. To get one representative of the equivalence 
class $[\{\sigma,\tau,s,t\}]$, we fix one sign $s_i$ in every 
cycle of $\sigma \circ\tau^{-1}$ to be $+1$. 

\begin{lemma}
There is a bijection between $\SS_{2R}$ and the equivalence classes of $\SS_{R}\times\SS_{R}\times\Z_2^R\times\Z_2^R$.
\end{lemma}

\proof{To construct the bijection we take a permutation 
$\pi\in\SS_{2R}$. We will relabel the indices $i=1,\dots,2R$ and 
call them $\alpha(i)=1,\dots,R,-R,\dots,-1$ (see figure \ref{Bijection}),
\begin{displaymath}
\alpha(i)=
\left\{
\begin{array}{ll}
i\qquad & \mathrm{if }\,\,i\leq R\\
i-(2R+1) & \mathrm{if }\,\,i>R
\end{array}
\right. .
\end{displaymath}
Define also the auxiliary ``rainbow'' permutation $e(i)=2R-i$ for which $e^2=\mathrm{id}$ and $e(\alpha(i))\equiv\alpha(e(i))=-\alpha(i)$.

\begin{figure}[ht!]\lbl{Bijection}
\centering
\includegraphics[scale=1]{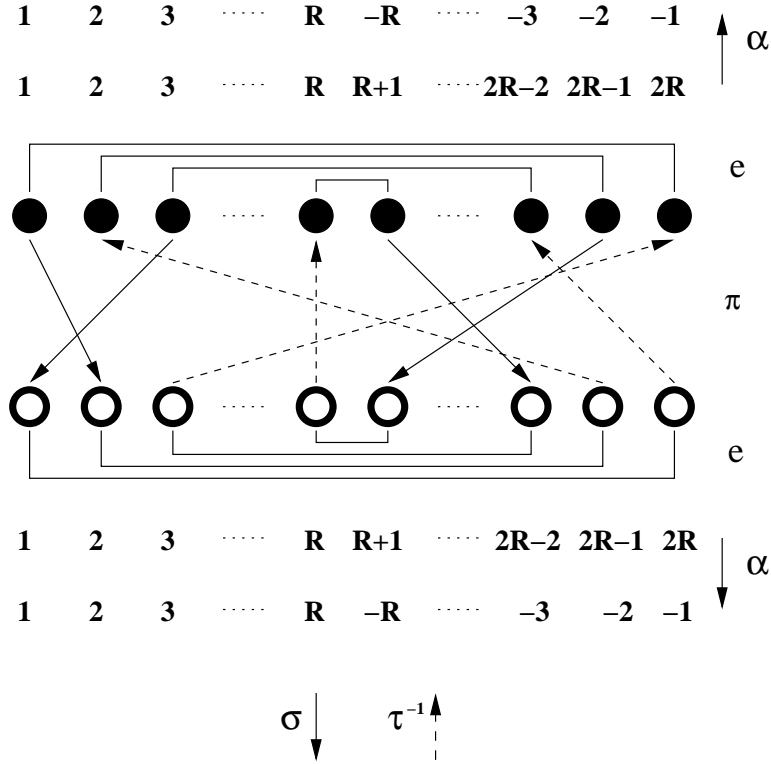}
\caption{Representation of the bijection. 
The set of lines, irrespective of their type (solid or dashed) 
represents $\pi\in\SS_{2R}$. %
Solid lines represent $\sigma$ and dashed ones represent $\tau$. 
The signs $s$, resp $t$,  are %
the signs at the origin, resp.the end, of the solid lines. 
The arrows, though redundant, are meant to help the reader follow the iterations 1-4 above.}
\end{figure}

With this new labelling we construct the permutations $\sigma$ and $\tau$ and the signs $s$ and $t$ as follows %
\begin{itemize}
\item{1.} {Begin with $i=1$.}
\item{2.} {Set $s(|\alpha(i)|)=\sgn(\alpha(i))$, then $j:=\pi(i)$ and set $\sigma(|\alpha(i)|)=|\alpha(j)|$}
\item{3.} {Set $t(|\alpha(j)|)=\sgn(\alpha(j))$, then $k=e\pi^{-1}e(j)$ and set $\tau^{-1}(|\alpha(j)|)=|\alpha(k)|$}
\item{4.} {If we do not close a cycle set $i=k$ and go back to 2.}
\end{itemize}
When we close a cycle of $\sigma \circ\tau^{-1}$, (for example, to close the first cycle we must find again $k=1$ at the end of step 3.), 
we must open a new cycle. To do so, look at which positive $\alpha$-type indices we have not used yet and choose the smallest one. Set $i$ equal to this 
value and restart from point 2. When there is no positive index left, 
the last cycle is completed and the tetrad $\{\sigma,\tau,s,t\}$ is constructed.
}

This procedure is illustrated in figure \ref{Bijection}: there, the permutation in $\SS_{2R}$ is $\pi=(1,2,R+2,2R,R+3,R+1,3)(R)(\cdots)$. Following the above rules we
determine from it  $\sigma=(1,2,R,3)(\cdots)$, $\tau=(1,3)(2)(R)(\cdots)$, 
$s=(+,-,+,\dots,-)$ and $t=(+,+,-,\dots,-)$. 

Note that the first $s$ sign of every cycle is positive by construction. Note 
also that 
conversely, constructing $\pi$ from the tetrad $\{\sigma,\tau,s,t\}$ can be done with the same kind of 
procedure: this is clearly seen in the example.  
By construction,  in this reverse operation,  $\pi$ depends only 
on the equivalence class $[\omega]$. 
Since the method is deterministic in both directions we have a bijection.


\subsection{Relation between \texorpdfstring{$\ovl{\cal M}^{(R)}$}{`bar-M'sup-R} and \texorpdfstring{${\cal M}^{(2R)}$}{M sup-R}}

Consider now the set of variables
\bea
\{\x\}_{2R}=&\{\x_1,\dots,\x_R,\x_{R+1}=-\x_R,\dots,\x_{2R}=-\x_1\}\,,&\quad \{\x\}_{R}=\{\x_1,\dots,\x_R\}\nonumber\\
\{\y\}_{2R}=&\{\y_1,\dots,\y_R,\y_{R+1}=-\y_R,\dots,\y_{2R}=-\y_1\}\,,&\quad \{\y\}_{R}=\{\y_1,\dots,\y_R\}\nonumber
\eea
and consider also the pairs of indices 
\bea
\pi,\,\pi^\prime&\in&\SS_{2R}\nonumber\\ 
\left[\{\sigma,\sigma^\prime,s,s'\}\right],
\left[\{\tau,\tau^\prime,t,t'\}\right]&\in&
\textrm{Equivalence Classes of}\,\,\SS_R\times\SS_R\times\Z_2^R\times\Z_2^R
\eea
where the indices are related through the bijection shown above. 
Using these definitions it is easy to verify that 
\bea
\ovl{\cal M}_{\phantom{(R)\,}\{\sigma,\sigma^\prime,s,s^\prime\}}^{(R)\,\{\tau,\tau^\prime,t,t^\prime\}} (\{\x\}_R,\{\y\}_R,\alpha,\beta)&=&
\left(\prod_{i=1}^R\left(\delta_{\sigma(i)\tau(i)}\delta_{s(i)t(i)}\delta_{s^\prime(\sigma(i))t^\prime(\sigma(i))}
+\frac{1}{s(i) \x_i+\alpha}\,\,\frac{1}{s^\prime(\sigma(i)) \y_{\sigma(i)}+\beta}\right)\right)\nonumber\\
&&\back\back\back\times\left(\prod_{i=1}^R\left(
\delta_{\sigma^\prime(i)\tau^\prime(i)}\delta_{s(i)t(i)}\delta_{s^\prime(\sigma^\prime(i))t^\prime(\sigma^\prime(i))}
+\frac{1}{s(i) \x_i-\alpha}\,\,\frac{1}{s^\prime(\sigma^\prime(i)) \y_{\sigma^\prime(i)}-\beta}\right)\right)\nonumber\\
&=&\left(\prod_{i=1}^{2R}\left(\delta_{\pi(i),\pi^\prime(i)}+\frac{1}{\x_i+\alpha}\frac{1}{\y_{\pi(i)}+\beta}\right)\right)\nonumber\\
&=&{\cal M}_{\pi,\pi^\prime}^{(2R)}(\{\x\}_{2R},\{\y\}_{2R},-\alpha,-\beta)\nonumber
\eea
{\it i.e.} the two  matrices encountered in the orthogonal/symplectic and
unitary cases are in fact the same. In this calculation, 
we have  used the bijection to reexpress the Kronecker $\delta$ symbols:
\bea
\delta_{\sigma(i),\tau(i)}\delta_{s(i),t(i)}\delta_{s^\prime(\sigma(i)),t^\prime(\sigma(i))}&=&
\delta_{\pi(\alpha^{-1}(s(i)i)),\pi^\prime(\alpha^{-1}(s(i)i))}\nonumber\\
\delta_{\sigma^\prime(i)\tau^\prime(i)}\delta_{s(i)t(i)}\delta_{s^\prime(\sigma^\prime(i))t^\prime(\sigma^\prime(i))}&=&
\delta_{\pi(\alpha^{-1}(-s(i)i)),\pi^\prime(\alpha^{-1}(-s(i)i))}\nonumber
\eea
and to relate the $\{\x\}_R$ and $\{\y\}_R$ variables with the $\{\x\}_{2R}$ and $\{\y\}_{2R}$ variables according to %
\beq
s(i) \x_i=\x_{\alpha^{-1}(s(i)i)}\,,\quad s^\prime(\sigma(i)) \y_{\sigma(i)}=\y_{\pi(\alpha^{-1}(s(i)i))}\,,\quad
-s^\prime(\sigma^\prime(i)) \y_{\sigma^\prime(i)}=\y_{\pi(\alpha^{-1}(-s(i)i))}.\nonumber
\eeq


\subsection{Relation between \texorpdfstring{$F^{\cal J}$}{F sup-J} and \texorpdfstring{$F^{U}$}{F sup-U}}

Finally consider the basis of correlation functions in the unitary case for $2R$ $\X$-type and $2R$ $\Y$-type resolvents. In particular consider the components $F_{\pi,e\pi e}^{U}(\{\x\}_{2R},\{\y\}_{2R},A,B)$. 
Call $[\omega]=[\{\sigma,\tau,s,t\}]$ %
the equivalence class of tetrads corresponding    to $\pi$.

The function $F_{\pi,e\pi e}^{U}(\{\x\}_{2R},\{\y\}_{2R},A,B)$ 
 can be constructed
from the kind of diagrams shown in figure \ref{Bijection} 
by following the $\pi e\pi^{-1}e$ cycles. These cycles are, by construction,
 the same ones we follow with $\sigma\tau^{-1}$, with the only difference 
that $\pi$ is a permutation of $2R$ elements instead of $R$. 
Because of this, $\pi e \pi^{-1} e$ contains two different representatives 
of the equivalence class $[\omega]$, and since the functions $F^{\cal J}$ 
are independent of the class representative, $F^{U}$ contains twice
the same function $F^{\cal J}$. We can write this as %
\bea
F_{\pi,e\pi e}^{U}(\{\x\}_{2R},\{\y\}_{2R},A,B)&=&
\left(F^{\cal J}_{\omega}(\{\x\}_R,\{\y\}_R,A,B)\right)^2\ .
\eea
The sign in \eq{BASIS3} is, however, not easy to read from $\pi$ and $e$.

\newpage


\end{document}